\documentclass [aps,
                prd,
                preprint,
                nofootinbib,
                superscriptaddress,
                floatfix,
                showkeys]
                {revtex4-2}

\usepackage[utf8]{inputenc}
\usepackage{amssymb}
\usepackage{amsmath}
\usepackage{graphicx}
\usepackage[export]{adjustbox}
\usepackage{hyperref}
\usepackage{caption}
\usepackage{subcaption}
\makeatletter
\newcommand\footnoteref[1]{\protected@xdef\@thefnmark{\ref{#1}}\@footnotemark}
\makeatother

\begin{document}

\title{A Numerical Rosenblatt Method for Forced Variable Independence}

\author{Radek Vav\v{r}i\v{c}ka}
\email{radek.vavricka@cern.ch}
\affiliation{Faculty of Mathematics and Physics, Charles University,
             Institute of Particle and Nuclear Physics,
             V~Hole\v{s}ovi\v{c}k\'{a}ch~2, 180\,00 Prague~8, Czech Republic}

\author{Tom\'a\v{s} S\'ykora}
\email{Tomas.Sykora@cern.ch}
\affiliation{Faculty of Mathematics and Physics, Charles University,
             Institute of Particle and Nuclear Physics,
             V~Hole\v{s}ovi\v{c}k\'{a}ch~2, 180\,00 Prague~8, Czech Republic}

\author{Matthias Schott}
\email{mschott@uni-bonn.de}
\affiliation{University of Bonn,
             Physikalisches Institut,
             Nußallee 12, 53115 Bonn}

\keywords{machine learning, classifier decorrelation, statistical independence,
          ABCD method, Rosenblatt transform, distance correlation,
          kernel density estimation, particle physics}

\begin{abstract}
Enforcing statistical independence between a predictor and a protected observable
is a common requirement in particle physics, fairness-aware machine learning,
medical imaging, and beyond.
We present a post-hoc numerical implementation of the Rosenblatt transform
that achieves this goal without retraining the original predictor.
Two complementary implementations are introduced:
Irregular Grid Interpolation (IRGI), based on recursive adaptive two-dimensional
binning, and Kernel Density Estimation (KDE), based on Gaussian kernel smoothing
of the conditional cumulative distribution function.
Both operate directly on a defining reference sample with no neural-network
training; for continuous densities the transform is exact, and in finite-sample
practice the residual distance correlation is small.
The method is validated on three examples of increasing specificity:
synthetic Gaussian distributions (controlled verification),
a CIFAR-10 image classification task (domain-agnostic demonstration), and
a realistic LHC dijet analysis using the LHC Olympics~2020 dataset (primary
particle-physics application).
In the latter, both IRGI and KDE outperform DisCo (distance correlation)
regularization in AUC preservation and training stability,
reducing the distance correlation coefficient by more than three orders of
magnitude relative to the untransformed classifier while maintaining
near-baseline discriminating power.
Utility for the ABCD data-driven background estimation method is confirmed
through a closure test achieving relative signal estimation errors of
order $10^{-6}$.
\end{abstract}

\maketitle

\section{Introduction}

\subsection{Background and Motivation}

Enforcing statistical independence between a predictor and a protected observable
is a recurring problem across many scientific and engineering domains.
In particle physics it is required for the ABCD data-driven background estimation
method described below.
In fairness-aware machine learning it arises as the requirement that a
decision-making algorithm be independent of sensitive demographic
attributes~\cite{Shimmin2017,Louppe2017}.
In medical imaging, diagnostic classifiers must be robust to nuisance variables
such as patient age or imaging hardware.
The same mathematical structure underlies all these applications: given a joint
distribution of a predictor $y$ and a protected observable $x$, transform $y$
to a new variable $\gamma$ that is statistically independent of $x$ for a
specified reference population.

This work presents a numerical implementation of the Rosenblatt
transform~\cite{Rosenblatt1952} that achieves this goal post-hoc —
without retraining the original predictor.
Two complementary implementations are introduced and compared against the
standard DisCo regularization approach~\cite{DisCoFever} in the particle physics
context, where the method is most thoroughly validated.
The treatment is, however, fully general, and the examples range from abstract
Gaussian distributions to image classification before arriving at the primary
high-energy physics application.

Statistical independence is quantified throughout using the Distance Correlation
Coefficient (DCC)~\cite{Szkely2007},\footnote{The DCC is the normalized
distance covariance, a measure of both linear and non-linear dependence that
vanishes if and only if the two variables are statistically independent\footnoteref {noteDCC}.}
taking values in $[0, 1]$ where $0$ indicates complete independence\footnote {\label {noteDCC}Save several pathological cases, where the DCC gives $0$ despite the condition of statistical independence not being met.}. For a two-dimensional sample represented by real-valued vectors $\left\lbrace \left( X_i, Y_i \right) \right\rbrace_{i=0}^{n}$, the distance covariance comes from double centered inter-element distances
\begin {equation}
\textnormal {dCov}^2 = \frac {1} {n^2} \sum_{i = 0} \sum_{j = 0} {\overline {\overline {||X_i - X_j||}}} \textnormal { } {\overline {\overline {||Y_i - Y_j||}}},
\end {equation}
where double centering of a two-dimensional array is defined from one- and two-dimensional averages as
\begin {equation}
\bar {\bar {a}}_{k, l} = a_ {k, l} - \bar {a}_ {k \cdot} - \bar {a}_ {\cdot l} + \bar {a}_ {\cdot \cdot}.
\end {equation}
Two variables $x$ and $y$ are considered quasi-independent when
$\mathrm{DCC}(x,y) \ll 1$.

\subsection{The ABCD Method and Statistical Requirements}

\begin{figure} [ht]
\begin {minipage} {0.55\textwidth}
\centering
\includegraphics [width=0.99\textwidth]{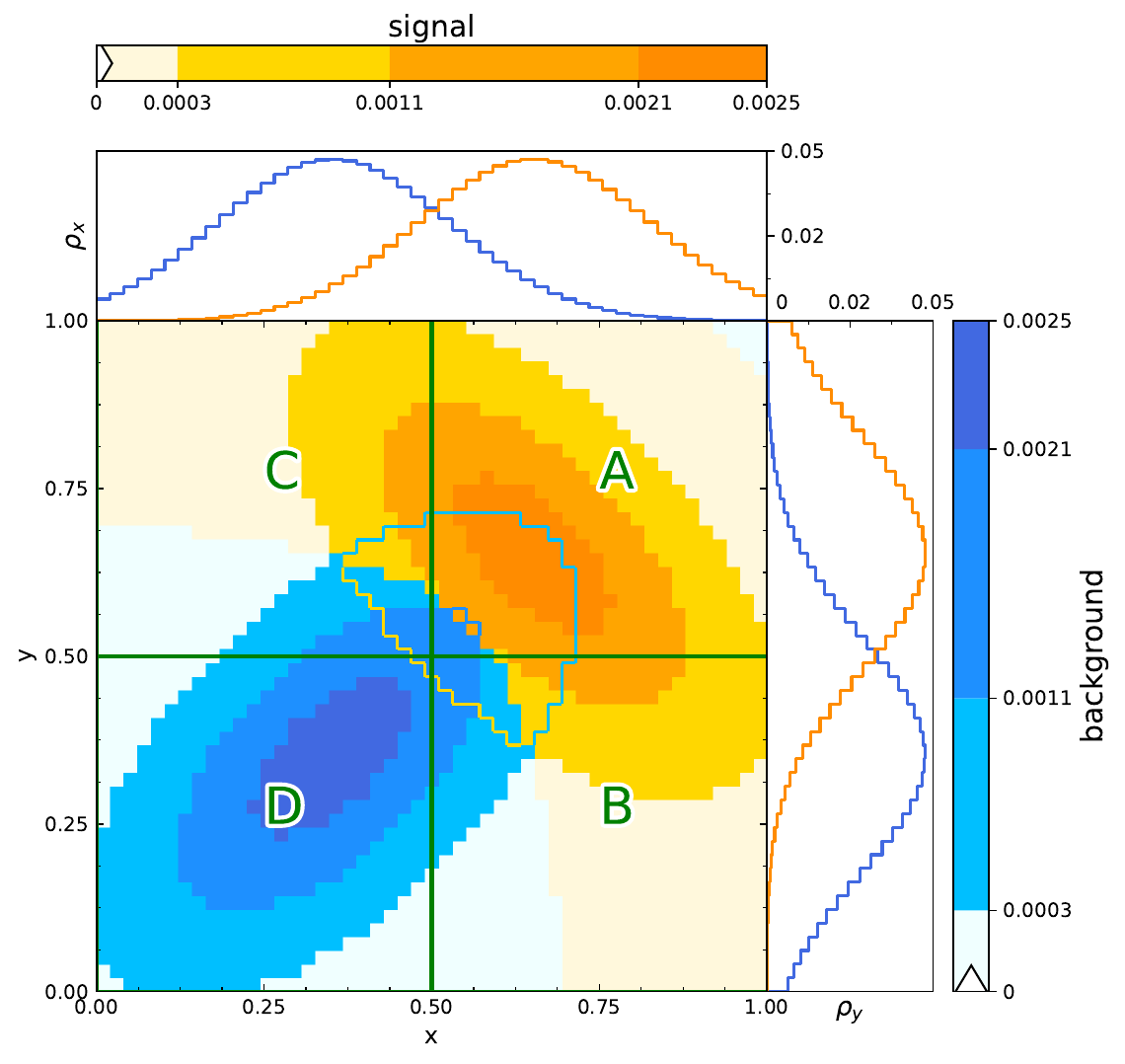}
\caption{Events in the $xy$ phase plane separated into regions ABCD.}
\label{fig:ABCD}
\end {minipage}
\begin {minipage} {0.44\textwidth}
\centering
    \begin{subfigure}[t]{0.99\textwidth}
    \centering
    \includegraphics[width=0.99\textwidth]{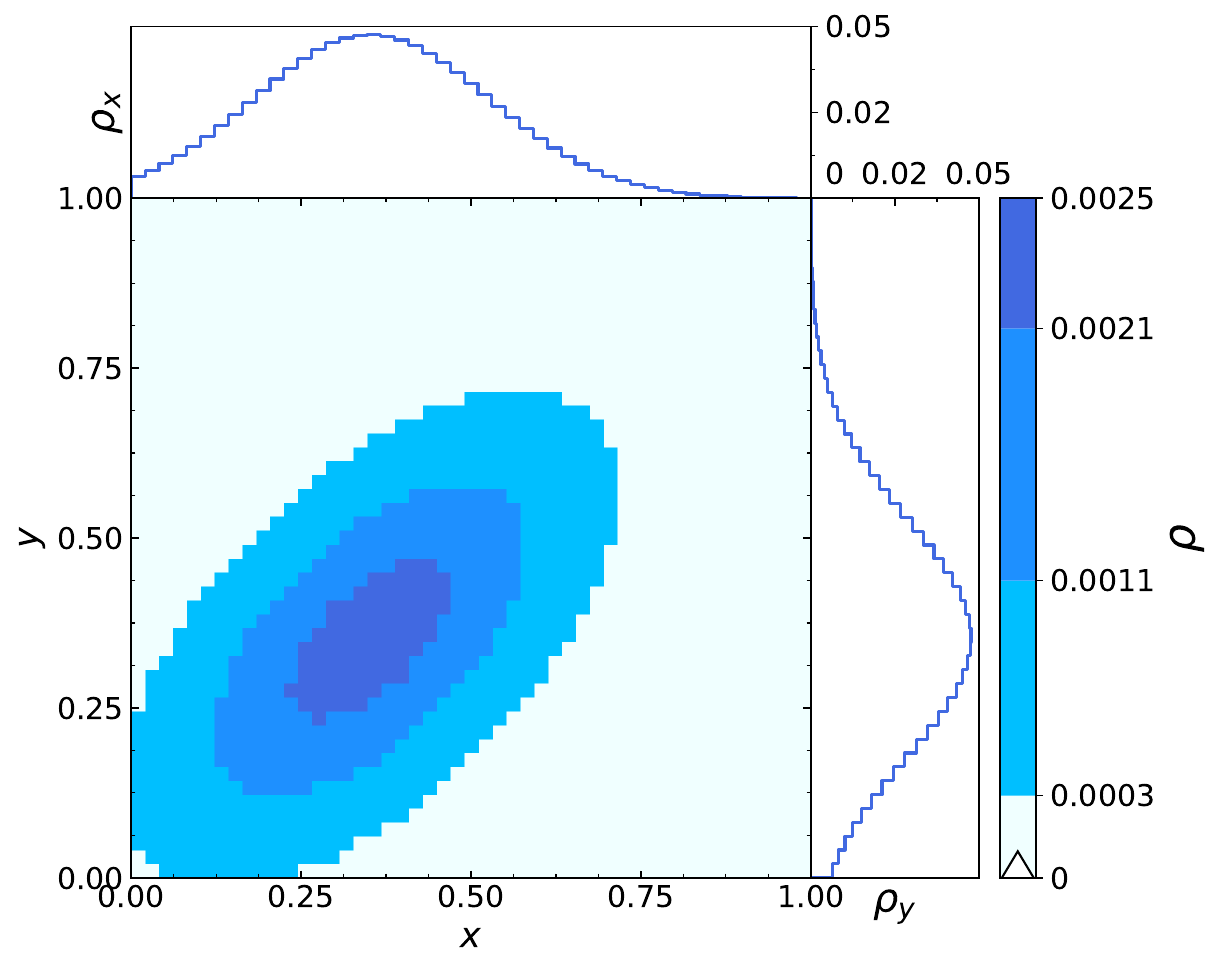}
    \caption{\small {Dependent variables,}}
    \end{subfigure}
\quad
    \begin{subfigure}[t]{0.99\textwidth}
    \centering
    \includegraphics[width=0.99\textwidth]{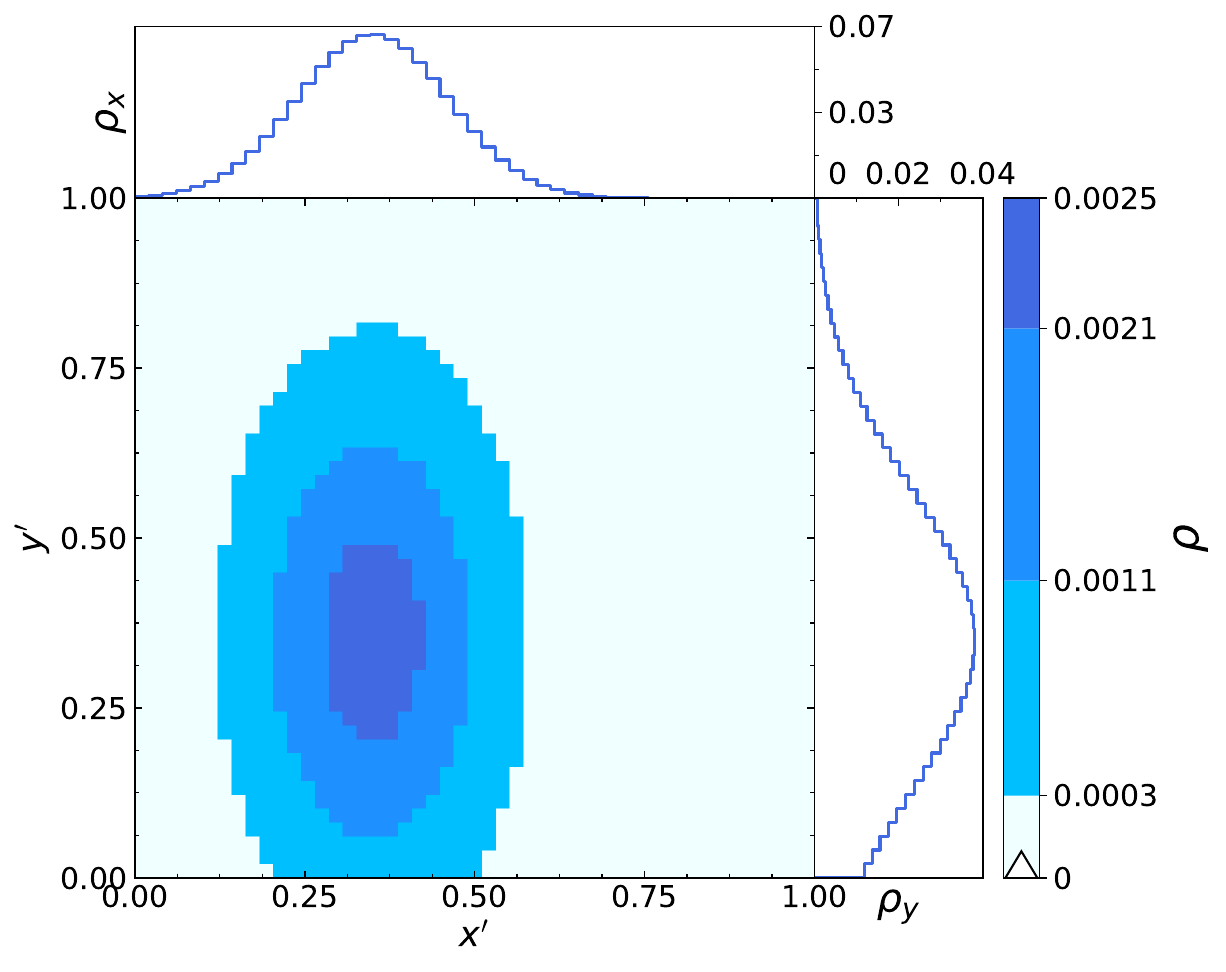}
    \caption{\small {independent variables.}}
    \end{subfigure}
    \nextfloat
\caption{Variable transformation toward independence through coordinate rotation.}
\label{fig:G2}
\end {minipage}
\end{figure}

Particle physics searches for phenomena beyond the Standard Model rely on event
analysis, in which the final state of each collision is classified as background
(known or perturbative physics) or signal (new physics).
In observable phase space, background-enriched regions (B, C, D) and a
signal-enriched region (A) can be defined, as shown in Fig.~\ref{fig:ABCD}.

Background processes are modeled by Monte Carlo simulations, but these
simulations may carry significant uncertainty in the signal region, rendering
direct background estimates unreliable.
The ABCD method~\cite{ABCDDisCo} addresses this limitation with a data-driven
approach: for observables $x$ and $y$ that are statistically independent under
the background hypothesis, the event counts in the four regions satisfy

\begin{equation}
\frac{N^\mathrm{bgr.}_A}{N^\mathrm{bgr.}_C} =
\frac{N^\mathrm{bgr.}_B}{N^\mathrm{bgr.}_D}.
\end{equation}

Measuring event counts in the background-enriched regions B, C, D — where
signal contamination is suppressed, $N^\mathrm{tot.}_X \approx N^\mathrm{bgr.}_X$
for $X \in \{B,C,D\}$ — then yields a background estimate for region A:

\begin{equation} \label{eq:sigAest}
N^\mathrm{sig.}_A \approx N^\mathrm{tot.}_A -
\frac{N^\mathrm{tot.}_B\, N^\mathrm{tot.}_C}{N^\mathrm{tot.}_D}.
\end{equation}

The fundamental obstruction to this approach is that observables are
\emph{not} statistically independent \textit{a priori}: all observables derive
from the same detected final state, so correlations are generically present.

Reference~\cite{ABCDDisCo} presents a widely used solution: for artificial neural
network (ANN) classifiers,\footnote{A classifier is a function mapping event
features to a score in $[0, 1]$ designed to separate two probability
distributions.} statistical independence from another observable is enforced as
an additional training constraint, using the DCC as a differentiable regularization
term~\cite{DisCoFever}.
This training-time approach has two drawbacks: the classifier's discriminating power
(quantified by the area under the receiver operating characteristic curve,
AUC~\cite{Fawcett2006}) decreases, and a non-zero DCC typically persists,
creating a trade-off between AUC and DCC.

This work proposes an alternative: given two observables, one is transformed
post-hoc (without classifier retraining) to achieve statistical quasi-independence
from the other for background-like events.

\subsection{Statistical Independence}

A multivariate probability density $\rho(x, y)$ describes the joint distribution
over real random variables $x$ and $y$.
Variables $x$ and $y$ are \textbf{independent} if and only if one-dimensional
marginal densities $f(x)$ and $g(y)$ exist such that $\rho(x,y)=f(x)\,g(y)$
(Fig.~\ref{fig:G2}).

Coordinate transformations can remove dependencies, analogous to finding
eigenvectors of a covariance matrix to reach statistically ``proper coordinates''
for $\rho$ (Fig.~\ref{fig:G2} shows a simple example: coordinate rotation).
The general transform achieving independence — the Rosenblatt transform — exists
for any absolutely continuous $\rho$, as demonstrated in Ref.~\cite{Rosenblatt1952};
it has also been employed for testing conditional independence~\cite{Song2009}.
The present work provides explicit, sample-efficient numerical implementations
of this transform targeting finite HEP event samples.

\subsection{Transformation via Conditional CDF}
\label{sec:transform}

The single-variable transformation $y \to \gamma$ toward independence from
variable $x$ is the normalized conditional cumulative distribution function:

\begin{equation} \label{eq:AM_GM}
\gamma (x, y) = \frac{\int^y_{-\infty} \mathrm{d} y'\; \rho (x, y')}
                     {\int^\infty_{-\infty} \mathrm{d} y''\; \rho (x, y'')},
\end{equation}

the $y$-primitive of $\rho$ normalized along the $y$-axis.
A rigorous proof that $\rho(x, \gamma)$ factorizes is provided in
Sec.~\ref{sec:math}.
The transform additionally renders $\gamma$ uniform on $[0, 1]$, and since
$\gamma$ is monotonically non-decreasing in $y$ at fixed $x$, it retains
classifier validity.

Variable $\gamma$, generated by $\rho$ as its ``proper coordinate,'' can also
serve as a classifier separating $\rho$ from a distinct density $\kappa(x, y)$,
provided the marginal $y$-distributions differ between the two densities.
When $\rho$ and $\kappa$ overlap minimally in the $xy$ plane, $\gamma$ may be
nearly flat in the $\kappa$-dominated region, compressing probability mass and
reducing separation power.
For the transform to be well-defined, $\rho$ must be absolutely continuous
(satisfied by any non-discrete density) and the marginal $\rho_X(x)$ must be
non-zero on the defining $x$-interval.

Extension to higher dimensions recasts $x$ as a vector $\vec{x}$ and applies
the transform conditionally; however, the independence proof does not generalize
straightforwardly in that case, and higher-dimensional applications are left for
future work.

\subsection{Related Work}
\label{sec:related}

Enforcing classifier-observable independence has been studied extensively
in high-energy physics.

\textit{Training-time methods} incorporate a dependence penalty into the loss
function.
Adversarial neural networks~\cite{Shimmin2017,Louppe2017} use a secondary
discriminator to penalize correlation with the protected observable, but suffer
from training instability.
DisCo~\cite{DisCoFever} adds the DCC as a regularization term, achieving
state-of-the-art decorrelation with simpler training dynamics.
Double-DisCo~\cite{Kasieczka2021} extends this to jointly optimize two
independent classifiers for direct ABCD automation, and has recently been applied
to LHC data~\cite{ABCDDisCo}.
All training-time methods incur some AUC penalty and require classifier retraining
when the protected observable changes.

\textit{Post-hoc transformation methods} modify the pre-trained classifier output
without retraining.
The earliest and most widely deployed is the Designed Decorrelated Tagger (DDT)
approach~\cite{Dolen2016} and its percentile-shift descendants used in
boosted-object searches at the LHC: the tagger is shifted by a conditional
quantile evaluated in bins of the protected variables, fixing the background
efficiency of a \emph{single} working point.
The conditional-CDF transform applied here extends this construction to all
working points simultaneously.
Conditional normalizing flows~\cite{Klein2022} learn an invertible mapping that
uniformizes the classifier output conditional on the protected observable,
exactly preserving the per-observable AUC.
Optimal transport methods~\cite{Algren2024} use convex neural solvers to remap
the score distribution, approaching normalizing-flow performance for binary
classification and outperforming DisCo for multi-class outputs.
The Rosenblatt transform has also been employed to \textit{test} (rather than
enforce) conditional independence~\cite{Song2009}.

The present work belongs to the post-hoc category and can be viewed as a
principled generalization of percentile-based decorrelation: IRGI replaces
the fixed grid and single-working-point shift of the DDT family with an
adaptive equal-occupancy grid and a full conditional-CDF map, KDE provides its
smooth (binning-free) limit, and the continuum independence proof of
Sec.~\ref{sec:math} supplies a formal guarantee that empirical flattening
procedures lack.
Its distinguishing features are: (i) direct numerical implementation of the
Rosenblatt transform using two complementary algorithms — IRGI and KDE — that
require only a finite background sample and no additional neural network training;
(ii) a closed-form independence guarantee for the background component; and
(iii) explicit validation in the ABCD background estimation context.

Although the comparison and primary validation are conducted in the particle
physics setting, the method is domain-agnostic.
Any application requiring a predictor to be independent of a protected
observable — including fairness-aware machine learning~\cite{Shimmin2017,Louppe2017},
survey calibration in astrophysics, and medical diagnostic classifiers —
can employ IRGI or KDE directly, as illustrated by the image classification
example in Sec.~\ref{sec:images}.

\section{Mathematical Foundation}
\label{sec:math}

For continuous probability densities satisfying the conditions stated below,
the Rosenblatt transform achieves \emph{exact} independence.
In numerical practice with finite samples the residual DCC is small but
non-zero; this is what we call quasi-independence throughout the paper.

Consider the probability density

\begin{equation*}
\rho (x, y): \mathbb{R}^2 \to \mathbb{R}_{0}^+.
\end{equation*}

The probability density $\rho$ contains complete information about variables
$x$ and $y$; the marginal distribution of $x$ is recovered by $y$-integration:

\begin{equation}
\begin{split}
\rho_X (x) &= \int_{-\infty}^{\infty} \mathrm{d} y \, \rho (x, y),\\
\rho_X (x) &: \mathbb{R} \to \mathbb{R}_{0}^+.
\end{split}
\end{equation}

Independence requires $\rho (x, y) = f (x)\, g (y)$ for marginal densities
$f : \mathbb{R} \to \mathbb{R}_{0}^+$ and $g : \mathbb{R} \to \mathbb{R}_{0}^+$,
with $f(x) = \rho_X(x)$ and $g(y) = \rho_Y(y)$.

Under the bijective differentiable variable transformation

\begin{equation*}
(\xi, \gamma) = \Gamma (x, y) : \mathbb{R}^2 \to \mathbb{R}^2,
\end{equation*}

the probability density transforms as

\begin{equation} \label{eq:TF}
\rho_{\Xi \Gamma} (\xi, \gamma) = \bigl|\det J_{\Gamma^{-1}} (\xi, \gamma)\bigr|\;
\rho_{X Y} \!\left(\Gamma^{-1} (\xi, \gamma)\right),
\end{equation}

where $J_{\Gamma^{-1}}$ is the Jacobian of the inverse transformation.

Consider the special case where $\Gamma$ transforms only $y$:

\begin{equation*}
(x, \gamma) = \Gamma (x, y), \qquad \gamma = \gamma (x, y).
\end{equation*}

The universal candidate for $\gamma$ is the normalized conditional CDF of
Eq.~\eqref{eq:AM_GM}.
At fixed $x$, $\gamma$ is monotonically non-decreasing in $y$ and maps
$\mathbb{R}$ into $[0, 1]$.
For $\Gamma$ to be bijective and differentiable, two conditions must hold:
(i) $\rho (x, y)$ is $x$-differentiable;
(ii) the marginal $\rho_X (x)$ is non-zero on the defining $x$-interval.

Substituting into Eq.~\eqref{eq:TF}:

\begin{equation} \label{eq:gGM}
\begin{split}
J_{\Gamma^{-1}} (x, \gamma) &=
  \begin{pmatrix}
    1 & 0 \\
    \dfrac{\partial y (x, \gamma)}{\partial x} &
    \left(\dfrac{\partial \gamma (x, y)}{\partial y}\bigg|_{(x,\gamma)}\right)^{-1}
  \end{pmatrix},\\[6pt]
\bigl|\det J_{\Gamma^{-1}} (x, \gamma)\bigr| &= \frac{\rho_X (x)}{\rho (x, y (x, \gamma))},\\[4pt]
\rho_{X \Gamma} (x, \gamma) &= \rho_X (x) \cdot \mathbf{1}_{[0, 1]}(\gamma).
\end{split}
\end{equation}

The resulting density $\rho_{X\Gamma}(x,\gamma) = \rho_X(x) \times \mathrm{Uniform}[0,1]$
is a product of marginals, confirming exact independence of $x$ and $\gamma$
and uniformity of $\gamma$.

Variable $\gamma$ is not uniquely determined: any bijective differentiable
$h: [0, 1] \to [0, 1]$ applied to $\gamma$ also yields an $x$-independent
classifier, and may be used to align $\gamma$ more closely with $y$ for
threshold applications.

\section{Numerical Implementation}
\label{sec:numerical}

The underlying background probability density is \textit{a priori} unknown and
represented by finite samples — paired vectors $(x_i, y_i)$ — with $N$
entries in the $xy$ plane.
Starting from a \textbf{defining background sample}, a transform
$T:(x, y) \to (x, \gamma)$ is constructed such that $x$ and $\gamma$ are
statistically quasi-independent for any sample drawn from the same underlying
background density.

Two implementations are developed in parallel, differing in how $\rho(x, y)$
is estimated from the sample.

\subsection{Irregular Grid Interpolation (IRGI)}

The first method, termed \textbf{IR}regular \textbf{G}rid \textbf{I}nterpolation
(IRGI), discretizes the probability density into a histogram matrix $H$ and
approximates the conditional CDF via cumulative sums — matrix $G$.
Irregular discretization accommodates varying event densities across phase space.
Given a defining sample ($N$ entries) and a finite rectangular domain, the IRGI
algorithm proceeds as follows:

\begin{enumerate}
\item Perform recursive bisection along the $x$-axis to depth $d + 1$ based on
the number of entries, yielding $2^{d+1} + 1$ edge positions indexed
$0, 1, 2, \ldots, 2^{d+1}$.
Bin centers are placed at even-indexed edges ($2^d + 1$ centers) and bin
boundaries at odd-indexed edges together with the two domain endpoints
($2^d + 2$ boundaries), giving $2^d + 1$ bins along $x$, each containing
approximately equal numbers of entries.
The two outermost bins are half-bins whose centers coincide with the domain
boundaries.

\item For each of the $2^d + 1$ $x$-bins, perform recursive bisection along the
$y$-axis to depth $d$ on the entries within that $x$-column.
This yields $2^d$ bins along $y$ per $x$-column, each containing approximately
equal numbers of entries.
Interpolation vertices are placed at $x$-bin centers and at all $y$-bin edges
within the corresponding $x$-column.
\end{enumerate}

The defining domain is thus subdivided into $(2^d + 1)\times 2^d$ irregular
two-dimensional bins.
Vertices are placed on the domain boundary so that interpolation is well-defined
across the entire phase space, which motivates the half-bin choice for the
outermost $x$-bins.

For illustration, see Fig.~\ref{fig:IRGI3} ($d = 3$), where black lines denote
bin edges and red dots the interpolation vertices.

The single free parameter of IRGI is the recursive bisection depth $d$.
Alternative binning algorithms may be used; optimality of the recursive bisection
strategy requires further investigation.
The core technique remains unchanged with any binning scheme that ensures at
least one entry per $x$-column.

With the selected binning, the defining background sample is histogrammed to
obtain the two-dimensional matrix $H$ — an $(m, n)$ matrix with $m$ columns
(one per $x$-bin) and $n$ rows (one per $y$-bin per column).
The cumulative sum along the $y$-axis gives the $(m, n + 1)$ matrix $G$:

\begin{equation}
G_{k l} = \sum_{l' = 0}^{l} H_{k l'}
\qquad \text{(with } G_{k 0} = 0 \text{)},
\end{equation}

normalized column-wise:

\begin{equation} \label{eq:Gtl}
\widetilde{G}_{k l} = G_{k l} / G_{k n}
\qquad \text{(requiring at least one entry per column)}.
\end{equation}

Matrix $\widetilde{G}$ serves as discrete approximation of $\gamma(x, y)$. For any real $\lambda \in [0, 1]$ (classifier cut value), selecting entries where $\widetilde{G}_{k l} \leq \lambda$ yields identical normalized distributions.

This is demonstrated by considering two arbitrary distinct columns indexed $k_1$ and $k_2$. Bin contents satisfying $\widetilde{G}_{k_1 l} \leq \lambda$ and $\widetilde{G}_{k_2 l} \leq \lambda$ are $G_{k_1 l_1}$ and $G_{k_2 l_2}$, where indices $l_1$ and $l_2$ correspond to maximal values $\widetilde{G}_{k_1 l_1}$ and $\widetilde{G}_{k_2 l_2}$ not exceeding $\lambda$ (well-defined due to $G$ and $\widetilde{G}$ being monotonically increasing along $y$). Apart from discretization errors, $\widetilde{G}_{k_1 l_1} = \widetilde{G}_{k_2 l_2} = \lambda$.

From Eq.~(\ref{eq:Gtl}), bin contents are $G_{k_1 l_1} = \lambda G_{k_1 n}$ and $G_{k_2 l_2} = \lambda G_{k_2 n}$, with ratio $G_{k_1 n} / G_{k_2 n}$ independent of $\lambda$. Any cut or binning along increasing $\widetilde{G}$ values yields identical normalized observable distributions.

For resolution finer than the $m, n + 1$ lattice, linear interpolation on the grid places interpolation vertices on $y$-bin edges and $x$-bin centers.

\subsection{Kernel Density Estimation (KDE)}

The second implementation uses \textbf{K}ernel \textbf{D}ensity \textbf{E}stimation
(KDE) to approximate $\rho(x,y)$ directly.
Given a defining background sample $\{(x_i, y_i)\}_{i=1}^N$, the numerical
$\gamma$ is

\begin{equation}
\gamma_{\sigma_r} (x, y) =
\frac{\displaystyle\sum_{i=1}^N H(y - y_i)\;
      \exp\!\left(-\dfrac{(x - x_i)^2}{2(\sigma_r \sigma_a)^2}\right)}
     {\displaystyle\sum_{i=1}^N
      \exp\!\left(-\dfrac{(x - x_i)^2}{2(\sigma_r \sigma_a)^2}\right)},
\end{equation}

where $H(t)$ is the Heaviside step function.
The typical length scale of the defining background distribution along the $x$-axis $\sigma_a$ (absolute sigma) is defined as

\begin{equation}
\sigma_a = \sqrt{\frac{2}{N(N-1)} \sum_{i<j} (x_i - x_j)^2},
\end{equation}

and $\sigma_r$ (relative sigma) is the single free parameter.
This expression implements Eq.~\eqref{eq:AM_GM} with the sample replacing the analytical density: for a fixed $\left( x, y \right)$,
all entries are assigned the weight of the Gaussian kernel of $|x - x'|$ and summed at $y'\leq y$, subsequently normalized.
A well-tuned $\sigma_r$ yields a smoother estimate than IRGI at the cost of higher per-event evaluation time.

\subsection{Computational Characteristics}

IRGI constructs the irregular grid with $\mathcal{O}(N\log N)$ complexity
(dominated by recursive sorting), after which each event is evaluated by
bilinear interpolation requiring $\mathcal{O}(\log m + \log n)$ operations
via binary search.
KDE requires no precomputation but its evaluation is $\mathcal{O}(N)$ per event,
giving $\mathcal{O}(MN)$ total for $M$ test events.
For the $N = M = 10^5$ samples used here both methods are practical;
for substantially larger datasets IRGI is preferred.
Approximate nearest-neighbor methods could reduce KDE evaluation cost
and are left as future work.

\subsection{Relation to the TMVA Rarity Transform}
\label{sec:tmva}

\begin{figure} [ht]
\centering
    \begin{subfigure}[t]{0.48\textwidth}
    \centering
    \includegraphics[width=0.99\textwidth]{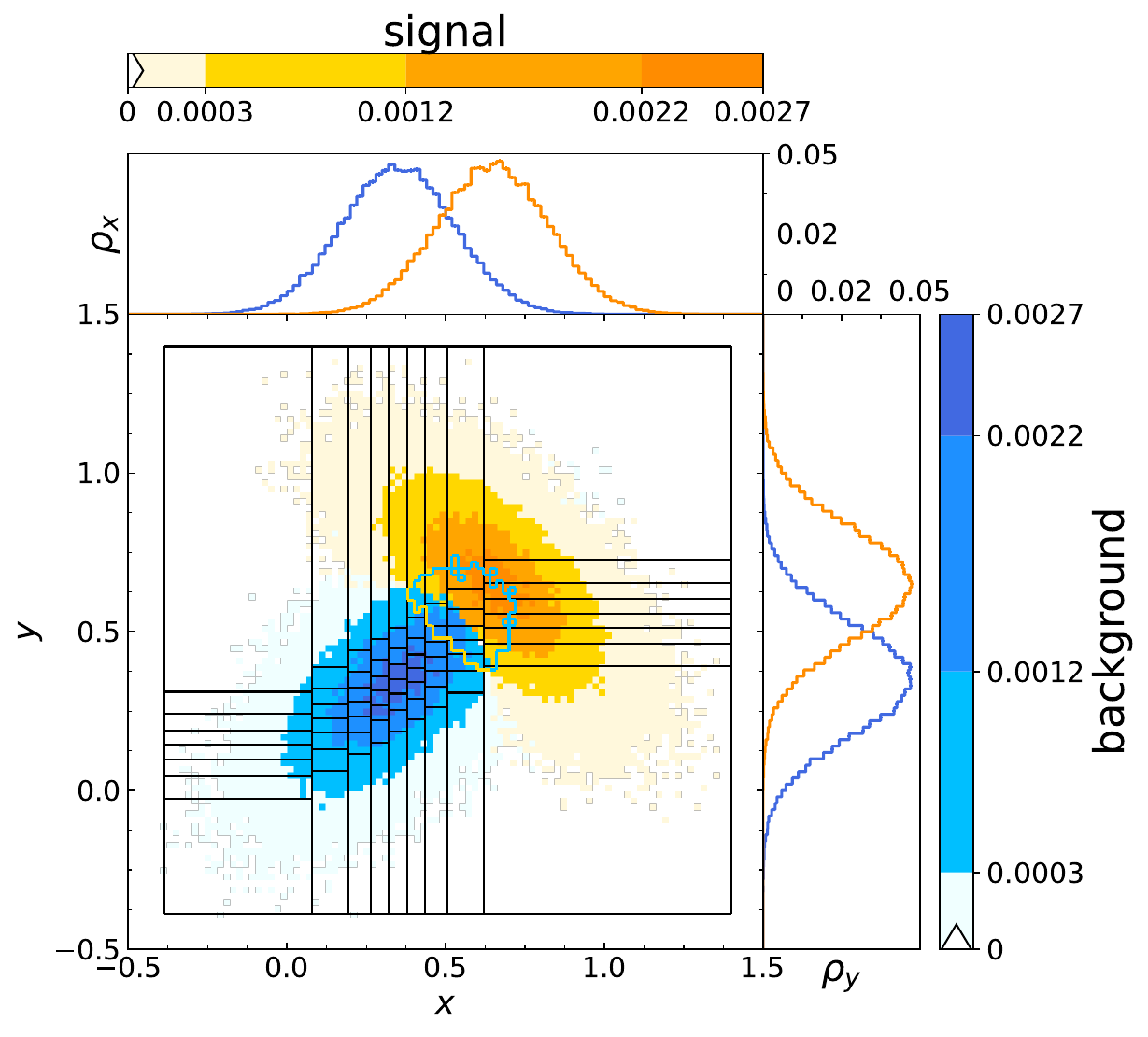}
    \caption{Irregular ($x$-local) binning of IRGI,}
    \end{subfigure}
\quad
    \begin{subfigure}[t]{0.48\textwidth}
    \centering
    \includegraphics[width=0.99\textwidth]{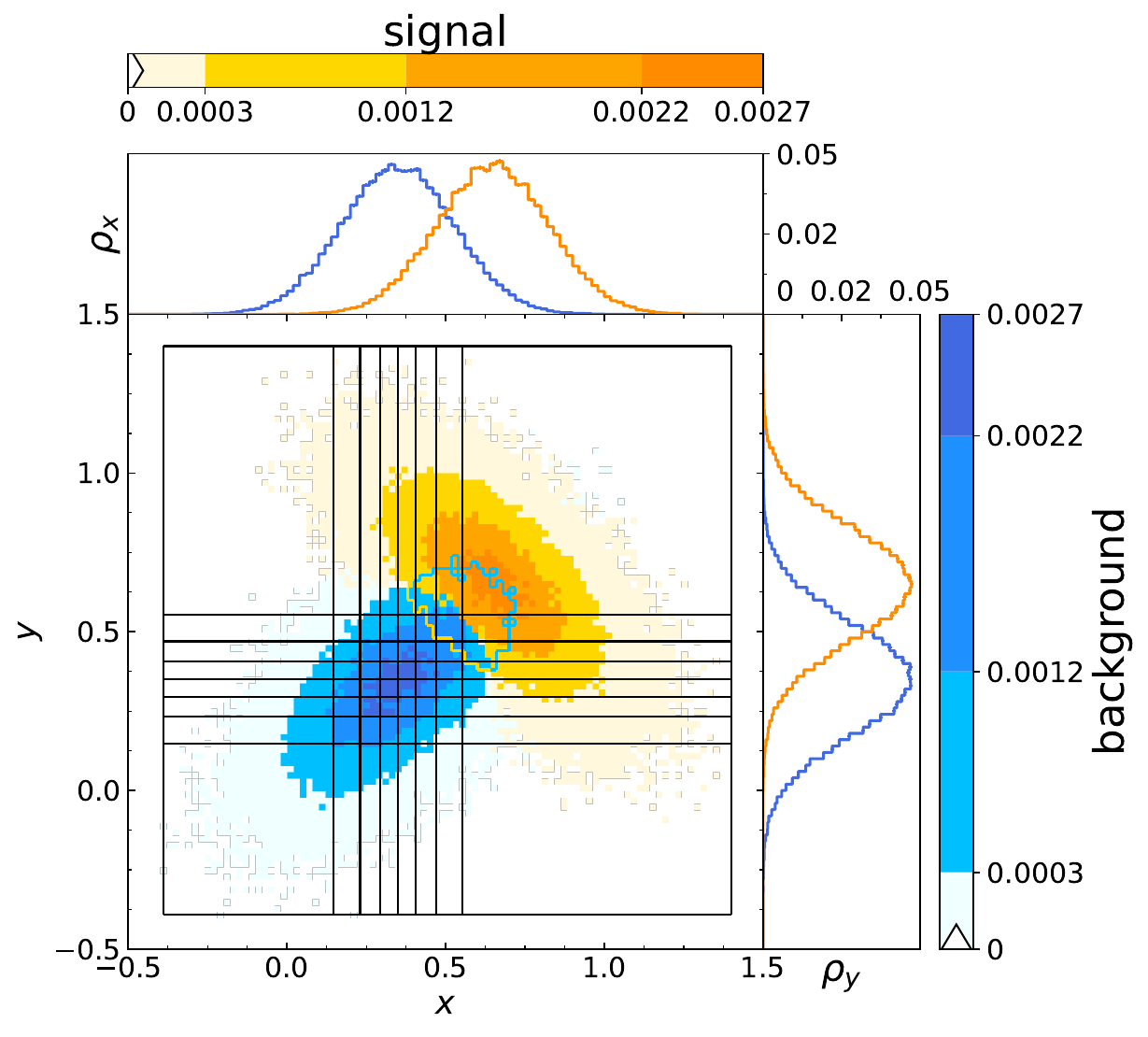}
    \caption{regular (single-axis) binning of the TMVA rarity transform.}
    \end{subfigure}
\caption{Two-dimensional adaptive binning of IRGI versus the one-dimensional
         binning underlying the TMVA rarity transform. The irregular binning is
         illustrative: it corresponds to $\gamma \approx i$ with $i$ the upper
         $y$-bin edge index; a different binning, or binning-free KDE, yields the
         same $\gamma$ in the continuum limit.}
\label{fig:vsTMVA}
\end{figure}

The present method superficially resembles the ``Uniform and Gaussian
transformation of variables'' in the TMVA toolkit~\cite{TMVA}
(Section~4.1.4).
Both employ integral transforms to uniformize a classifier distribution,
but they differ fundamentally:
TMVA operates in \emph{one dimension only} (projecting onto a single axis),
then applies linear decorrelation.
The present technique performs $x$-\emph{local} uniformization, adapting the
transform to the full two-dimensional density at each $x$-slice, which removes
non-linear correlations.
This difference is illustrated in Fig.~\ref{fig:vsTMVA}:
the irregular binning of IRGI (left) adapts to the local data density
two-dimensionally, whereas TMVA's regular binning (right) uniformizes
marginally along a single axis and cannot capture higher-order correlations.

\section{Results}
\label{sec:results}

Three examples are presented in order of increasing complexity and specificity.
The first (Sec.~\ref{sec:blobs}) uses synthetic Gaussian distributions to
introduce notation and verify the method in a fully controlled setting where the
underlying density is known.
The second (Sec.~\ref{sec:images}) applies IRGI and KDE to an image
classification problem from computer vision — a domain entirely unrelated to
particle physics — demonstrating that the method is not HEP-specific and can be
deployed wherever a predictor must be decorrelated from a protected observable.
The third (Sec.~\ref{sec:hep}) addresses the primary application: a QCD
dijet versus $W'$ boson classification, including a full comparison against DisCo
and a quantitative ABCD closure test.

\subsection{Abstract Example: Gaussian Blobs}
\label{sec:blobs}

The first example uses synthetic paired random vectors sampled from the same
tilted Gaussian densities as in the ABCD illustration (Fig.~\ref{fig:ABCD}).
Samples consist of $10^5$ background entries (defining background sample),
$10^5$ accompanying signal entries, another $10^5$ background entries (testing
sample), and $10^5$ additional signal entries.

\begin{figure} [ht]
\centering
    \begin{subfigure}[t]{0.48\textwidth}
    \centering
    \includegraphics[width=0.99\textwidth]{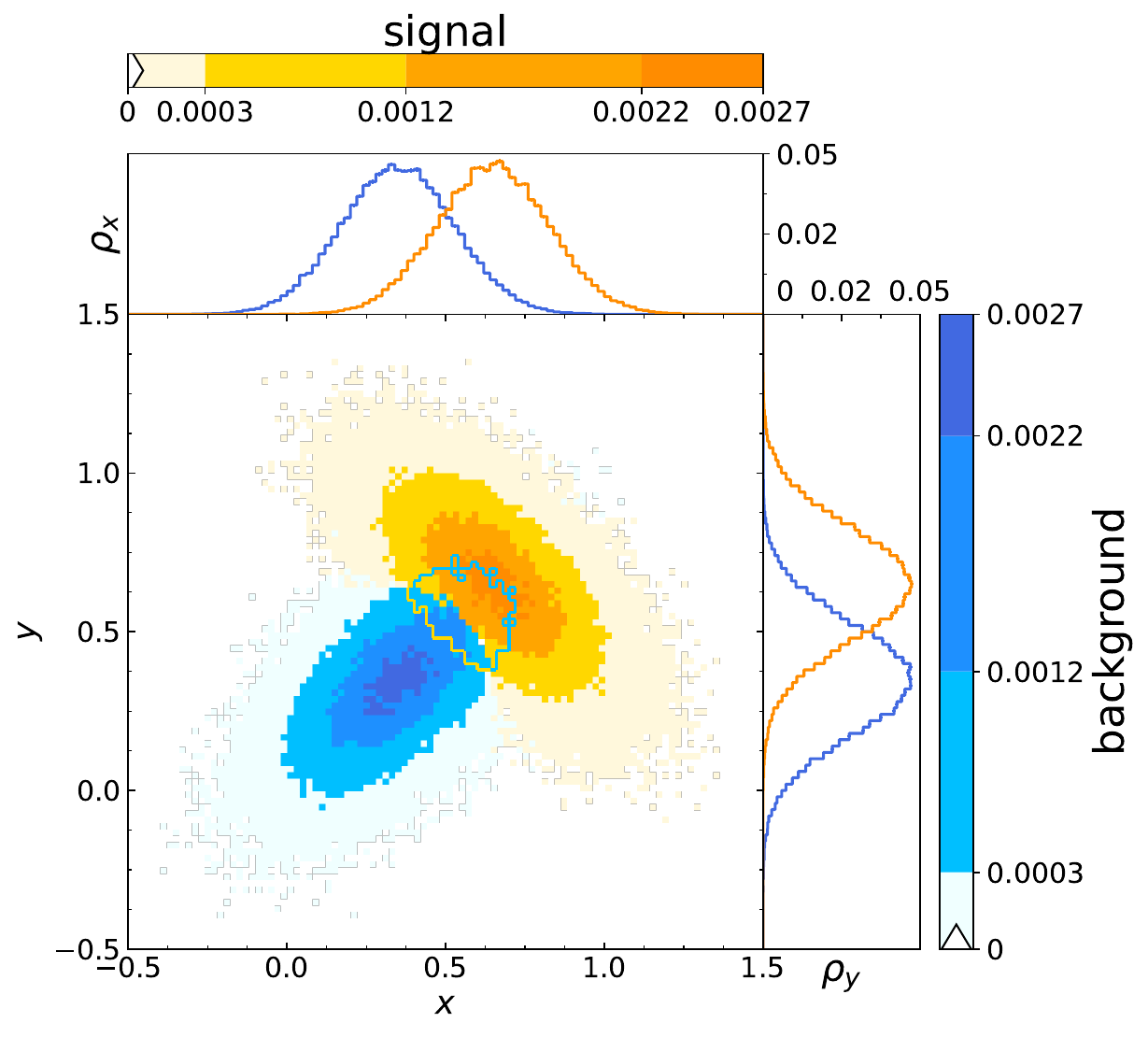}
    \caption{$10^5$ entries each of defining background and accompanying signal samples,}
    \end{subfigure}
\quad
    \begin{subfigure}[t]{0.48\textwidth}
    \centering
    \includegraphics[width=0.99\textwidth]{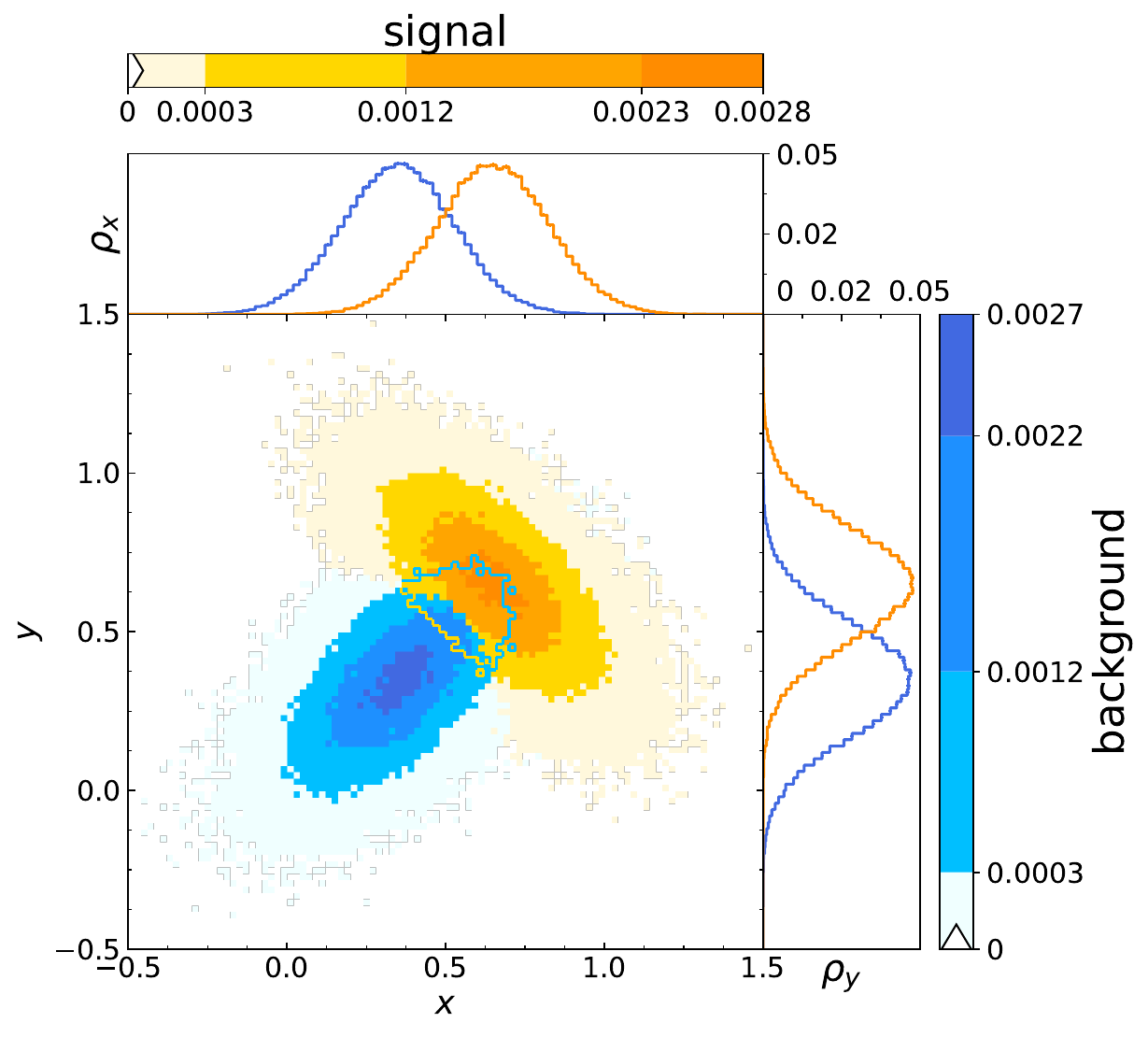}
    \caption{$10^5$ entries each of testing background and accompanying signal samples.}
    \end{subfigure}
\caption{Defining and testing samples drawn from identical underlying probability densities.}
\label{fig:Blob}
\end{figure}

\begin{figure} [ht]
\centering
    \begin{subfigure}[t]{0.48\textwidth}
    \centering
    \includegraphics[width=0.99\textwidth]{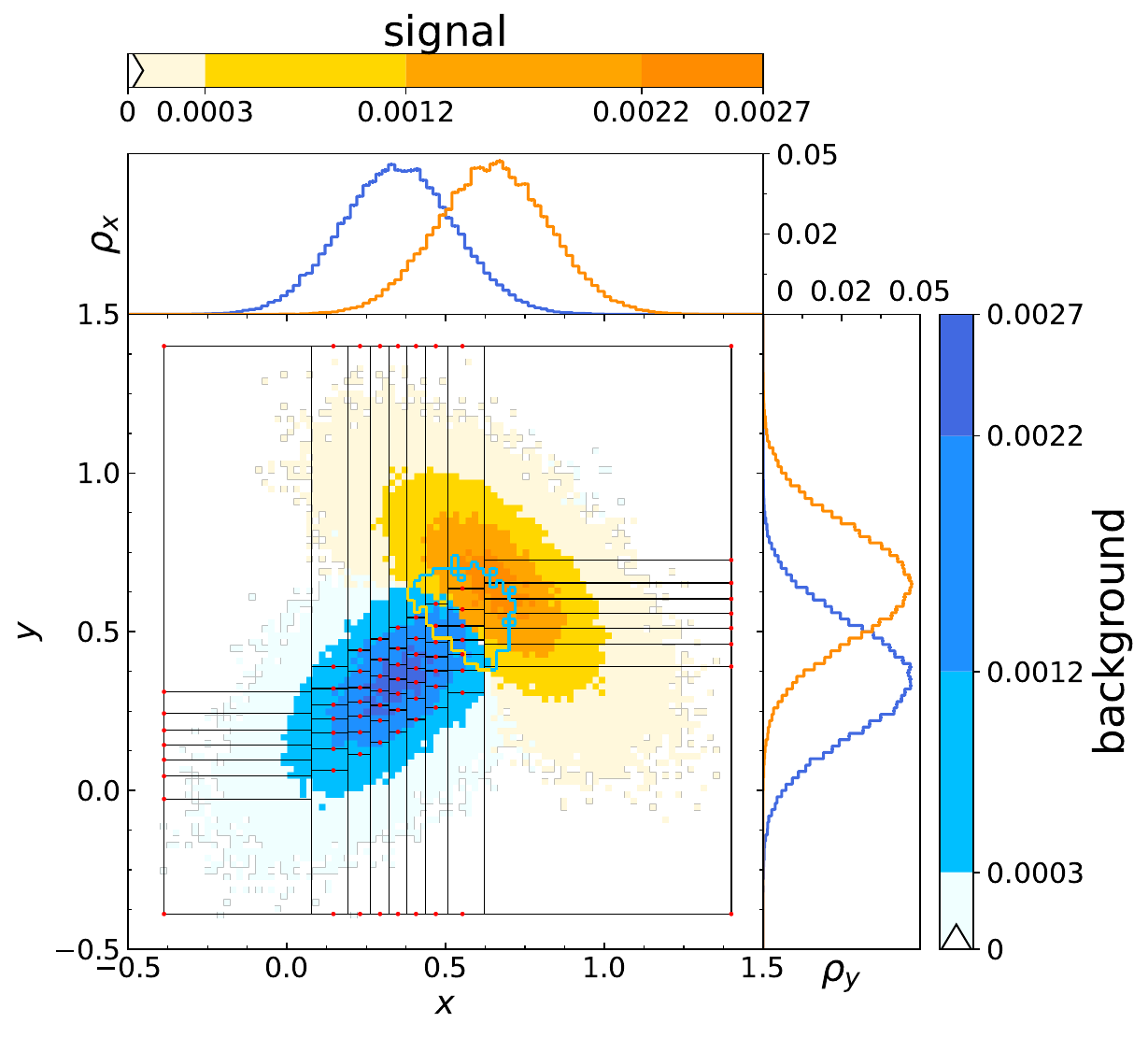}
    \caption{Defining background (and accompanying signal) sample with irregular grid for IRGI parameter $d = 3$,}
    \end{subfigure}
\quad
    \begin{subfigure}[t]{0.48\textwidth}
    \centering
    \includegraphics[width=0.99\textwidth]{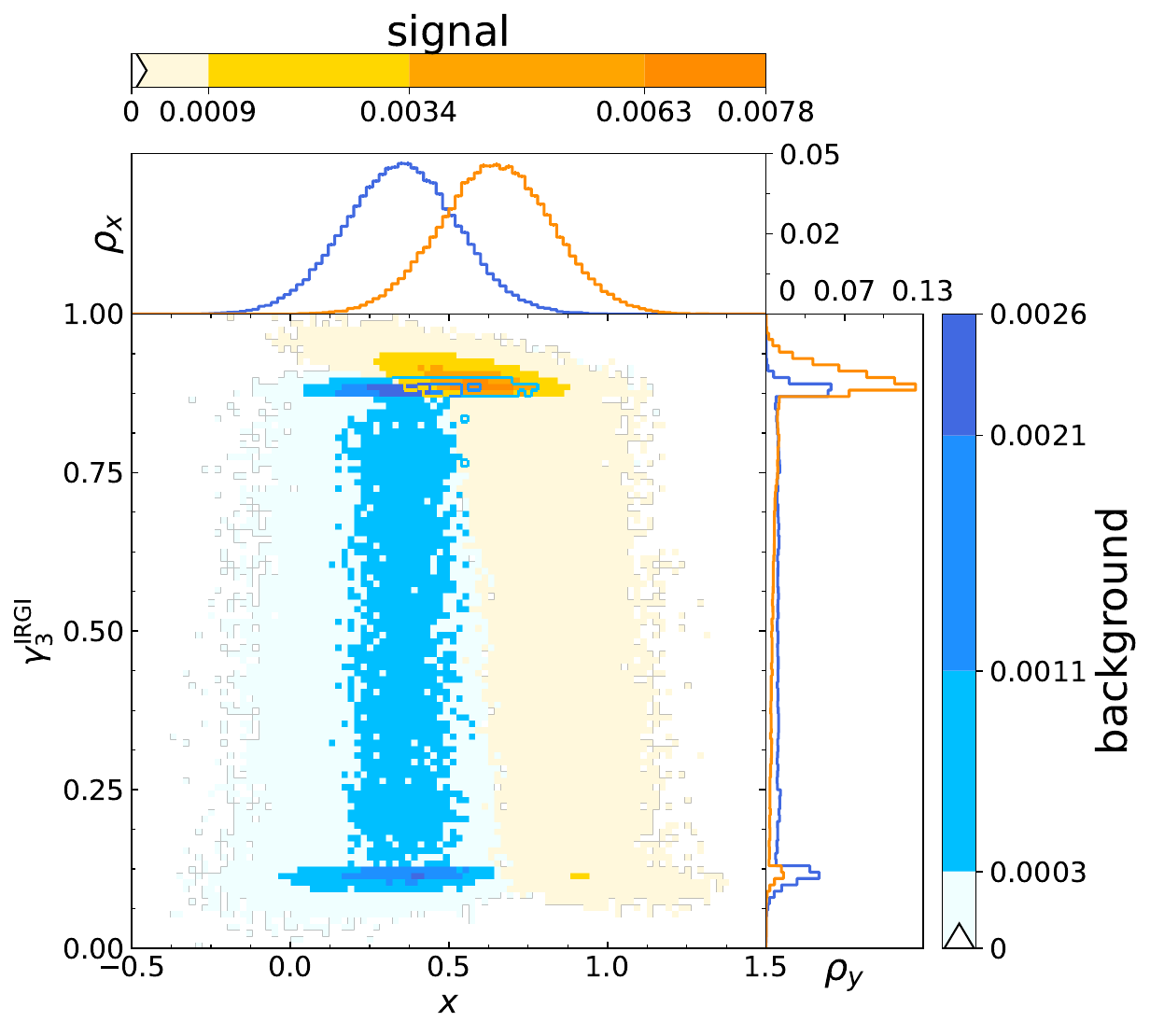}
    \caption{IRGI-modified classifier $\gamma_3$ applied to testing background (and accompanying signal) sample.}
    \end{subfigure}
\caption{Irregular grid construction (black lines) and interpolation vertices (red) for $d=3$, chosen for visual clarity; right: $\gamma_3$ applied to testing samples.}
\label{fig:IRGI3}
\end{figure}

\begin{figure} [ht]
\centering
    \begin{subfigure}[t]{0.48\textwidth}
    \centering
    \includegraphics[width=0.99\textwidth]{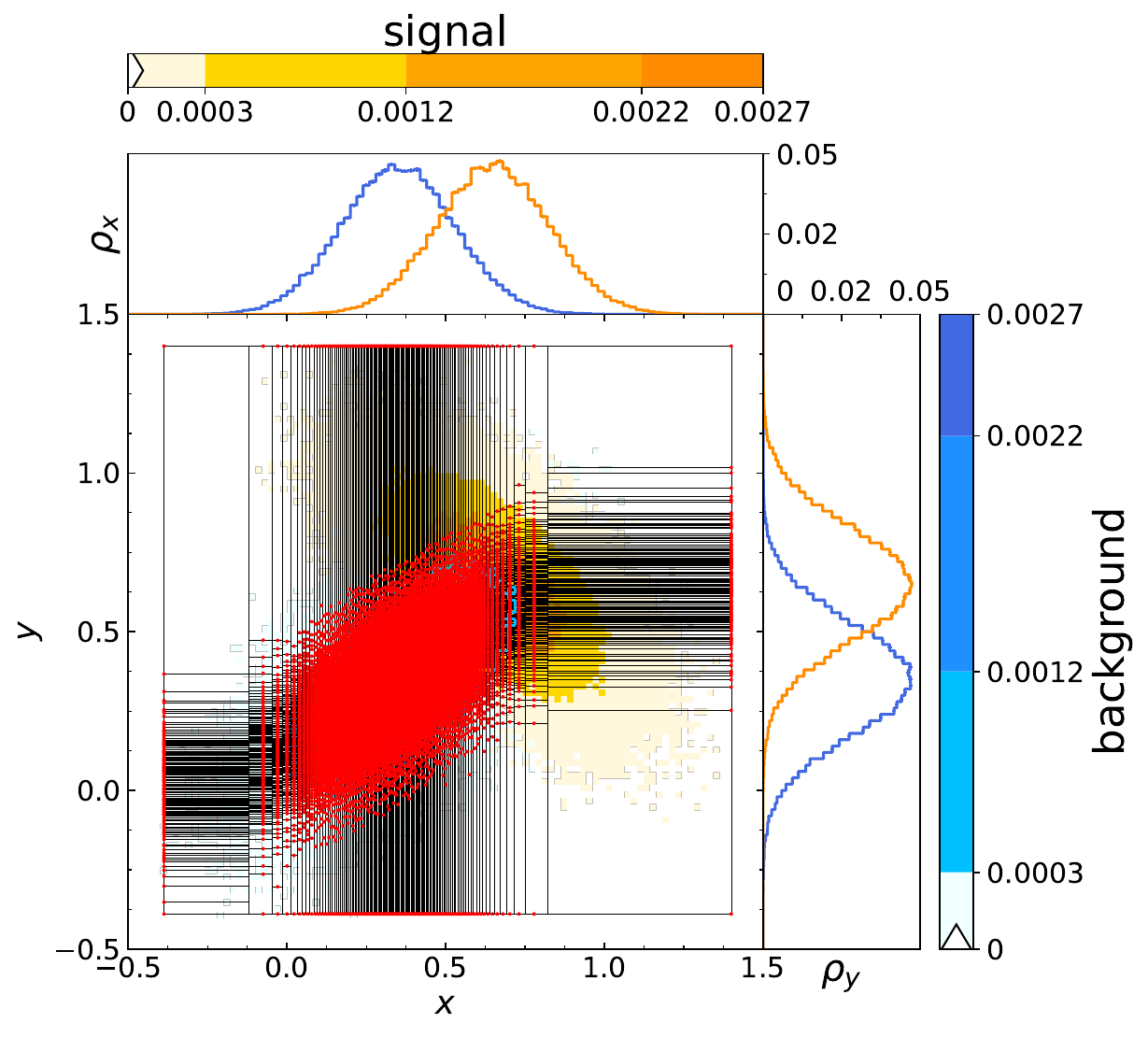}
    \caption{Defining background (and accompanying signal) sample with irregular grid for IRGI parameter $d = 7$,}
    \end{subfigure}
\quad
    \begin{subfigure}[t]{0.48\textwidth}
    \centering
    \includegraphics[width=0.99\textwidth]{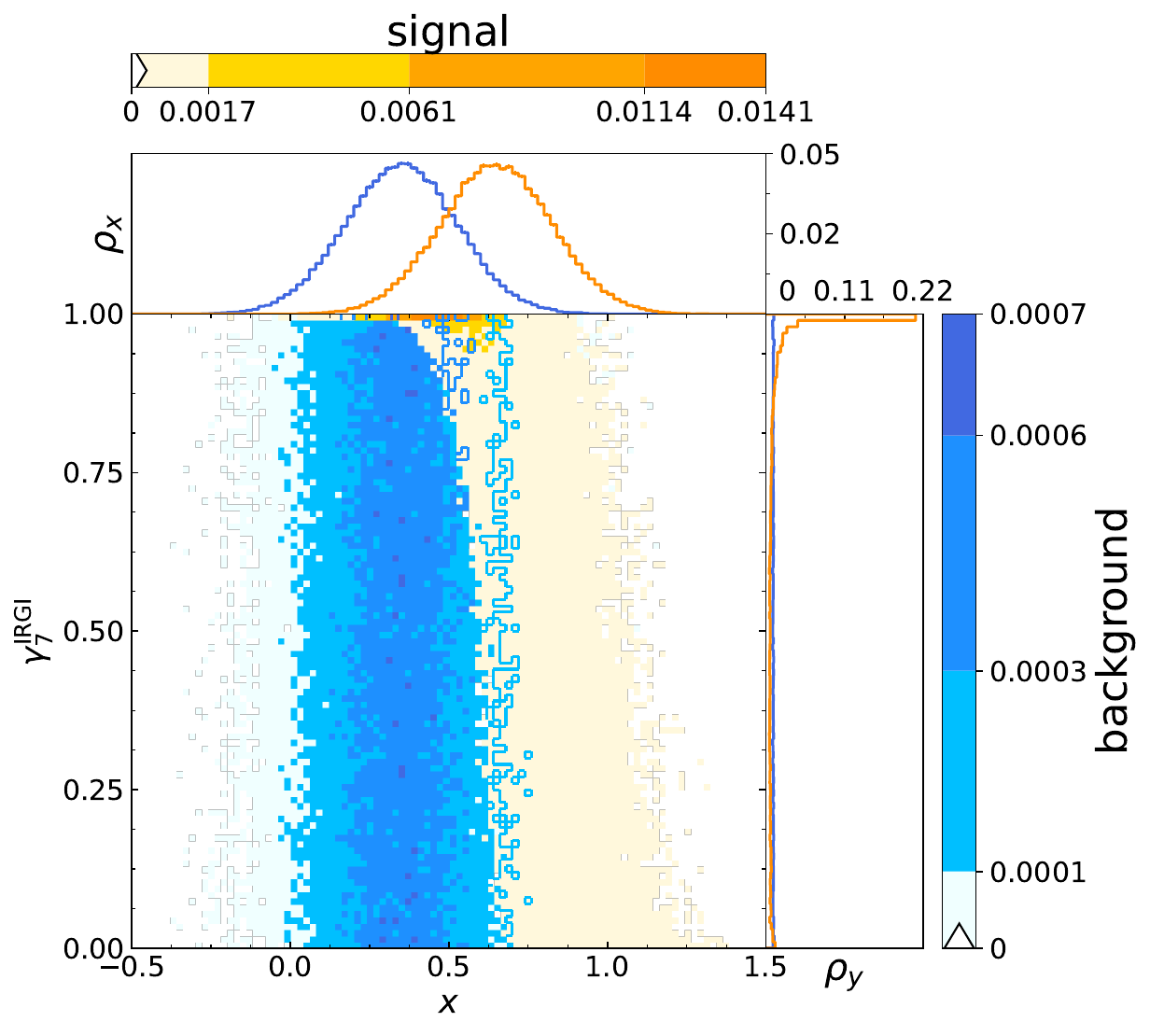}
    \caption{IRGI-modified classifier $\gamma_7$ applied to testing background (and accompanying signal) sample.}
    \end{subfigure}
\caption{As Fig.~\ref{fig:IRGI3} for the DCC-optimal parameter $d=7$.}
\label{fig:IRGI7}
\end{figure}

\begin{figure} [ht]
\centering
\includegraphics [width=0.81\textwidth]{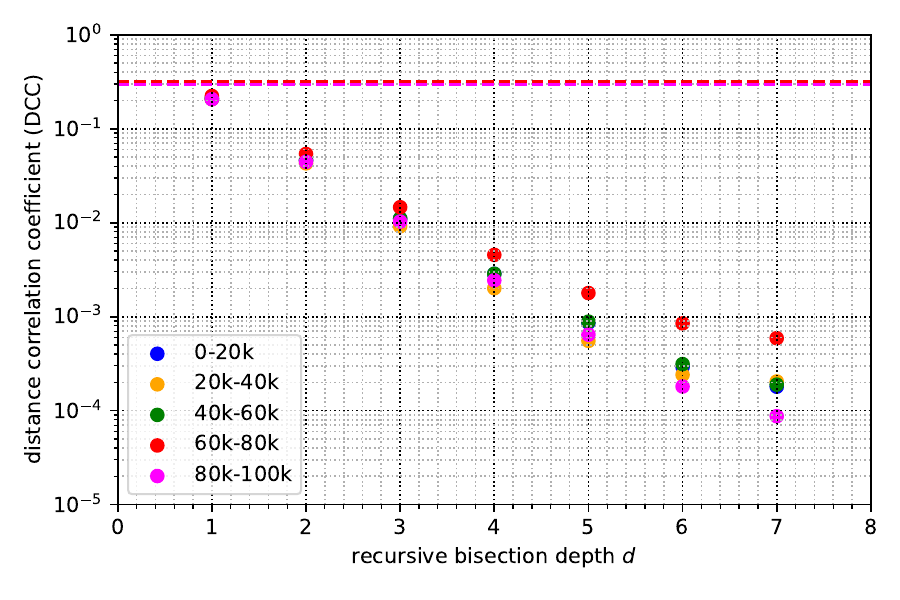}
\caption {DCC of IRGI $(x, \gamma_d)$ versus $d$ per subsample; colored dashed lines show per-subsample $(x, y)$ DCC.}
\label{fig:DCC_irrg}
\end{figure}

Fig.~\ref{fig:IRGI3} shows the irregular grid and interpolation vertex construction for the defining background sample, with derived numerical function $\gamma_3$ applied to testing samples. Parameter $d = 3$ is chosen for visual clarity. Fig.~\ref{fig:IRGI7} shows the same for DCC-optimal $d = 7$. Additional figures with $d$ varying from 1 to 7 (maximal recursive bisection depth) appear in Appendix~\ref{apx:d1to7}, demonstrating improved $\gamma$ distribution uniformity with increasing $d$. DCC variation with $d$ appears in Fig.~\ref{fig:DCC_irrg} and Table~\ref{tab:blob}.

\begin{figure} [ht]
\centering
    \begin{subfigure}[t]{0.48\textwidth}
    \centering
    \includegraphics[width=0.99\textwidth]{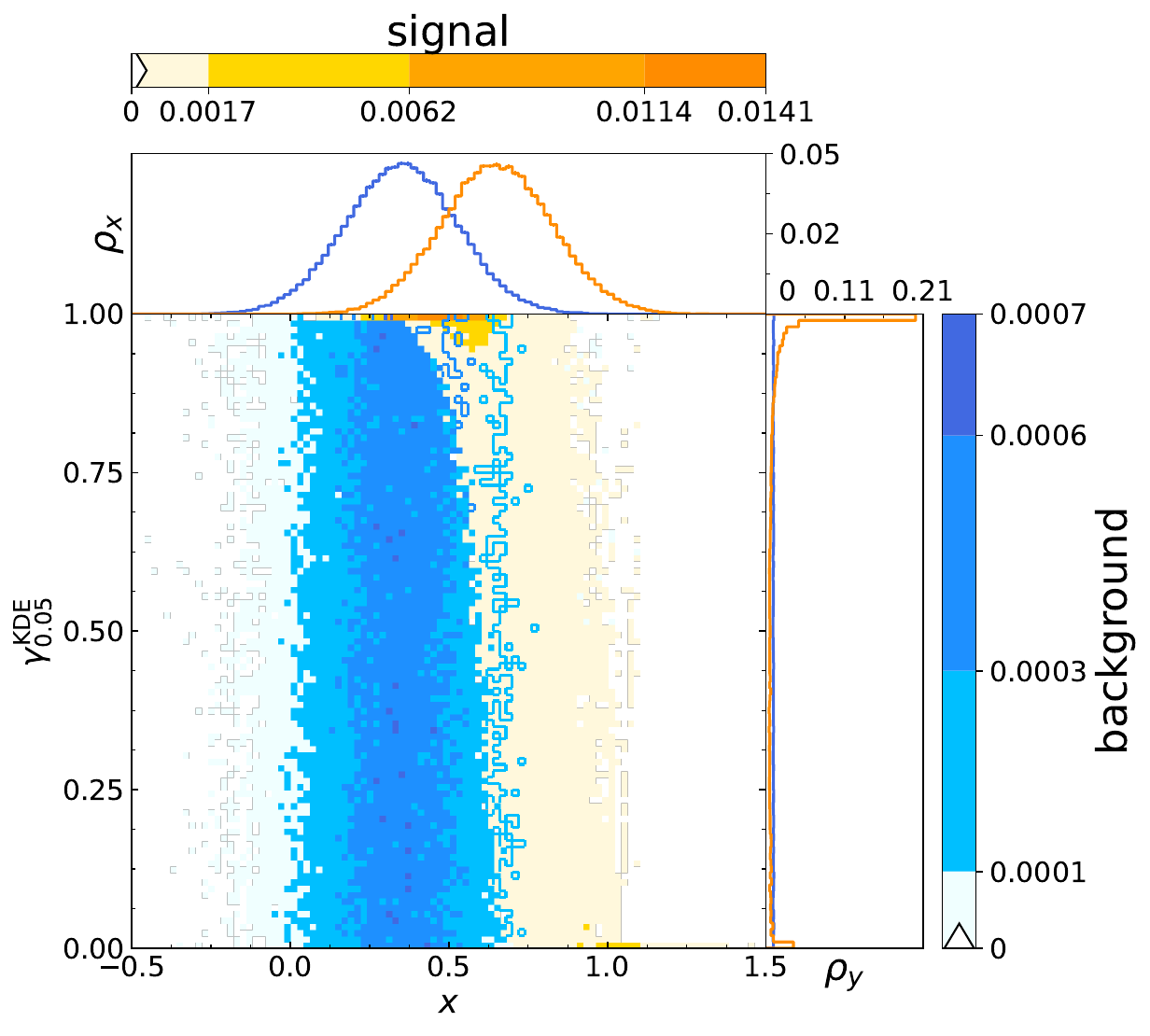}
    \caption{KDE $\gamma_{0.05}$,}
    \end{subfigure}
\quad
    \begin{subfigure}[t]{0.48\textwidth}
    \centering
    \includegraphics[width=0.99\textwidth]{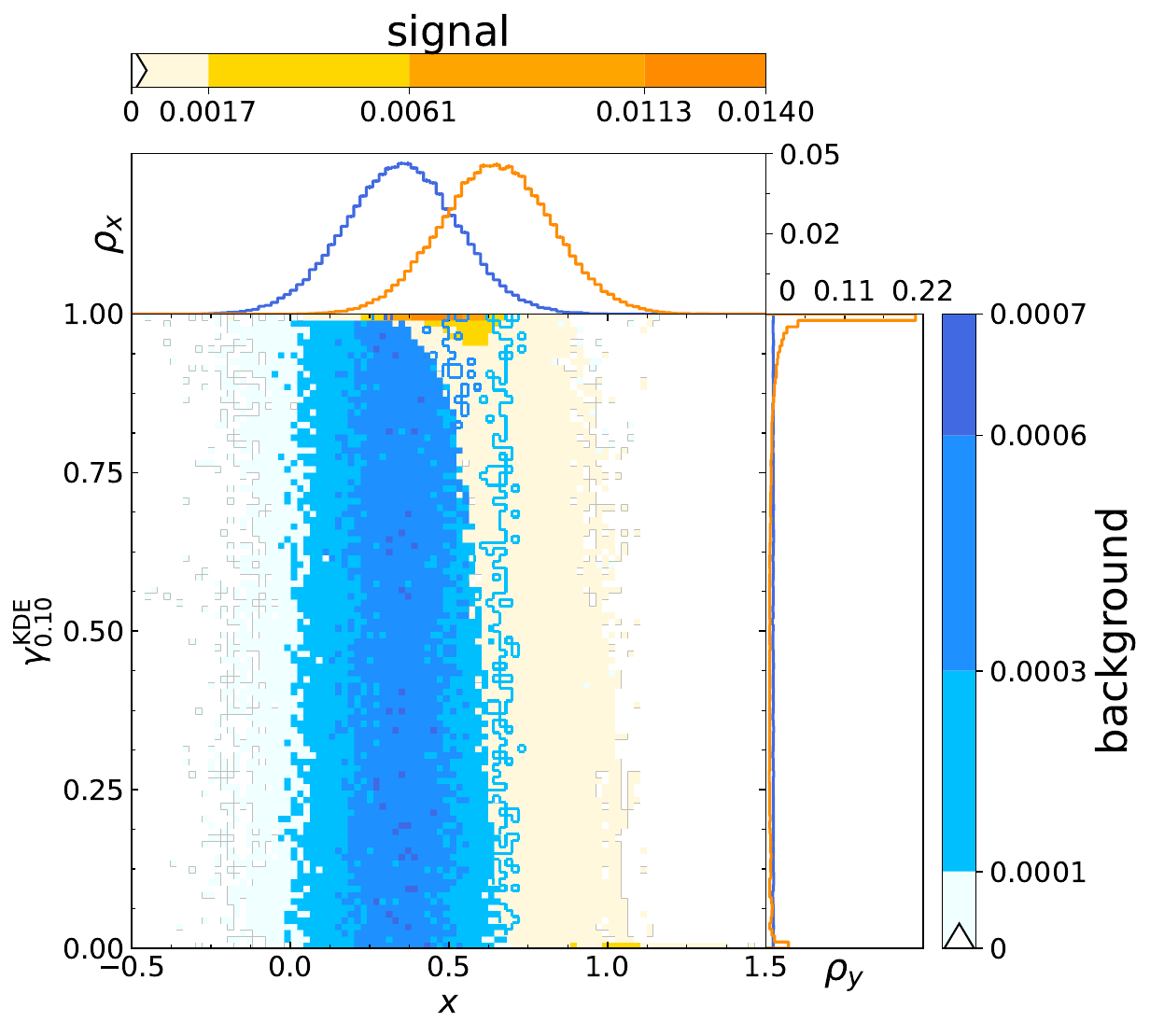}
    \caption{KDE $\gamma_{0.10}$.}
    \end{subfigure}
\caption{KDE $\gamma_{\sigma_r}$ with parameter $\sigma_r = 0.05$ (near DCC-optimal) and $\sigma_r = 0.1$.}
\label{fig:KDE1}
\end{figure}

\begin{figure} [ht]
\centering
    \begin{subfigure}[t]{0.48\textwidth}
    \centering
    \includegraphics[width=0.99\textwidth]{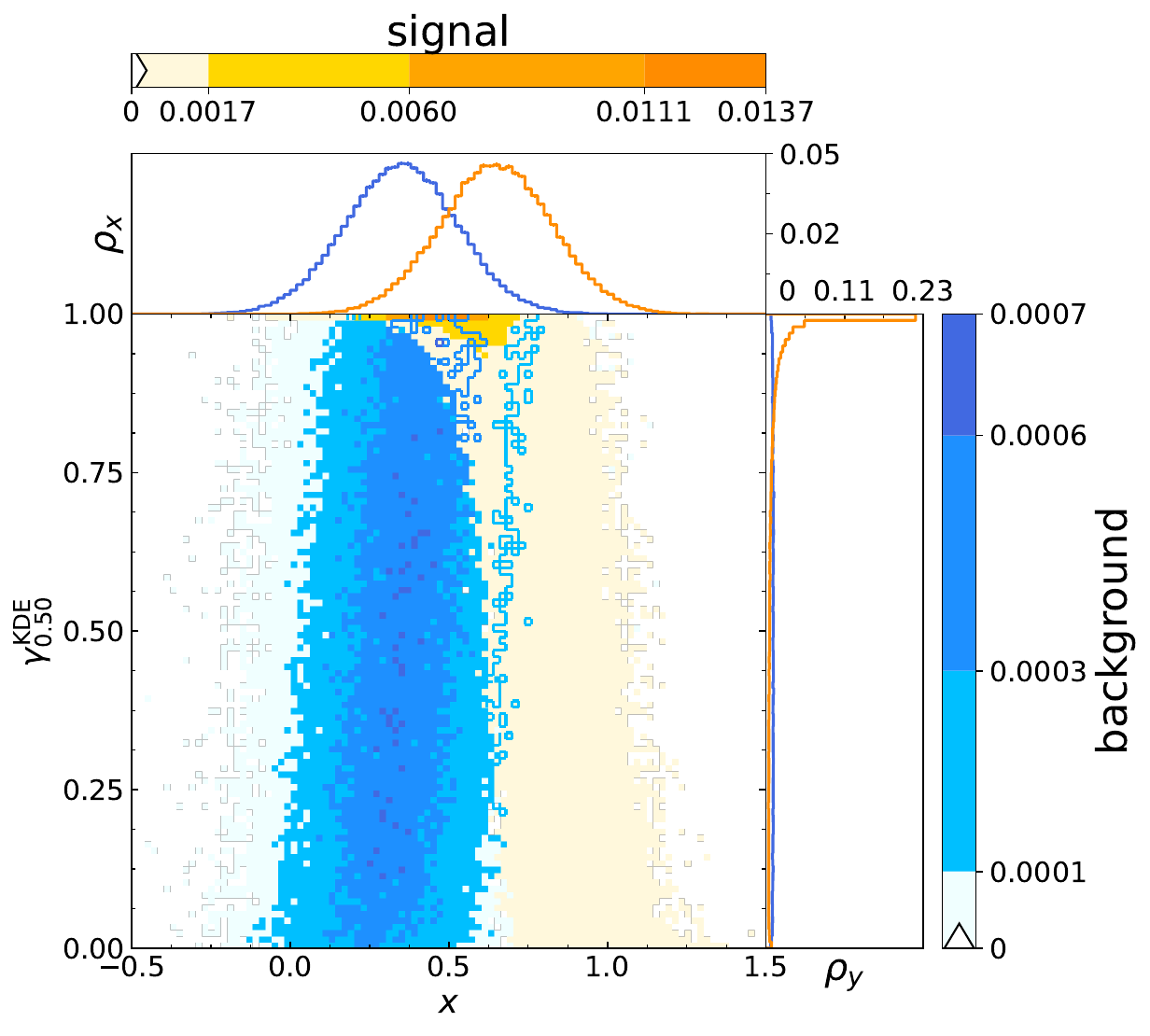}
    \caption{KDE $\gamma_{0.50}$,}
    \end{subfigure}
\quad
    \begin{subfigure}[t]{0.48\textwidth}
    \centering
    \includegraphics[width=0.99\textwidth]{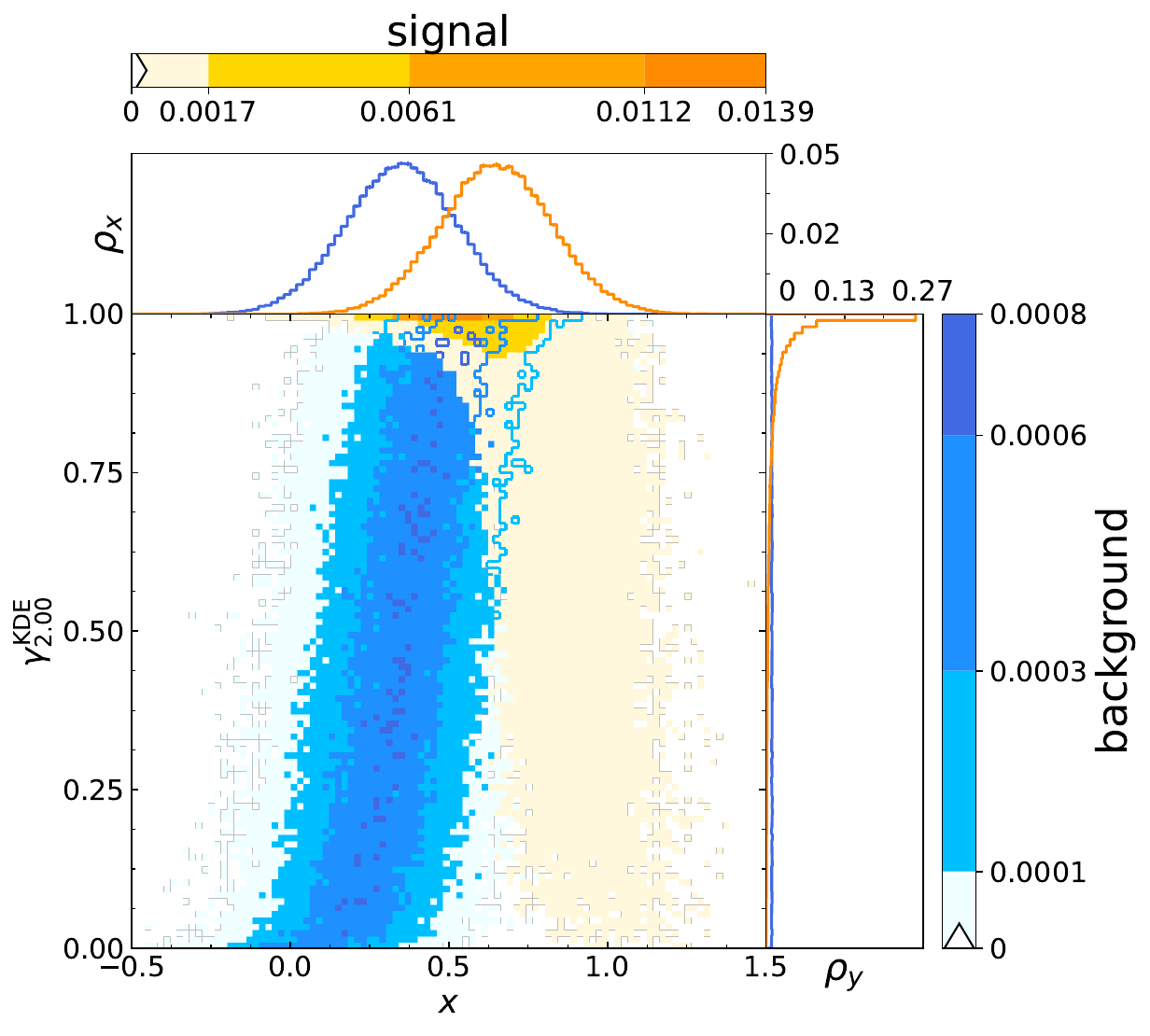}
    \caption{KDE $\gamma_{2.00}$.}
    \end{subfigure}
\caption{KDE $\gamma_{\sigma_r}$ with parameter $\sigma_r = 0.5$ (significantly less DCC-optimal) and $\sigma_r = 2$ (excessive Gaussian kernel width).}
\label{fig:KDE2}
\end{figure}

\begin{figure} [ht]
\centering
\includegraphics [width=0.81\textwidth]{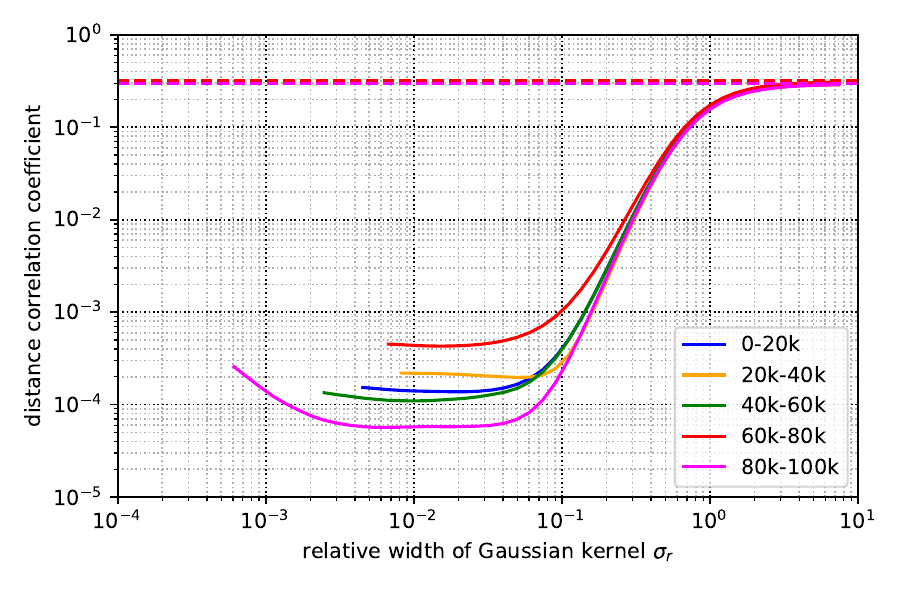}
\caption{DCC of KDE $(x, \gamma_{\sigma_r})$ versus $\sigma_r$ per subsample;
         colored dashed lines show per-subsample $(x, y)$ DCC.}
\label{fig:DCC_KI}
\end{figure}

Figs.~\ref{fig:KDE1} and~\ref{fig:KDE2} demonstrate how $\gamma_{\sigma_r}$ varies
with $\sigma_r$ across four representative values.
As $\sigma_r$ increases beyond the optimal range, the DCC of $(x, \gamma_{\sigma_r})$
rises, indicating degraded decorrelation due to over-smoothing.
DCC variation with $\sigma_r$ is summarized in Fig.~\ref{fig:DCC_KI} and
Table~\ref{tab:blob}.

To study DCC variation across both techniques, the testing background sample is
divided into five equal sub-samples of $2\times10^4$ events.
This serves two purposes: DCC evaluation scales quadratically in sample size,
and the sub-sample spread provides a qualitative indication of finite-sample
uncertainty.
For each sub-sample, $(x, y)$-DCC is computed as an upper bound and compared with
$(x, \gamma_d)$-DCC and $(x, \gamma_{\sigma_r})$-DCC for varying free parameters.

Fig.~\ref{fig:DCC_irrg} shows the IRGI DCC as a function of $d$ per sub-sample.
The DCC decreases monotonically up to the maximum $d = 7$; further bisection is
not possible with $N = 10^5$ defining entries.

Fig.~\ref{fig:DCC_KI} shows the KDE DCC as a function of $\sigma_r$.
For very small $\sigma_r$, the numerics become unstable (excessively narrow
kernels contribute effectively zero weight for distant events); for
$\sigma_r \gtrsim 1$, the kernels become too wide to retain useful information
and the DCC returns to its unmodified value.
All sub-samples show a broad plateau of near-minimal DCC for
$\sigma_r \approx 0.01$--$0.08$.

Per-sub-sample DCC minima for both techniques are compared in
Table~\ref{tab:blob}.
In each sub-sample, a KDE parameter $\sigma_r$ can be found that matches or
surpasses the corresponding IRGI minimum; both techniques reduce the DCC to the
same order of magnitude (factor $\sim 1500$ reduction from the unmodified value).

\begin{table}[ht]
\caption{Minimum DCC values per sub-sample for IRGI (at optimal $d$) and KDE
(at optimal $\sigma_r$), compared with the unmodified $(x,y)$-DCC.}
\label{tab:blob}
\begin{ruledtabular}
\begin{tabular}{lccccc}
 & 0--20k & 20k--40k & 40k--60k & 60k--80k & 80k--100k \\
\hline
DCC $(x, y)$           & 0.306 & 0.299 & 0.300 & 0.317 & 0.298 \\
DCC $(x, \gamma_d)$    & 0.00018 & 0.00020 & 0.00019 & 0.00059 & 0.00009 \\
DCC $(x, \gamma_{\sigma_r})$ & 0.00014 & 0.00020 & 0.00011 & 0.00043 & 0.00006 \\
\end{tabular}
\end{ruledtabular}
\end{table}

\subsection{Illustrative Example: Image Classification}
\label{sec:images}

\begin{figure} [ht]
\begin {minipage} {0.47\textwidth}
\centering
    \begin{subfigure}[t]{0.99\textwidth}
    \centering
    \includegraphics[width=0.99\textwidth]{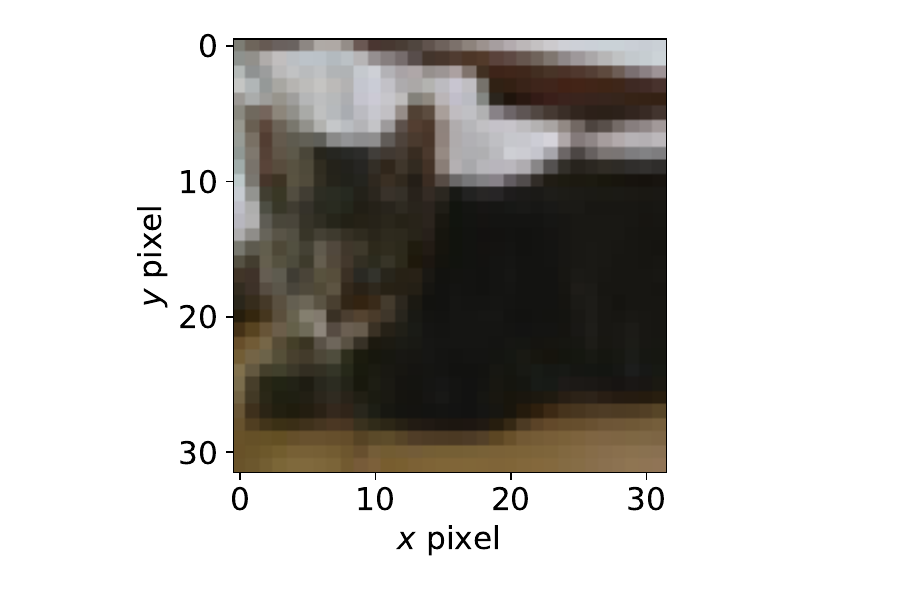}
    \caption{A $32\times 32$ pixel cat image,}
    \end{subfigure}
\quad
    \begin{subfigure}[t]{0.99\textwidth}
    \centering
    \includegraphics[width=0.99\textwidth]{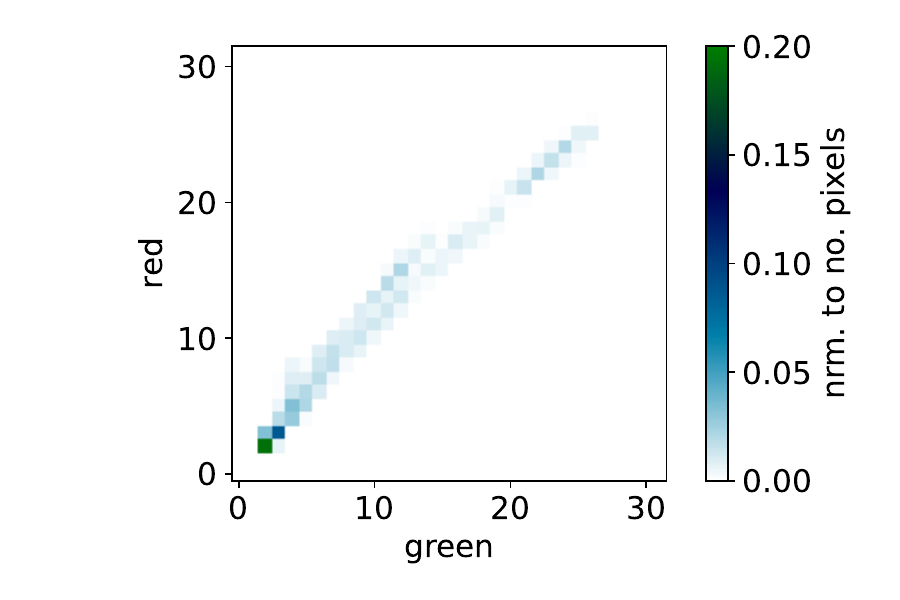}
    \caption{its $32\times 32$ red-green color histogram representation.}
    \end{subfigure}
\caption{Example entry from the cat-image (reference) class.}
\label{fig:RGCat}
\end {minipage}
\begin {minipage} {0.47\textwidth}
\centering
    \begin{subfigure}[t]{0.99\textwidth}
    \centering
    \includegraphics[width=0.99\textwidth]{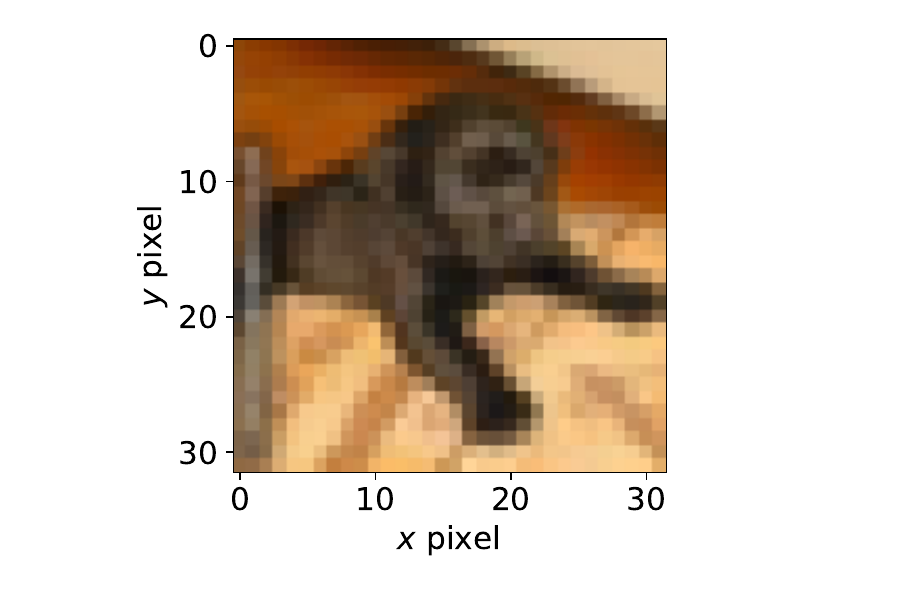}
    \caption{A $32\times 32$ pixel dog image,}
    \end{subfigure}
\quad
    \begin{subfigure}[t]{0.99\textwidth}
    \centering
    \includegraphics[width=0.99\textwidth]{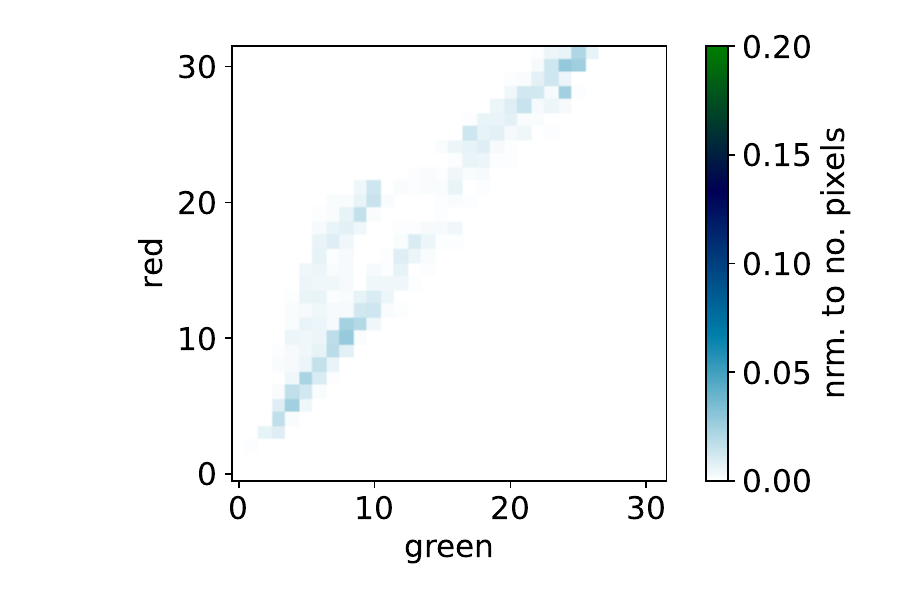}
    \caption{its $32\times 32$ red-green color histogram representation.}
    \end{subfigure}
\caption{Example entry from the dog-image (target) class.}
\label{fig:RGDog}
\end {minipage}
\end{figure}

To demonstrate that the method is not specific to particle physics, a second
validation uses the CIFAR-10 animal image dataset~\cite{cif}
(Figs.~\ref{fig:RGCat} and~\ref{fig:RGDog}), with cat images cast as the
reference class (``background'') and dog images as the target class (``signal'').
The protected observable $x$ is an image color score and the classifier $y$
is a CNN trained on the raw images — a setup that has no particle physics
content whatsoever, confirming that the method requires only a reference sample
and a scalar observable pair $(x, y)$.

The network accepts $32\times32$ RGB matrices, with three convolutional layers
(stride 3, pooling 2, no padding) followed by a single dense layer,
trained with TensorFlow~\cite{tensorflow2015-whitepaper}.
Both categories are split into 3000 training and 3000 validation events;
the training set serves as the defining background sample.

\begin{figure} [ht]
\centering
\includegraphics [width=0.64\textwidth]{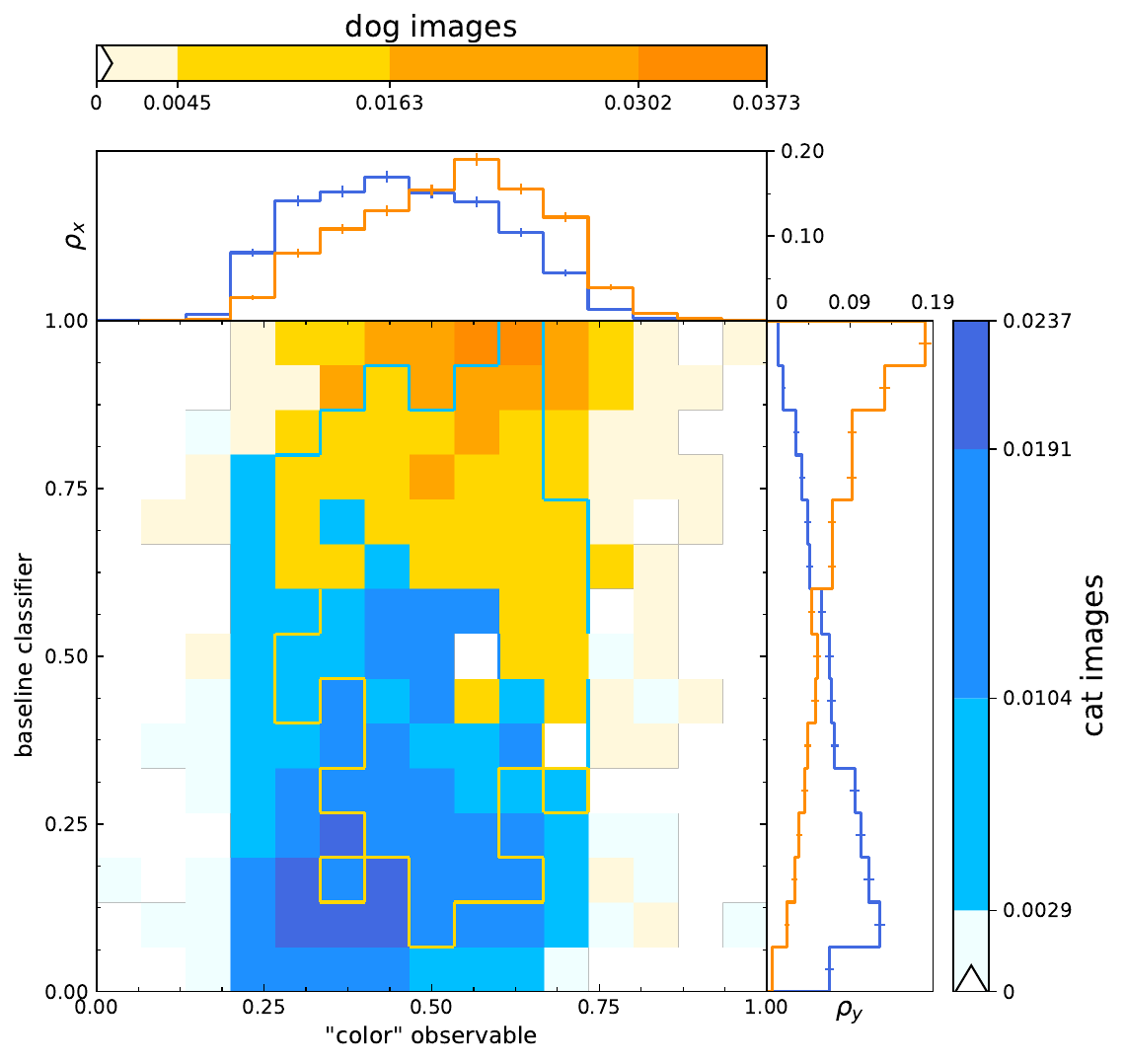}
\caption{Cat and dog images in the classifier ($y$) vs.\ color-observable ($x$) plane.}
\label{fig:CD_Y}
\end{figure}

The color observable $x$ is constructed as follows.
Each image is represented as a color histogram in the red--green plane
(Figs.~\ref{fig:RGCat} and~\ref{fig:RGDog}).
A separate CNN is trained on these color histograms; its scalar output serves
as $x$.
Classifier $y$ and color observable $x$ are shown in Fig.~\ref{fig:CD_Y};
both lie in $[0, 1]$.
Requiring $\gamma \perp\!\!\!\perp x$ means that selecting images with an
atypical color profile for their class will not bias the classifier output.

\begin{figure} [ht]
\centering
\includegraphics [width=0.81\textwidth]{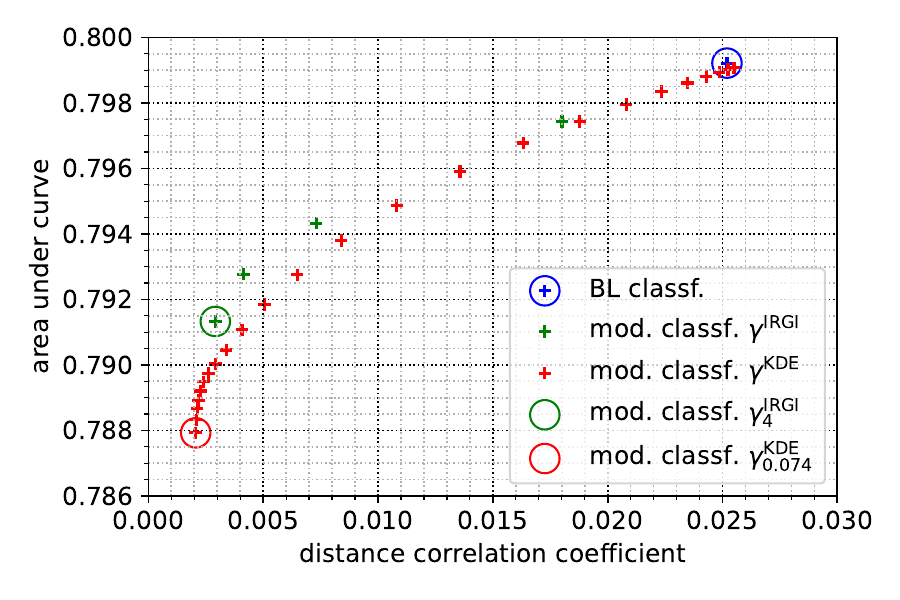}
\caption{IRGI and KDE modified classifiers in the DCC--AUC plane relative to the
         baseline. Classifiers of interest are marked with circles.}
\label{fig:CD_DA}
\end{figure}

To obtain decorrelated classifiers, IRGI and KDE are applied with varying free
parameters.
Unlike the purely synthetic case, changes in discriminating power (AUC) are also
monitored.
Fig.~\ref{fig:CD_DA} shows the modified classifiers in the DCC/AUC plane.
For IRGI, $d \in \{1, 2, 3, 4\}$; the lowest DCC is achieved at $d = 4$.
For KDE, $\sigma_r$ ranges from $e^{-9}$ to $e^{2}$ (logarithmic spacing);
the optimal value is $\sigma_r \approx 0.09$.

\begin{figure} [ht]
\centering
\includegraphics [width=0.81\textwidth]{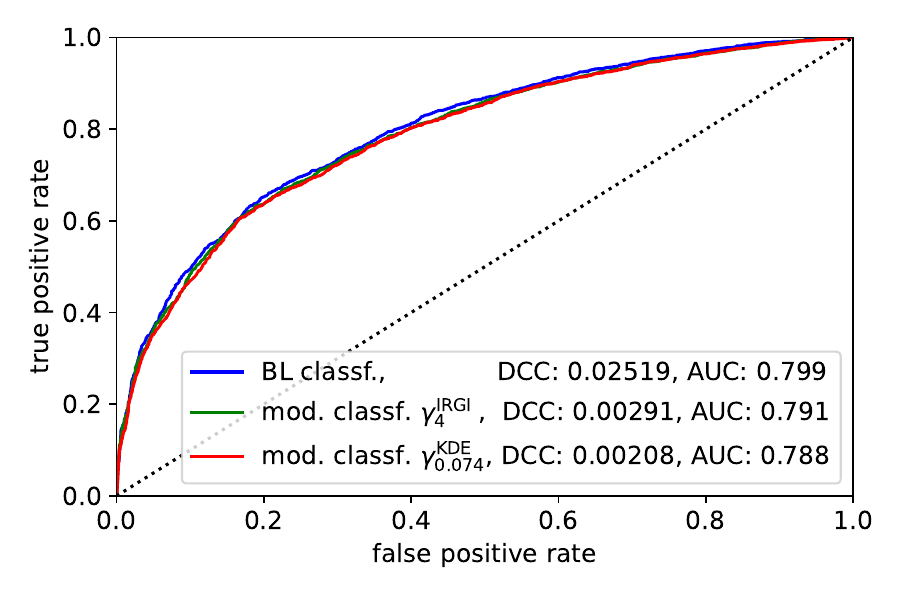}
\caption{ROC curves for the baseline and DCC-optimal IRGI and KDE classifiers.}
\label{fig:CD_ROC}
\end{figure}

Fig.~\ref{fig:CD_ROC} presents ROC curves for the baseline classifier,
IRGI-modified $\gamma_4$, and KDE-modified $\gamma_{0.09}$.
The AUC decrease is marginal in both cases.
DCC minimization is less complete than in the synthetic example, as expected
for a dataset two orders of magnitude smaller.

\begin{figure} [ht]
\centering
    \begin{subfigure}[t]{0.48\textwidth}
    \centering
    \includegraphics[width=0.99\textwidth]{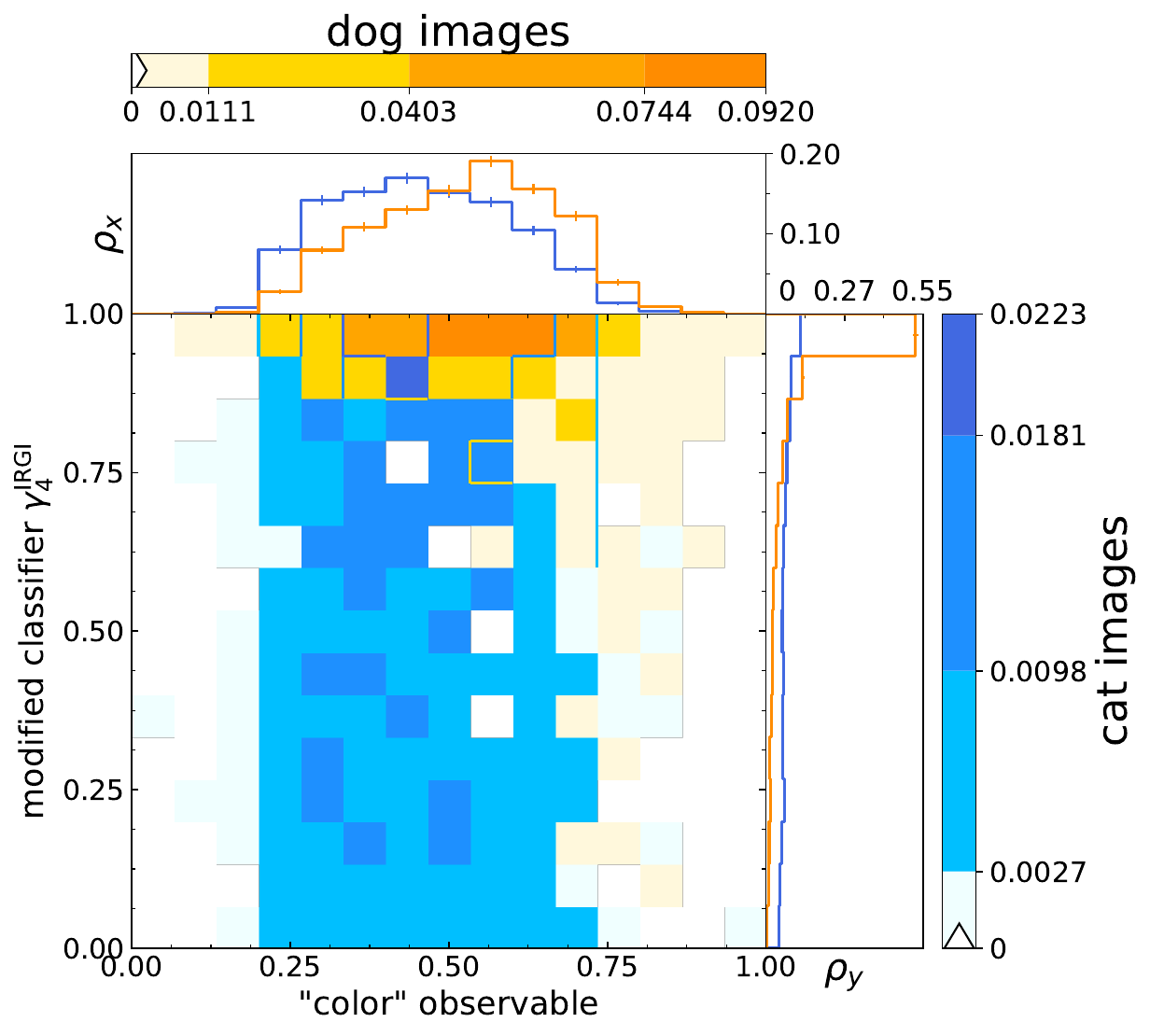}
    \caption{IRGI-modified classifier $\gamma_4$,}
    \end{subfigure}
\quad
    \begin{subfigure}[t]{0.48\textwidth}
    \centering
    \includegraphics[width=0.99\textwidth]{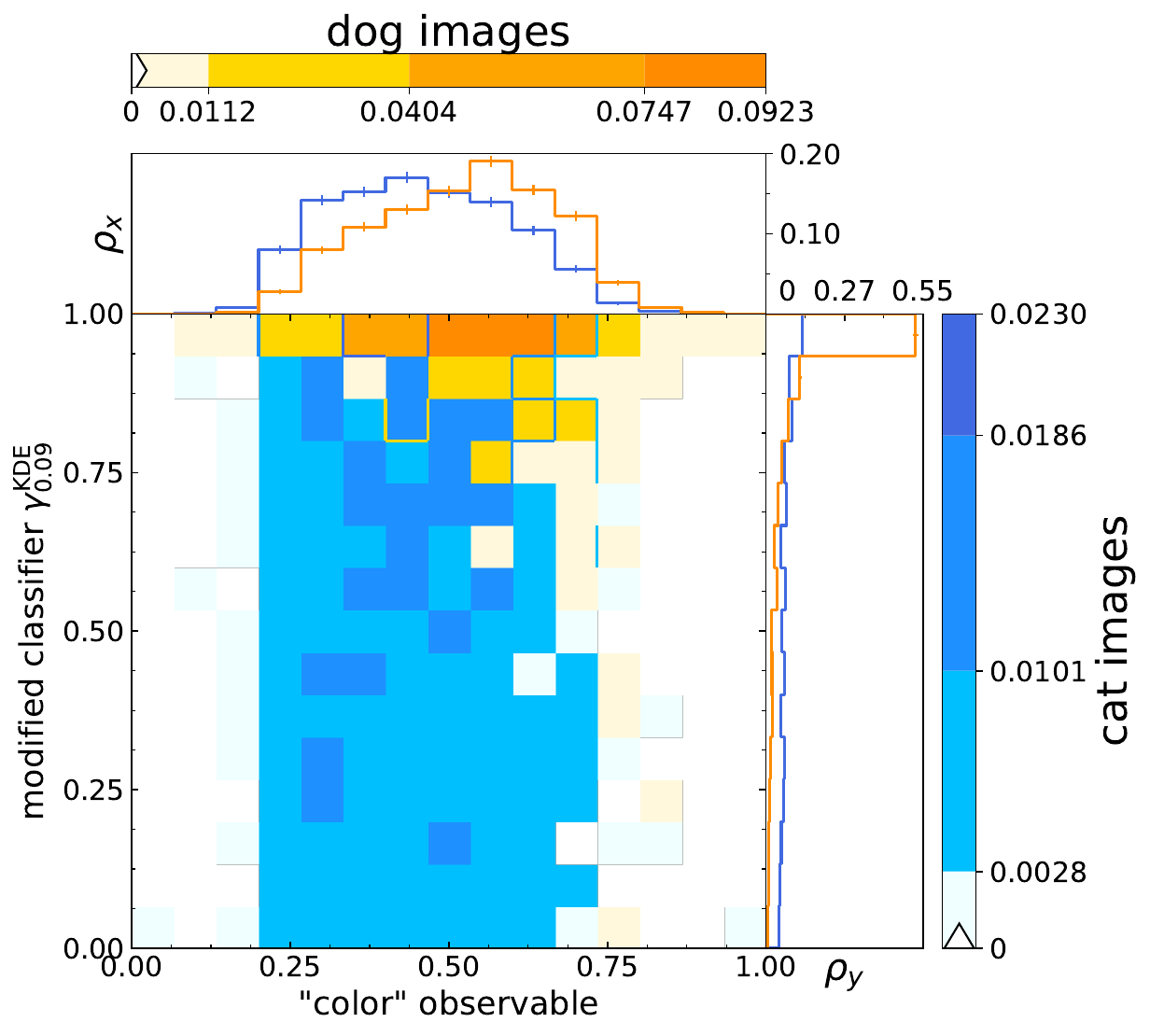}
    \caption{KDE-modified classifier $\gamma_{0.09}$.}
    \end{subfigure}
\caption{Cat and dog image samples in the color-observable vs.\
         quasi-independent classifier plane.}
\label{fig:GK}
\end{figure}

Fig.~\ref{fig:GK} shows the cat and dog samples in the color-observable versus
quasi-independent classifier plane for both implementations.

\subsection{High-Energy Physics Application}
\label{sec:hep}

\begin{figure} [ht]
\centering
    \begin{subfigure}[t]{0.48\textwidth}
    \centering
    \includegraphics[width=0.99\textwidth]{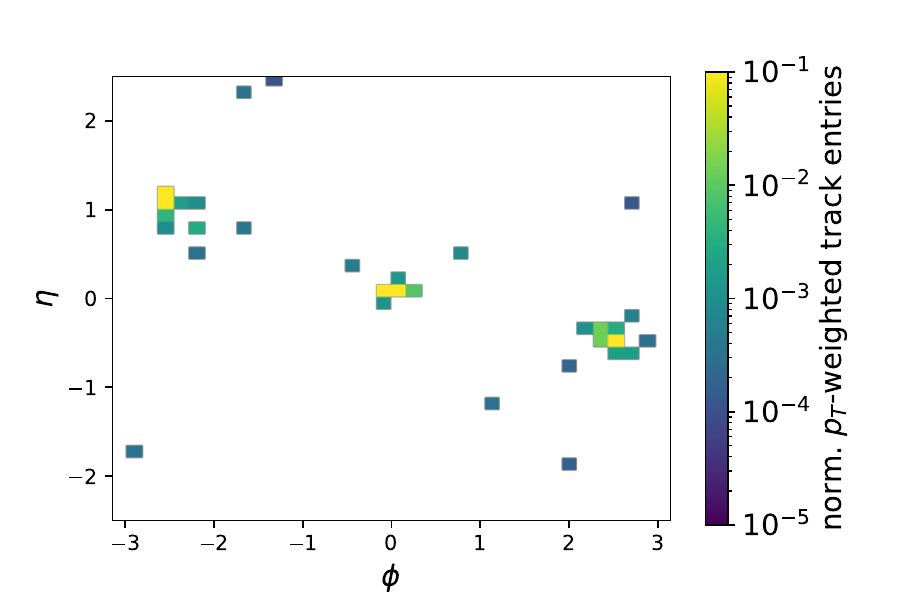}
    \caption{Random QCD dijet background event,}
    \end{subfigure}
\quad
    \begin{subfigure}[t]{0.48\textwidth}
    \centering
    \includegraphics[width=0.99\textwidth]{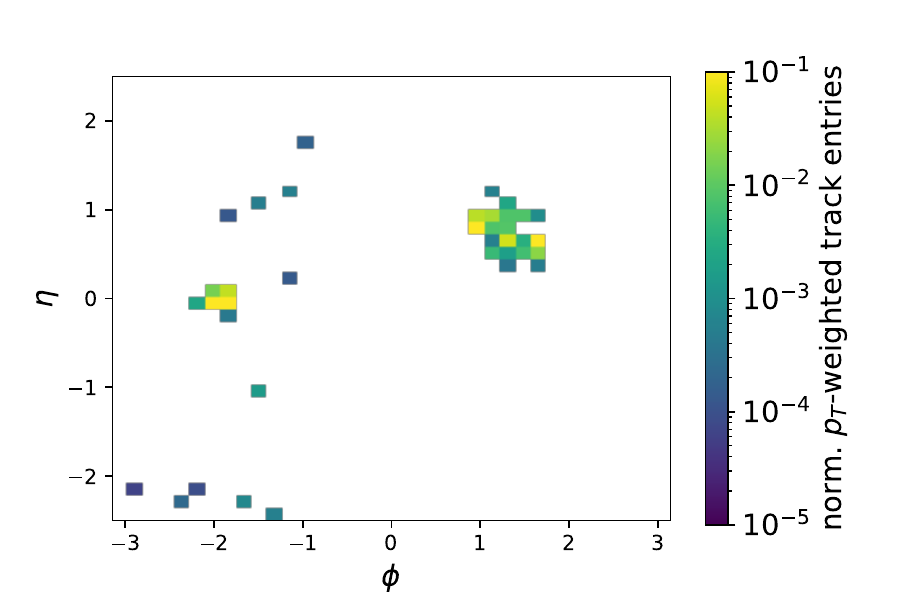}
    \caption{random $W'$ signal event.}
    \end{subfigure}
\caption{Final-state tracks binned in the $\eta\phi$ plane (unrolled cylindrical surface).}
\label{fig:TrImag}
\end{figure}

The primary target application is a search for the $W'$ resonance in QCD dijet
events using the LHC Olympics 2020 dataset~\cite{LHC_Olympics}.
The protected observable $x$ is the invariant mass of the leading dijet pair
(the ``bump-hunt'' variable~\cite{Nachman2020}).
The classifier $y$ is trained with a CNN on per-event track images, following the
approach of Ref.~\cite{Nachman2020}.
Track images are two-dimensional histograms of final-state tracks in the
$\eta\phi$ plane, treating all tracks as ultrarelativistic
(Fig.~\ref{fig:TrImag}); CNNs are well-suited to extract event-level features
from such representations.
All neural network training uses TensorFlow~\cite{tensorflow2015-whitepaper}.

The full dataset comprises $10^5$ signal and $10^6$ background events,
reflecting the lower signal cross-section.
These are partitioned as follows:
45k events per class for CNN training;
25k events per class as a CNN validation sample and as the defining background
sample for IRGI and KDE;
20k events per class as a free-parameter optimization sample
(minimizing DCC, maximizing AUC);
and 100k background plus 10k signal events as the final testing sample for
ABCD demonstration.

\begin{figure}[ht]
  \centering
  \includegraphics
    [width=0.59\textwidth, keepaspectratio, valign=b]{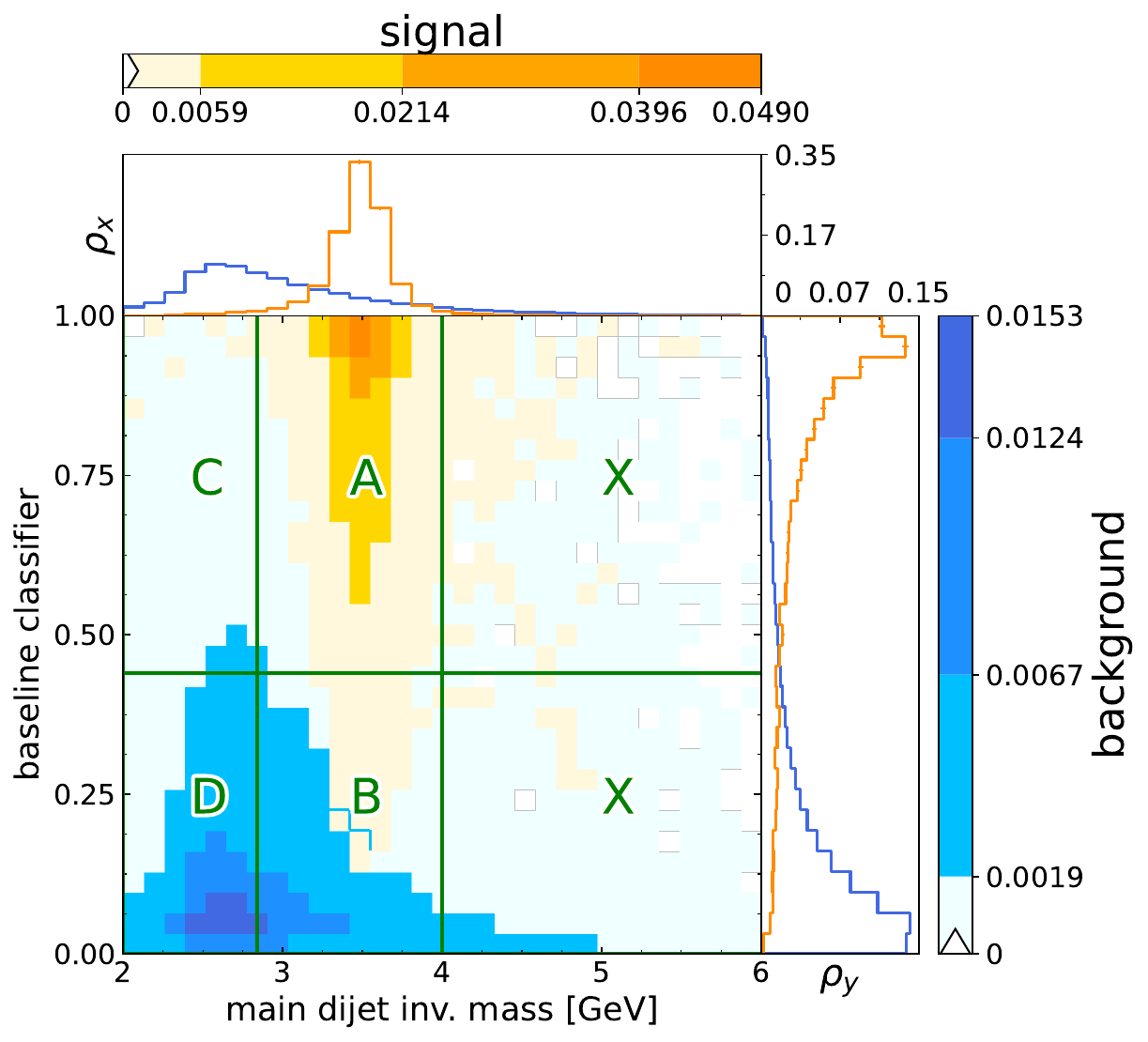}
  \includegraphics
    [width=0.388\textwidth, keepaspectratio, valign=b]{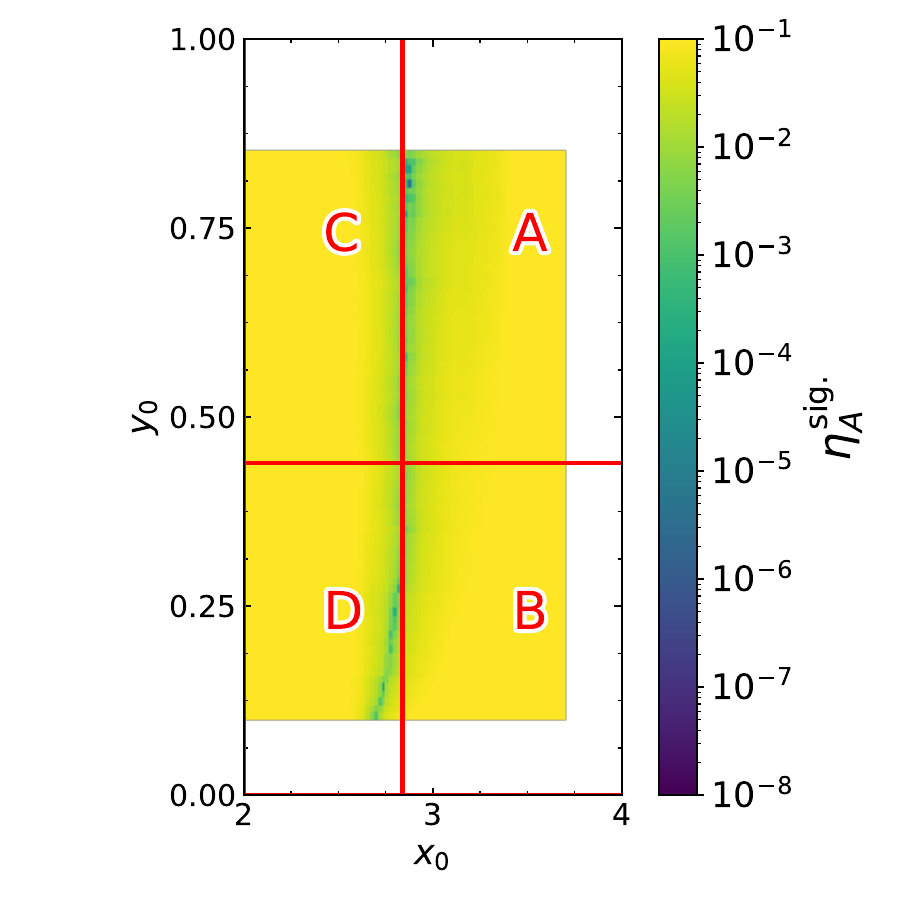}
  \caption{Left: testing sample in the observable vs.\ baseline-classifier plane;
           right: $|\eta^\mathrm{sig.}_A|$ as a function of $(x_0, y_0)$;
           the optimal cut point is marked.}
  \label{fig:RW_2D}
\end{figure}

Testing samples in the observable/classifier plane are shown in
Fig.~\ref{fig:RW_2D}.

The objective is to compare DisCo~\cite{DisCoFever,ABCDDisCo} with the proposed
IRGI and KDE implementations in terms of DCC and AUC on the validation sample,
seeking to minimize DCC while maximizing AUC.
Due to sensitivity of CNN classifiers to the random initialization of network
weights, training is repeated 10 times for each configuration: 10 runs for
unconstrained classifiers and 10 runs for each DisCo strength
$\alpha \in \{0, 0.5, 1.0, 1.5, \ldots, 10.0\}$ (21 values in steps of 0.5;
$\alpha$ is the DCC prefactor in the composite loss function).
Training uses batch size $10\,000$ (largest feasible for the DCC term to be
visible to the network), 30 epochs with early stopping.
The network architecture accepts $36\times36$ matrices through three convolutional
layers (stride 12, no pooling, no padding) and one dense layer.

\begin{figure} [ht]
\centering
\includegraphics [width=0.99\textwidth]{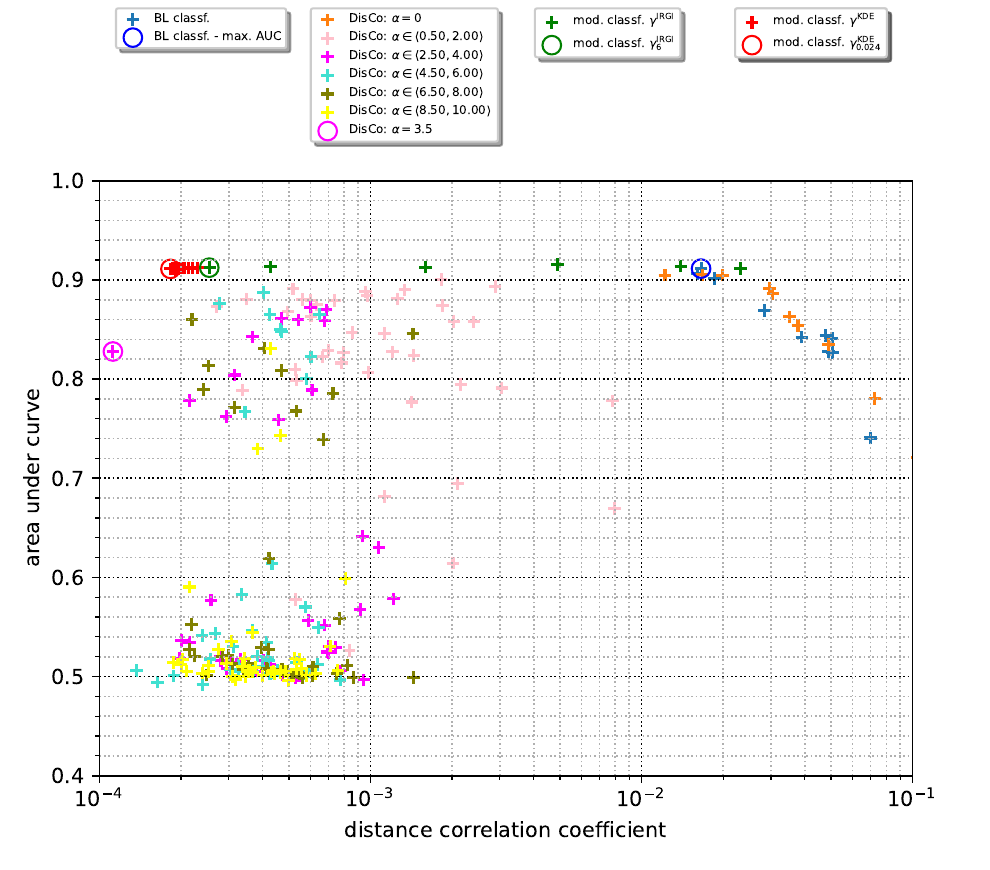}
\caption{
Each method occupies a characteristic region of the DCC-AUC plane, every point is a unique classifier; 10 classifiers are independently trained with the baseline (BL) CNN classifier method; 210 classifiers are independently trained with the $\textnormal{DisCo}_\alpha$ method, 10 per $\alpha \in \left\lbrace 0, 0.5, 1,... 10 \right\rbrace$. The best performing BL classifier (highest AUC, lowest DCC), and the best performing DisCo classifier ($\textnormal{AUC} > 0.8$, lowest DCC), are further circled. Other classifiers give a sense to the randomness involved in the training, and show a tendency towards collapse of the DisCo classifiers down to the pathological AUC value of $0.5$ (useless classifier). IRGI- and KDE-modified classifiers are constructed from the best performing BL classifier, with the free parameters varying over a finite range, retaining near-baseline AUC across the full range of achieved DCC values. The best performing IRGI- and KDE-modified classifiers (lowest DCC) are also further circled.}
\label{fig:WP_DA}
\end{figure}

Fig.~\ref{fig:WP_DA} shows all trained classifiers in the DCC--AUC plane.
Baseline classifiers (BL) show appreciable AUC spread due to random
initialization; the topmost baseline is chosen as the reference.
DisCo classifiers at $\alpha = 0$ reproduce baseline behavior; increasing $\alpha$
shifts classifiers toward lower DCC but at significant AUC cost and with high
instability — many runs collapse to AUC $\approx 0.5$.
IRGI and KDE are applied to modify only the single best baseline classifier,
with parameters $d \in \{1,\ldots,10\}$ and
$\sigma_r \in \{0.005, 0.0145, 0.024, \ldots, 0.1\}$.
Optimal parameters are $d = 6$ and $\sigma_r = 0.024$.
For the DisCo comparison, the classifier with DCC lower than IRGI and KDE at
maximal AUC is selected: this is the $\alpha = 3.5$ run — notably the only
DisCo classifier with comparable DCC that does not degrade to AUC $\approx 0.5$.

\begin{figure} [ht]
\centering
\includegraphics [width=0.81\textwidth]{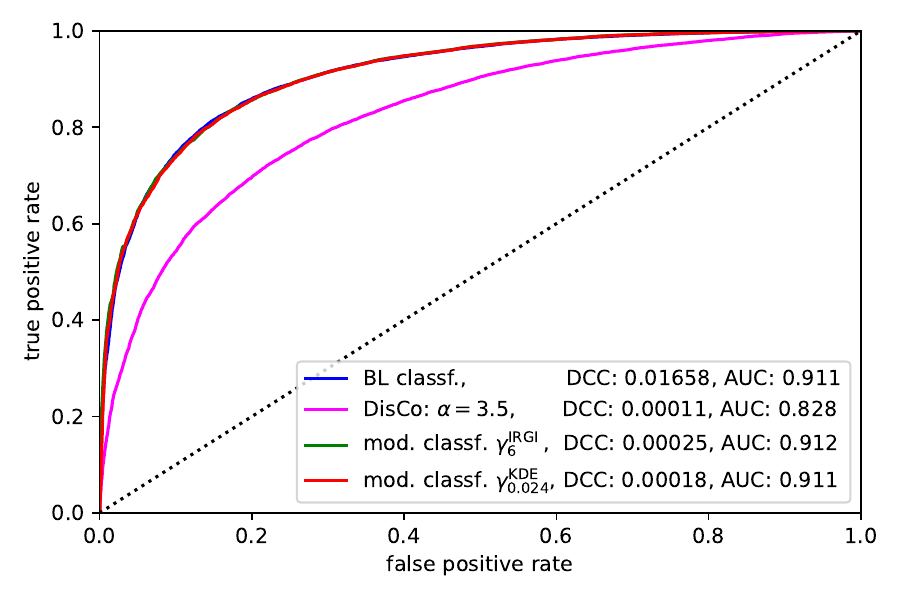}
\caption{ROC curves for the baseline, DisCo ($\alpha=3.5$), IRGI ($\gamma_6$),
         and KDE ($\gamma_{0.024}$) classifiers.}
\label{fig:WP_ROC}
\end{figure}

ROC curves for all classifiers are shown in Fig.~\ref{fig:WP_ROC}.

Both IRGI and KDE show minimal AUC loss from the best baseline classifier
regardless of the free parameter value, in contrast to DisCo classifiers,
which exhibit the AUC--DCC trade-off and require multiple training repetitions
per $\alpha$ value.

\paragraph{ABCD closure.}
Events with primary dijet invariant mass above $4\,\mathrm{TeV}$ are discarded due to
background and signal tailing.
The cut point $(x_0, y_0)$ dividing the phase plane into ABCD quadrants is varied:

\begin{equation}
\begin{split}
A:& \; x \geq x_0,\; y \geq y_0,\\
B:& \; x \geq x_0,\; y <   y_0,\\
C:& \; x <   x_0,\; y \geq y_0,\\
D:& \; x <   x_0,\; y <   y_0,
\end{split}
\end{equation}

over the grid $x_0 \in [2.0, 3.7]\,\mathrm{TeV}$,
$y_0 \in [0.1, 0.85]$,
each discretized to 100 bins.
The relative signal non-closure in region $A$ is

\begin{equation}
\eta^\mathrm{sig.}_A = \frac{N^\mathrm{sig.}_A}{N^\mathrm{tot.}_A}
- \left(1 - \frac{N^\mathrm{tot.}_B N^\mathrm{tot.}_C}
               {N^\mathrm{tot.}_A N^\mathrm{tot.}_D}\right),
\end{equation}

where Eq.~\eqref{eq:sigAest} provides the background estimate.
This measures non-closure of the signal estimate rather than the background
estimate; the two are equivalent up to sign when signal contamination outside
region $A$ is negligible.

\begin{figure} [ht]
\centering
    \begin{subfigure}[t]{0.48\textwidth}
    \centering
    \includegraphics[width=0.99\textwidth]{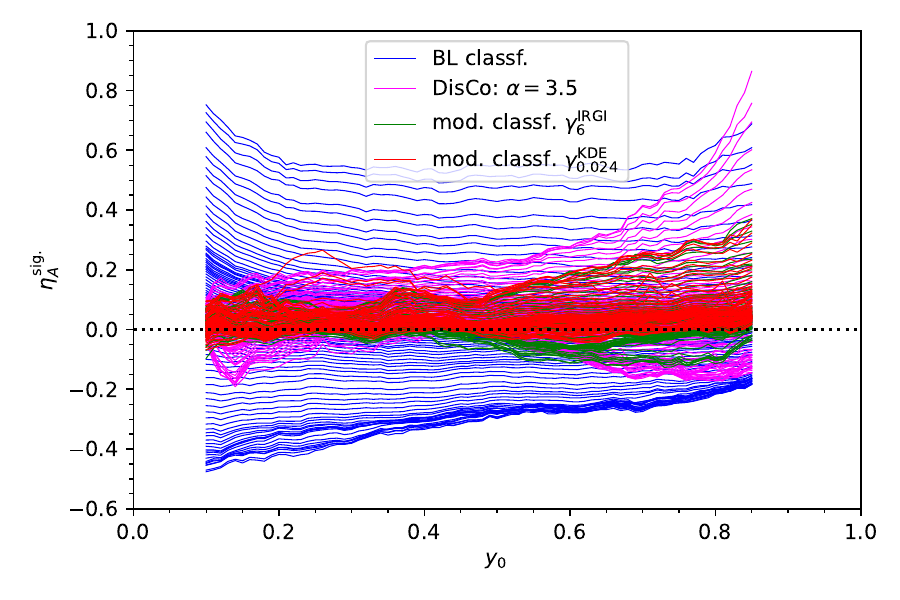}
    \caption{$\eta^\mathrm{sig.}_A$ versus $y_0$; each line corresponds to an $x_0$ value,}
    \end{subfigure}
\quad
    \begin{subfigure}[t]{0.48\textwidth}
    \centering
    \includegraphics[width=0.99\textwidth]{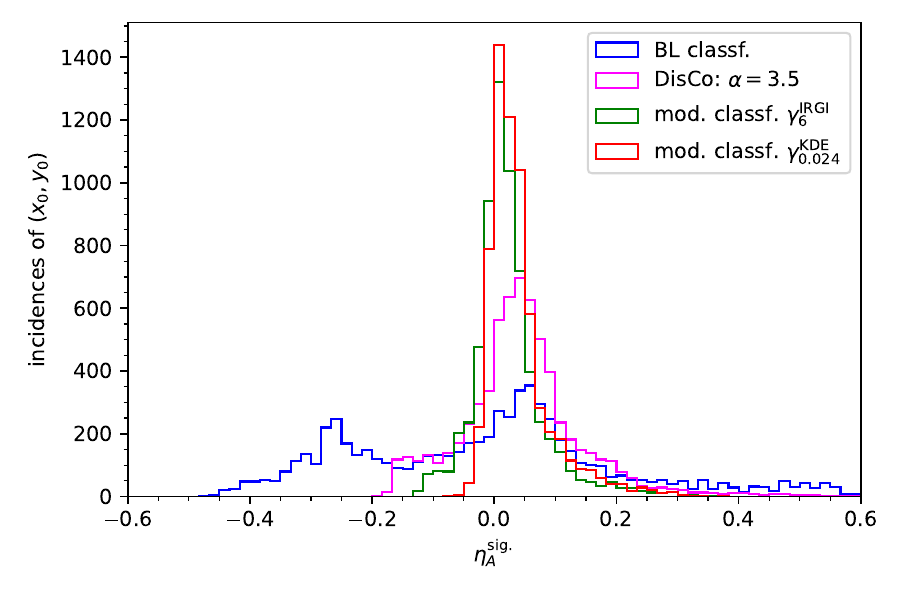}
    \caption{histogram of $\eta^\mathrm{sig.}_A$ for all $(x_0, y_0)$ points.}
    \end{subfigure}
\caption{Relative error in signal estimation for different $(x_0, y_0)$ and distinct classifiers.}
\label{fig:x0y0}
\end{figure}

For each classifier, the $(x_0, y_0)$ minimizing $|\eta^\mathrm{sig.}_A|$ is
identified; the minimum value quantifies ABCD performance.
Note that this reports the \emph{best-case} closure at the optimal cut point;
the histogram of $\eta^\mathrm{sig.}_A$ over all cut points (Fig.~\ref{fig:x0y0},
right) characterizes robustness.

For the baseline classifier, Fig.~\ref{fig:RW_2D} (left) shows the testing sample
in the observable/classifier plane; (right) shows $|\eta^\mathrm{sig.}_A|$ over~the~$(x_0, y_0)$ grid,
with the optimal cut marked.

\begin{figure}[ht]
  \centering
  \includegraphics
    [width=0.59\textwidth, keepaspectratio, valign=b]{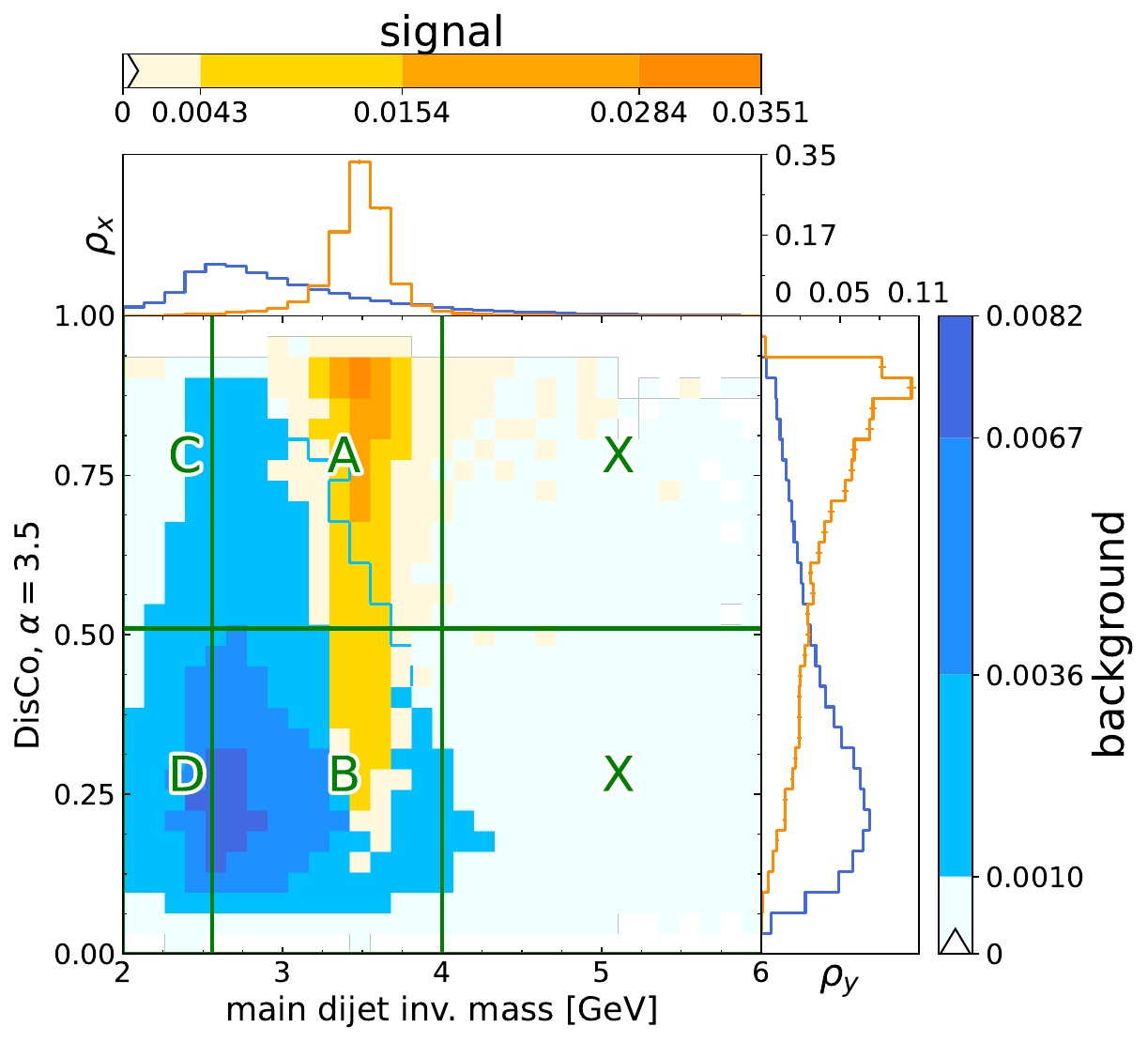}
  \includegraphics
    [width=0.388\textwidth, keepaspectratio, valign=b]{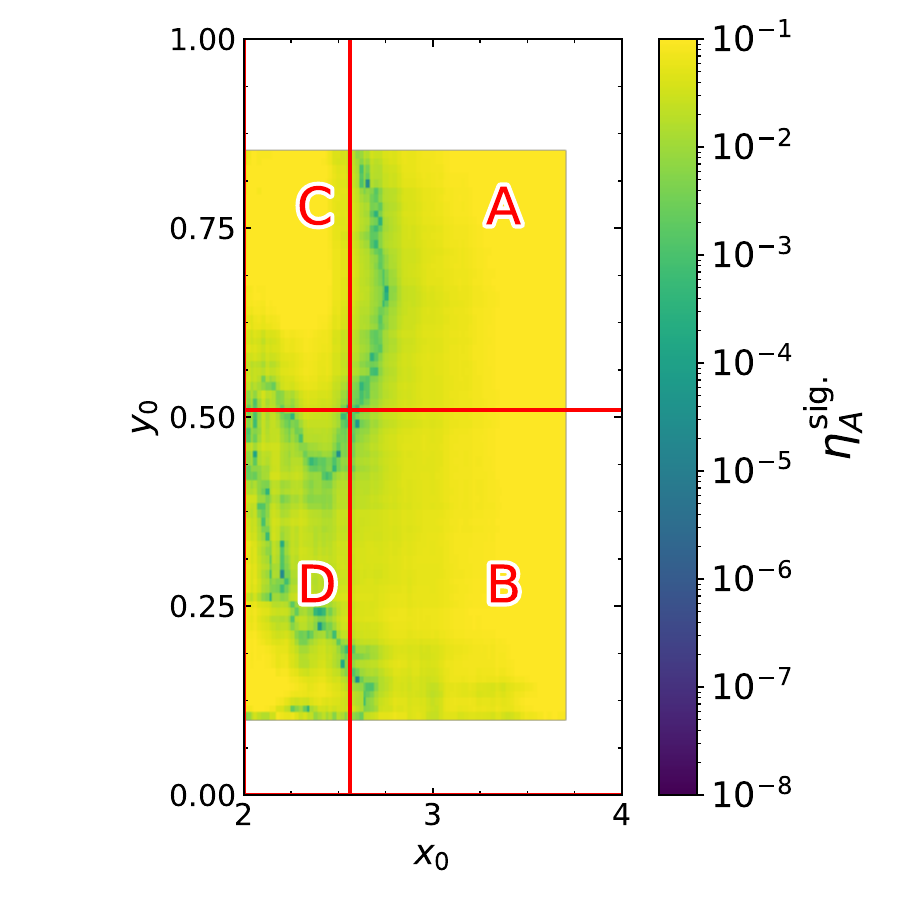}
  \caption{As Fig.~\ref{fig:RW_2D} for the DisCo classifier.}
  \label{fig:DC_2D}
\end{figure}

\begin{figure}[ht]
  \centering
  \includegraphics
    [width=0.59\textwidth, keepaspectratio, valign=b]{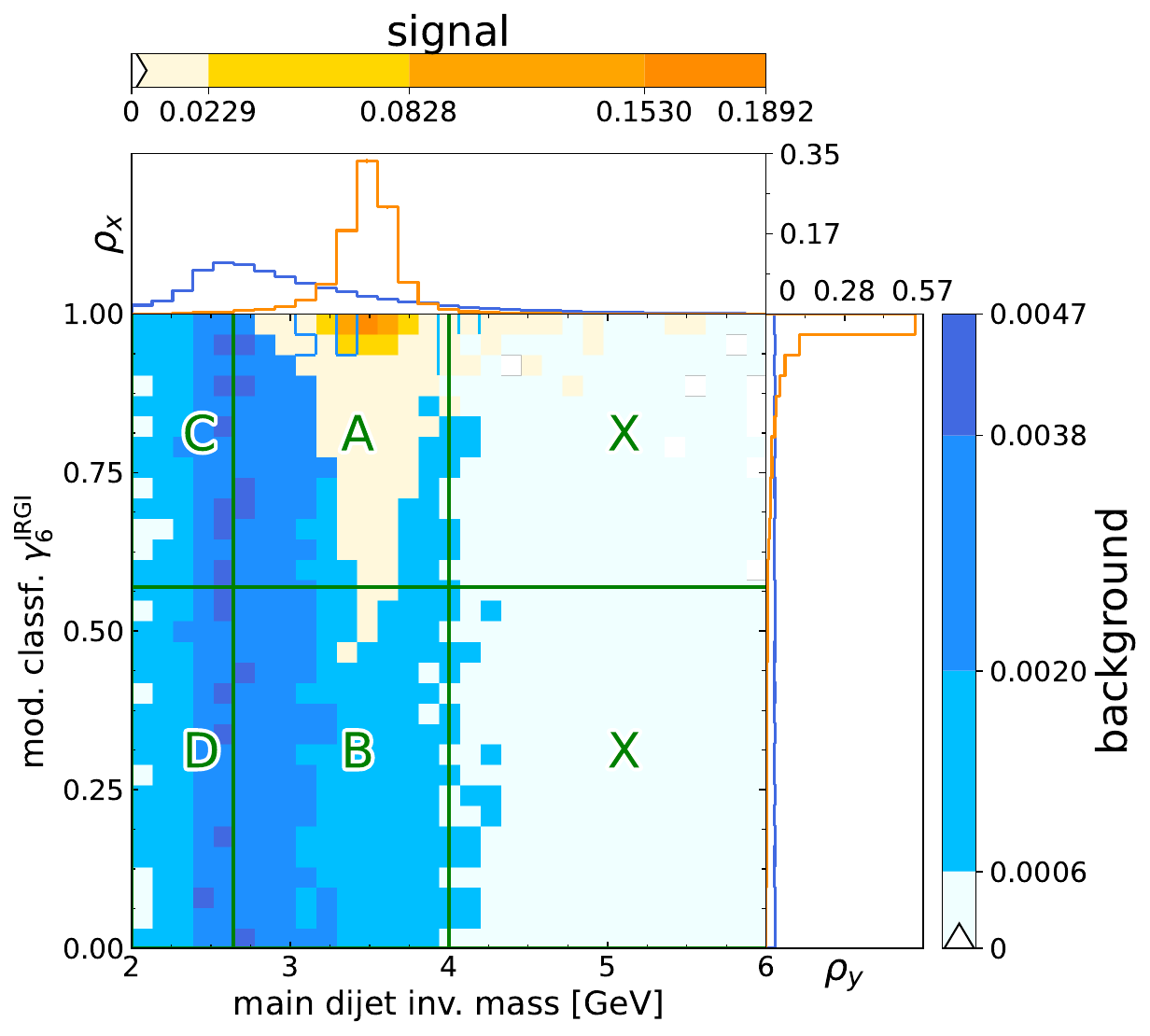}
  \includegraphics
    [width=0.388\textwidth, keepaspectratio, valign=b]{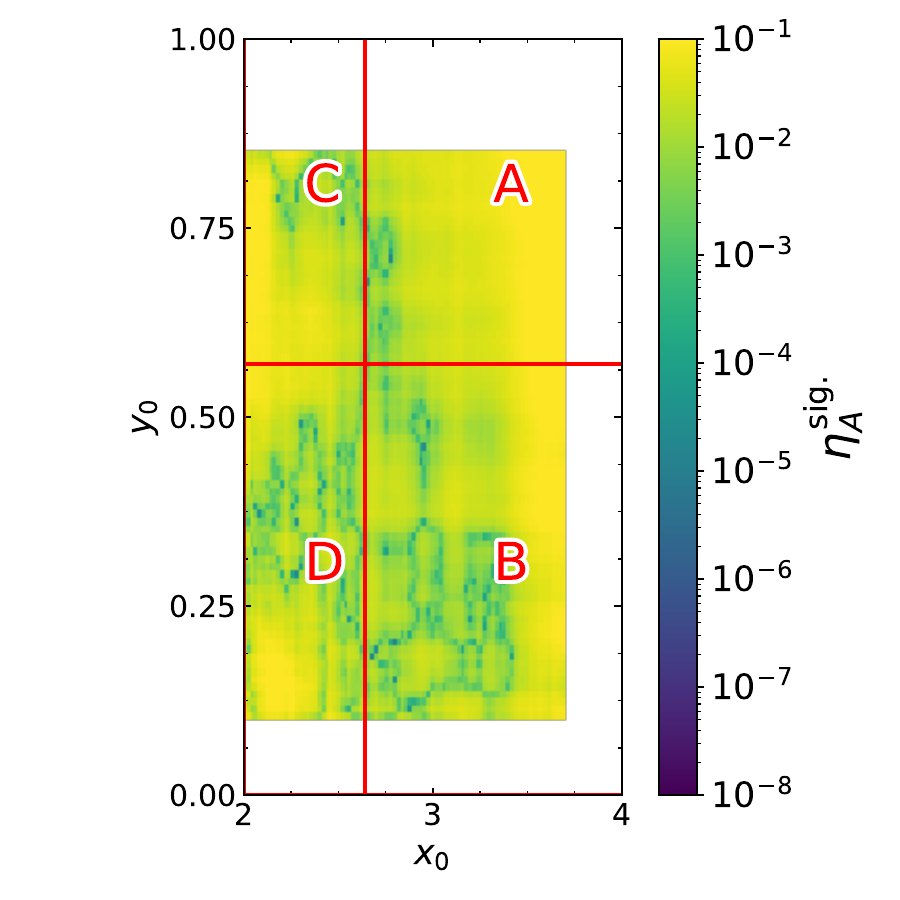}
  \caption{As Fig.~\ref{fig:RW_2D} for the IRGI-modified classifier $\gamma_6$.}
  \label{fig:GM_2D}
\end{figure}

\begin{figure}[ht]
  \centering
  \includegraphics
    [width=0.59\textwidth, keepaspectratio, valign=b]{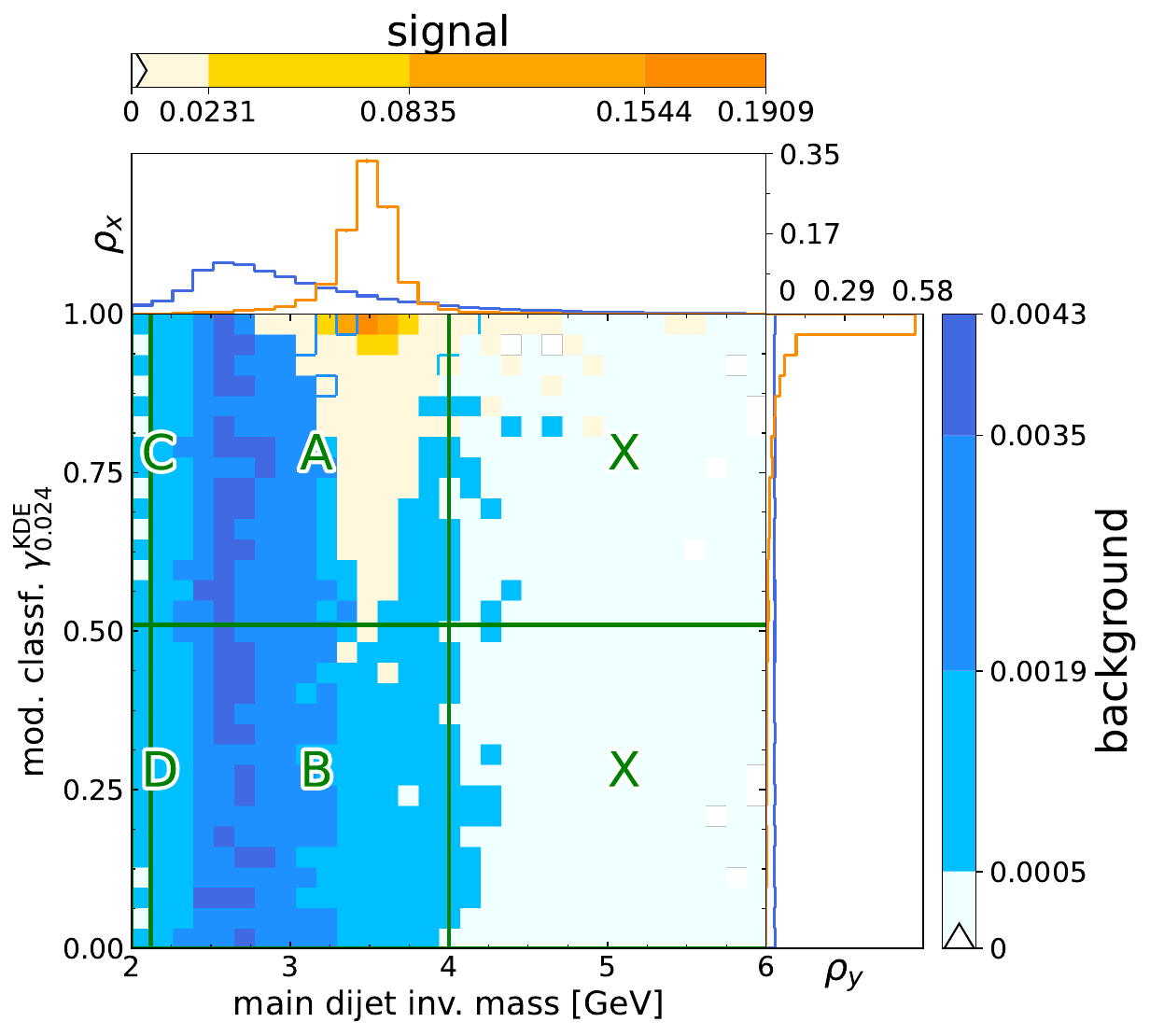}
  \includegraphics
    [width=0.388\textwidth, keepaspectratio, valign=b]{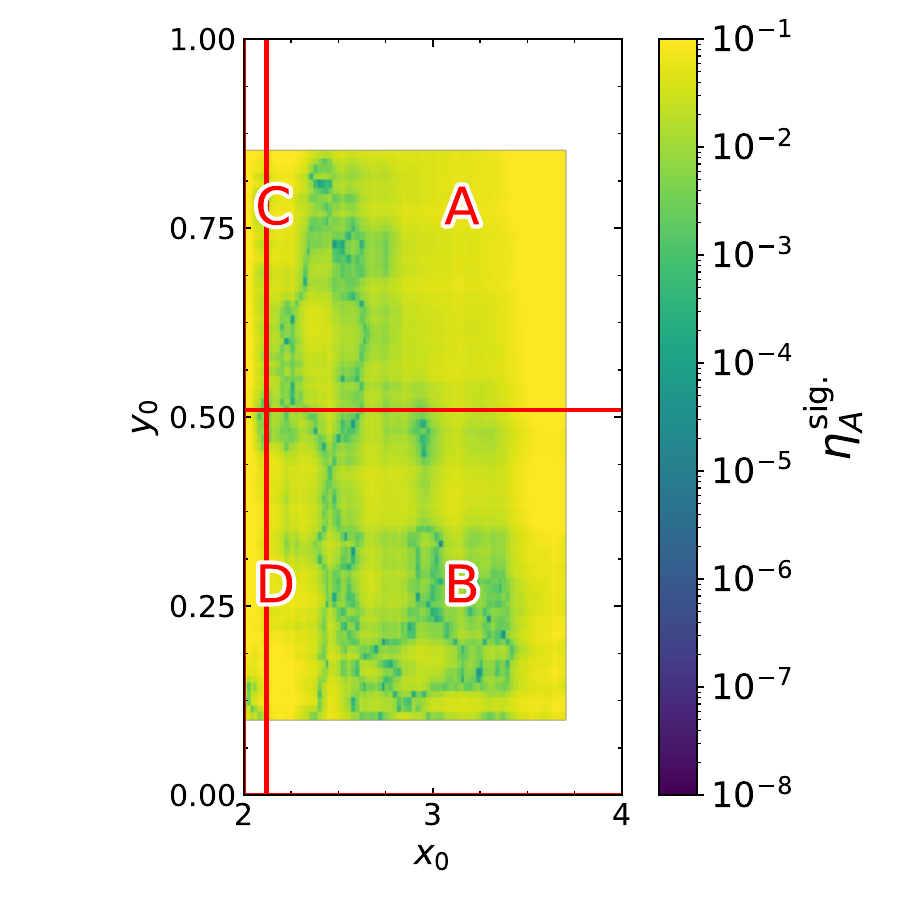}
  \caption{As Fig.~\ref{fig:RW_2D} for the KDE-modified classifier $\gamma_{0.024}$.}
  \label{fig:KM_2D}
\end{figure}

Analogous figures for the DisCo, IRGI, and KDE classifiers appear in
Figs.~\ref{fig:DC_2D}, \ref{fig:GM_2D}, and~\ref{fig:KM_2D}, respectively.

ABCD performance at the optimal cut point is summarized in
Table~\ref{tab:WP}.

\begin{table}[ht]
\caption{ABCD region $A$ signal estimation at the optimal $(x_0, y_0)$ cut for
         each classifier.
         The optimal cut is defined as the one minimizing $|\eta^\mathrm{sig.}_A|$.}
\label{tab:WP}
\begin{ruledtabular}
\begin{tabular}{lcccc}
 & Baseline & DisCo ($\alpha=3.5$) & IRGI $\gamma_6$ & KDE $\gamma_{0.024}$ \\
\hline
$\eta^\mathrm{sig.}_A$  & $-6.3\!\times\!10^{-6}$ & $-1.9\!\times\!10^{-5}$ & $2.0\!\times\!10^{-6}$ & $-4.9\!\times\!10^{-6}$ \\
Total events  & $15{,}597$ & $23{,}283$ & $34{,}609$ & $52{,}472$ \\
True signal   & $7{,}973$  & $7{,}236$  & $9{,}237$  & $9{,}446$ \\
Signal estimate & $7{,}973.1$ & $7{,}236.5$ & $9{,}236.9$ & $9{,}446.3$ \\
\end{tabular}
\end{ruledtabular}
\end{table}
\clearpage

Table~\ref{tab:WP} shows that, at the optimal cut, all four classifiers achieve
$|\eta^\mathrm{sig.}_A| \sim 10^{-5}$--$10^{-6}$.
The IRGI-modified $\gamma_6$ provides the lowest non-closure ($2.0\times 10^{-6}$),
approximately three times smaller than the baseline.
The DisCo classifier at $\alpha = 3.5$ achieves the highest non-closure
($1.9\times 10^{-5}$), one order of magnitude worse than the baseline — further
evidence of the instability induced by DisCo training at large $\alpha$.
The principal advantage of classifier quasi-independence is not the absolute
minimum non-closure but the substantially reduced sensitivity to the ABCD cut
choice (Fig.~\ref{fig:x0y0}); the fraction of $(x_0, y_0)$ cut points yielding
$|\eta^\mathrm{sig.}_A|$ below any fixed tolerance is substantially larger for
IRGI and KDE than for the baseline or DisCo.

\section{Conclusions}
\label{sec:conclusions}

A post-hoc numerical implementation of the Rosenblatt transform for enforcing
statistical quasi-independence between a classifier and a protected observable
has been presented.
Two complementary implementations are provided: Irregular Grid Interpolation
(IRGI), based on adaptive recursive bisection and bilinear interpolation, and
Kernel Density Estimation (KDE), based on Gaussian kernel smoothing of the
conditional CDF.
Both require only a finite defining background sample and impose no constraints
on classifier training.

The method was validated on three examples.
For synthetic tilted-Gaussian distributions, both IRGI and KDE reduce the DCC
by more than three orders of magnitude (Table~\ref{tab:blob}), with KDE
achieving marginally lower residual DCC.
In the LHC Olympics 2020 QCD/$W'$ benchmark — the primary target application
— IRGI and KDE match or surpass DisCo in residual DCC while preserving
near-baseline AUC and exhibiting far superior training stability
(Figs.~\ref{fig:WP_DA} and~\ref{fig:WP_ROC}).
ABCD closure errors of order $10^{-6}$ are achieved for both post-hoc methods,
with the IRGI result approximately three times better than the baseline and
DisCo at $\alpha = 3.5$ performing an order of magnitude worse
(Table~\ref{tab:WP}).
The decisive practical advantage is the reduced sensitivity to the ABCD cut
choice: IRGI- and KDE-modified classifiers yield $|\eta^\mathrm{sig.}_A| \ll 1$
over a far larger fraction of the $(x_0, y_0)$ parameter space than the
unmodified or DisCo-trained classifier.

\textit{Limitations.}
The current formulation handles only a scalar protected variable $x$.
Extension to a vector $\vec{x}$ would require sequential application along each
dimension, for which the independence proof does not straightforwardly generalize.
The method assumes a background defining sample free of signal contamination;
robustness under contamination has not been quantified here.
In the HEP application, the CNN validation sample and the IRGI/KDE defining
sample coincide (25k events each); this is a practical choice driven by dataset
size, but a fully independent partition would be preferable.
Finally, DCC values reported in Table~\ref{tab:blob} are point estimates;
formal uncertainty propagation through the Rosenblatt transform is deferred to
future work.

\textit{Future directions.}
Planned extensions include:
(i) multi-dimensional decorrelation via sequential or joint application;
(ii) a data-driven procedure for free parameter selection without a held-out
validation sample;
(iii) uncertainty propagation for use in profile-likelihood fits;
and (iv) application within a full LHC analysis using real collision data.

\textit{Broader applicability.}
Although validated in the particle physics context, the IRGI and KDE
implementations require only a reference sample and a scalar observable pair
$(x, y)$, and are therefore immediately applicable in any domain where a
predictor must be decorrelated from a protected attribute.
Potential applications include fairness-aware classifiers (protected attributes:
gender, age, ethnicity), medical diagnostic models (protected attributes:
patient age, imaging protocol), and photometric redshift estimators in
astrophysics (protected attribute: apparent magnitude).
Code and documentation will be made available upon publication.

\begin{acknowledgments}
The authors gratefully acknowledge support from project LM2023040.
They thank Marek Omelka (Faculty of Mathematics and Physics, Charles University,
Department of Probability and Mathematical Statistics) for pointing out the
existing Rosenblatt transform~\cite{Rosenblatt1952}, which the authors had
independently derived.
\end{acknowledgments}

\paragraph*{Open access.}
This article is submitted for publication under open access.
The authors request publication under a Creative Commons Attribution~4.0
International (CC-BY~4.0) license, permitting unrestricted use, distribution,
and reproduction in any medium provided the original work is properly cited.
As a high-energy physics methods paper, it is eligible for fee-free gold open
access through the SCOAP$^3$ initiative in participating journals.

\paragraph*{Data and code availability.}
The LHC Olympics~2020 dataset~\cite{LHC_Olympics} and the CIFAR-10
dataset~\cite{cif} are publicly available.
Implementations of the IRGI and KDE transforms will be made available in a
public repository upon publication.

\bibliography{reference}

\appendix

\section{IRGI Technique: Complete Parameter Sweep}
\label{apx:d1to7}

\begin{figure} [ht]
\centering
    \begin{subfigure}[t]{0.48\textwidth}
    \centering
    \includegraphics[width=0.99\textwidth]{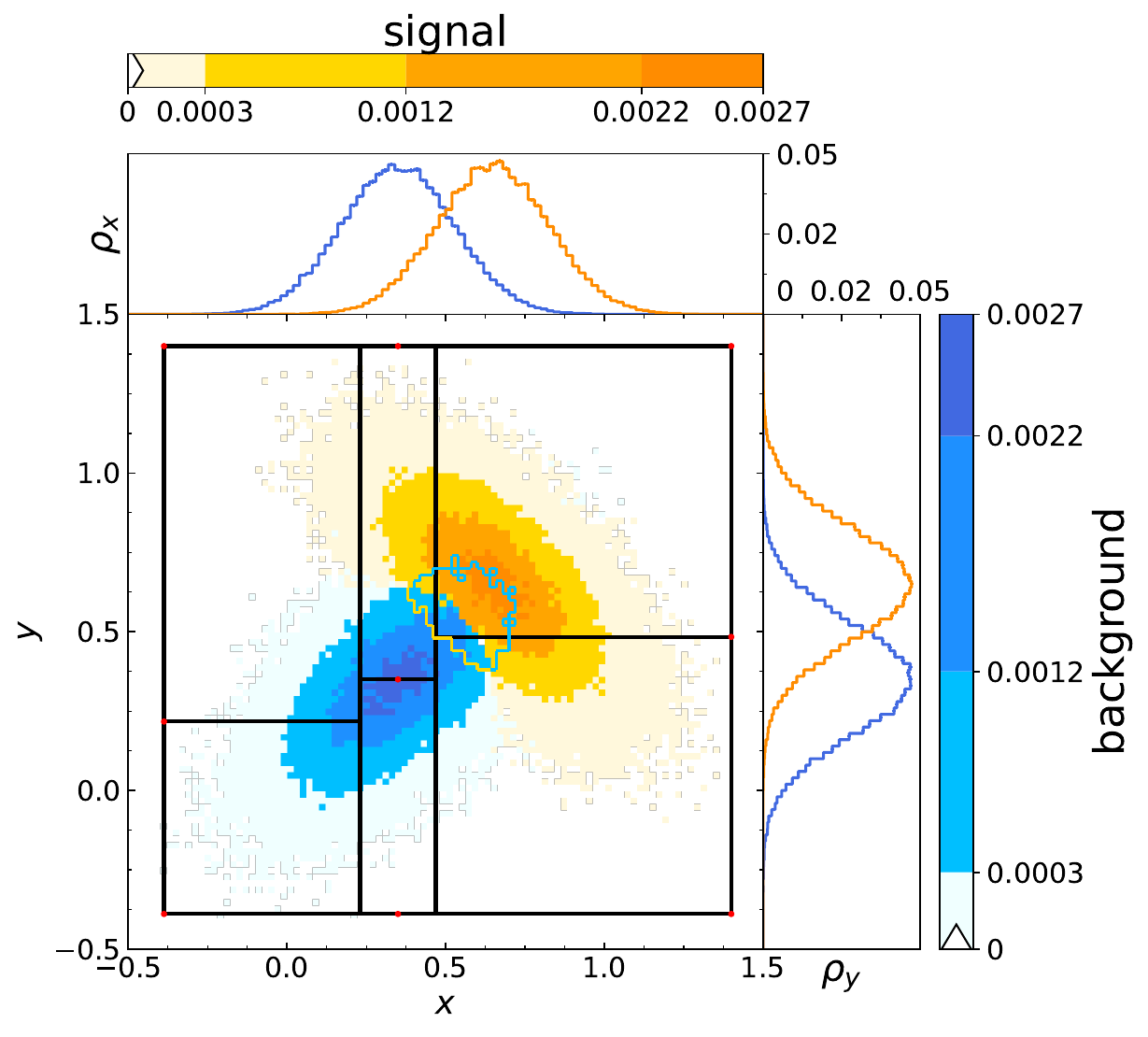}
    \caption{Defining background sample with irregular grid for $d = 1$,}
    \end{subfigure}
\quad
    \begin{subfigure}[t]{0.48\textwidth}
    \centering
    \includegraphics[width=0.99\textwidth]{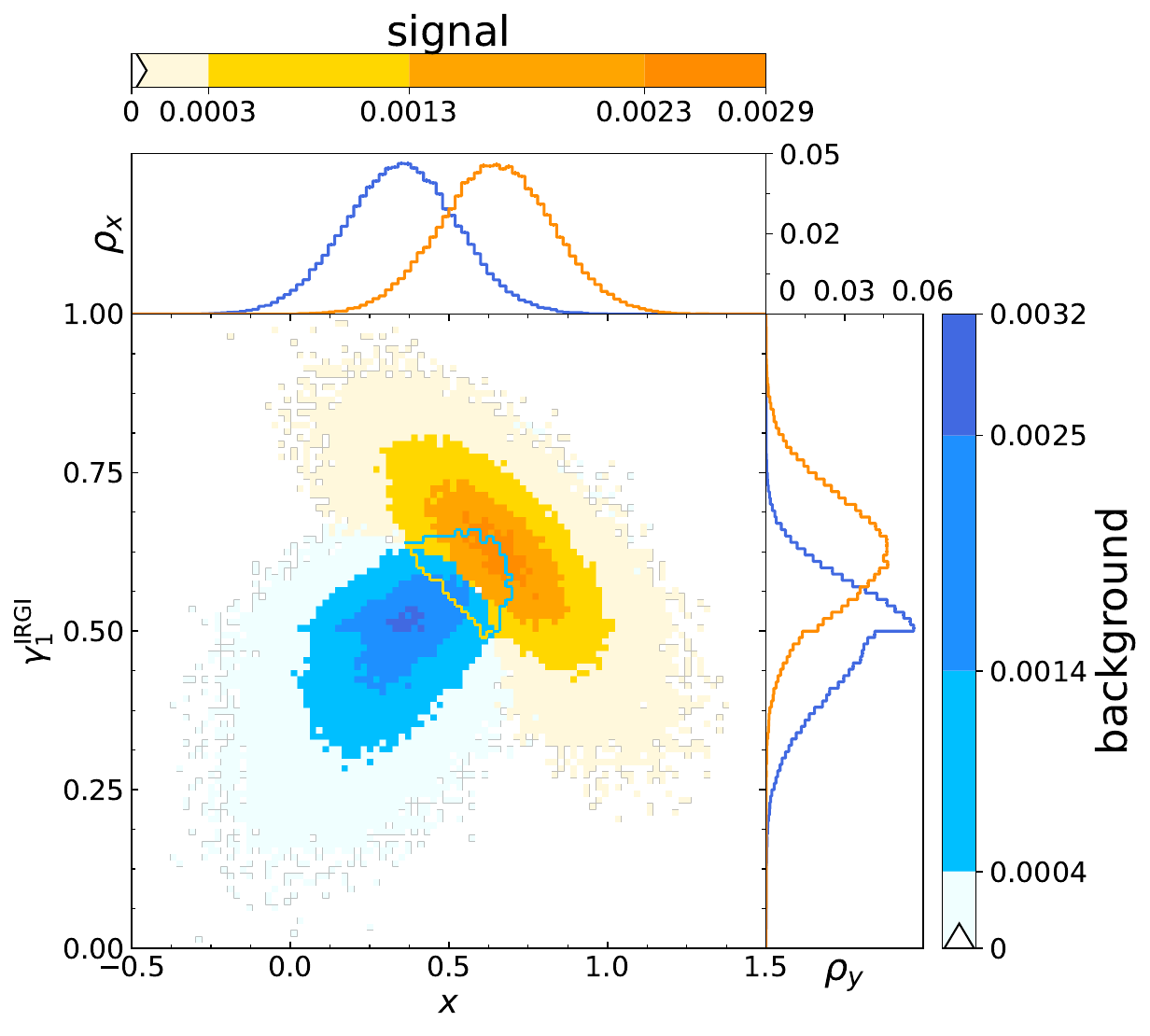}
    \caption{IRGI-modified classifier $\gamma_1$ applied to testing sample.}
    \end{subfigure}
\caption{IRGI technique, $d=1$.}
\label{fig:IRGI1A}
\end{figure}

\begin{figure}
\centering
    \begin{subfigure}[t]{0.48\textwidth}
    \centering
    \includegraphics[width=0.99\textwidth]{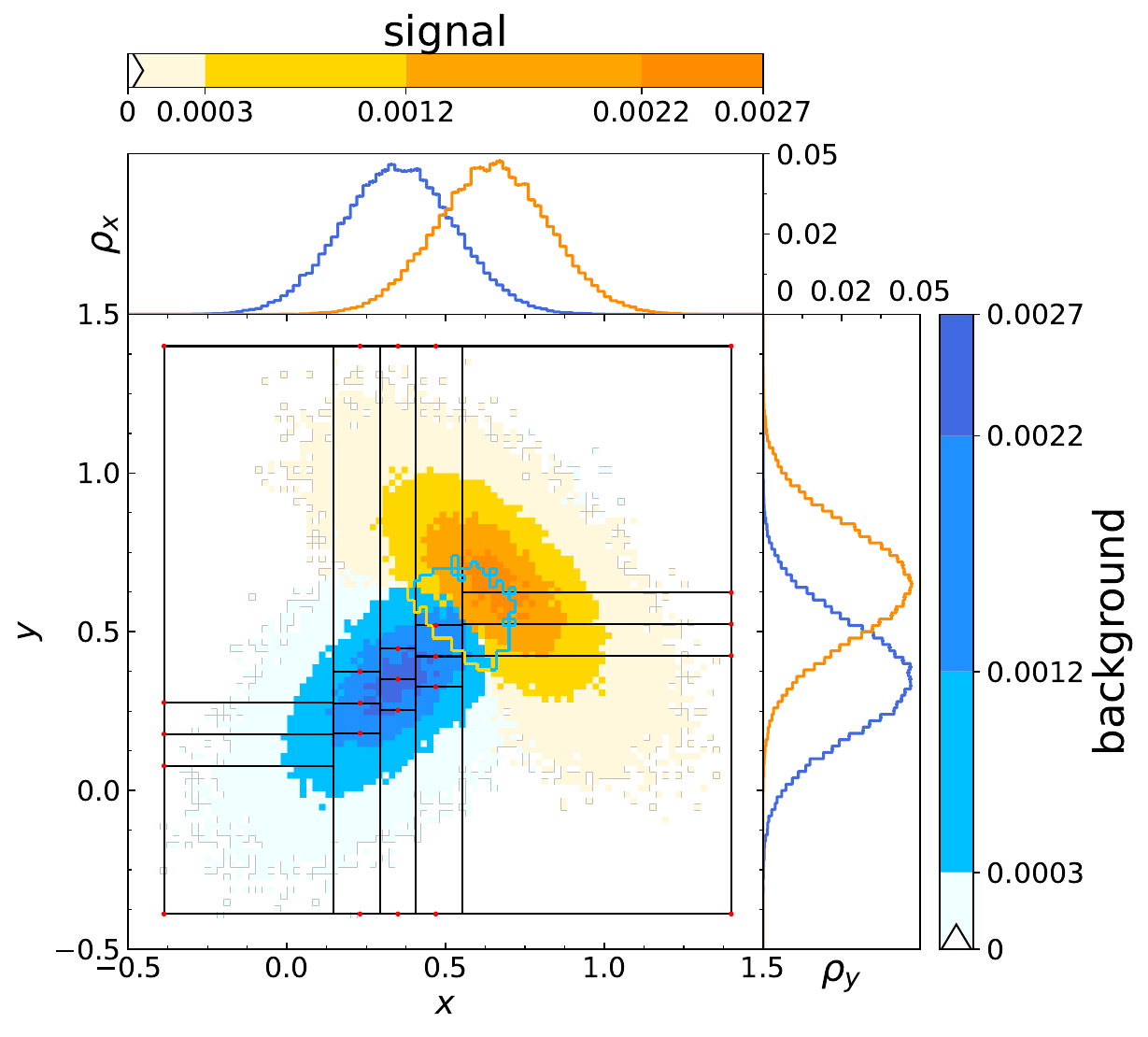}
    \caption{Defining background sample with irregular grid for $d = 2$,}
    \end{subfigure}
\quad
    \begin{subfigure}[t]{0.48\textwidth}
    \centering
    \includegraphics[width=0.99\textwidth]{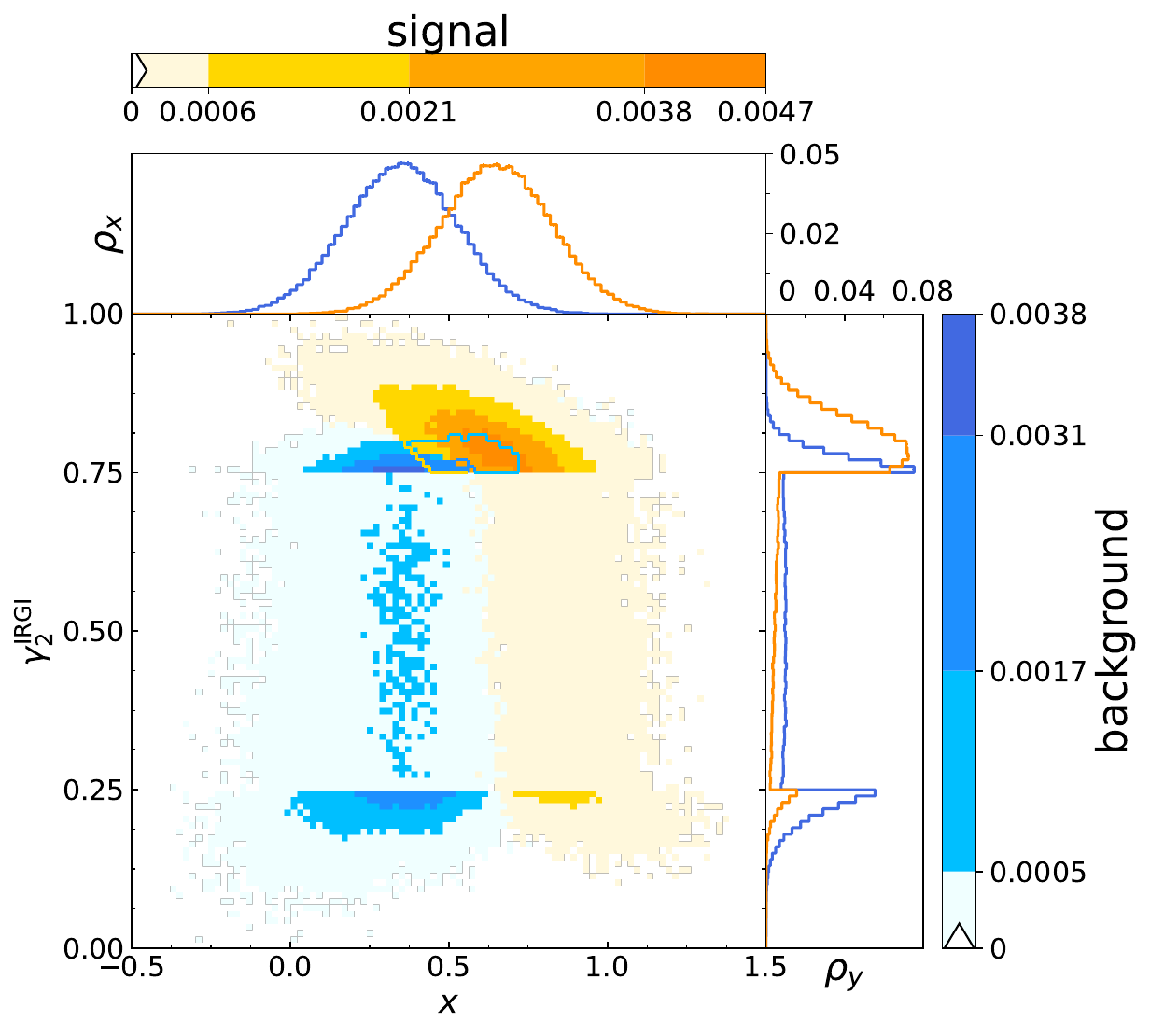}
    \caption{IRGI-modified classifier $\gamma_2$ applied to testing sample.}
    \end{subfigure}
\caption{IRGI technique, $d=2$.}
\label{fig:IRGI2A}
\end{figure}

\begin{figure}
\centering
    \begin{subfigure}[t]{0.48\textwidth}
    \centering
    \includegraphics[width=0.99\textwidth]{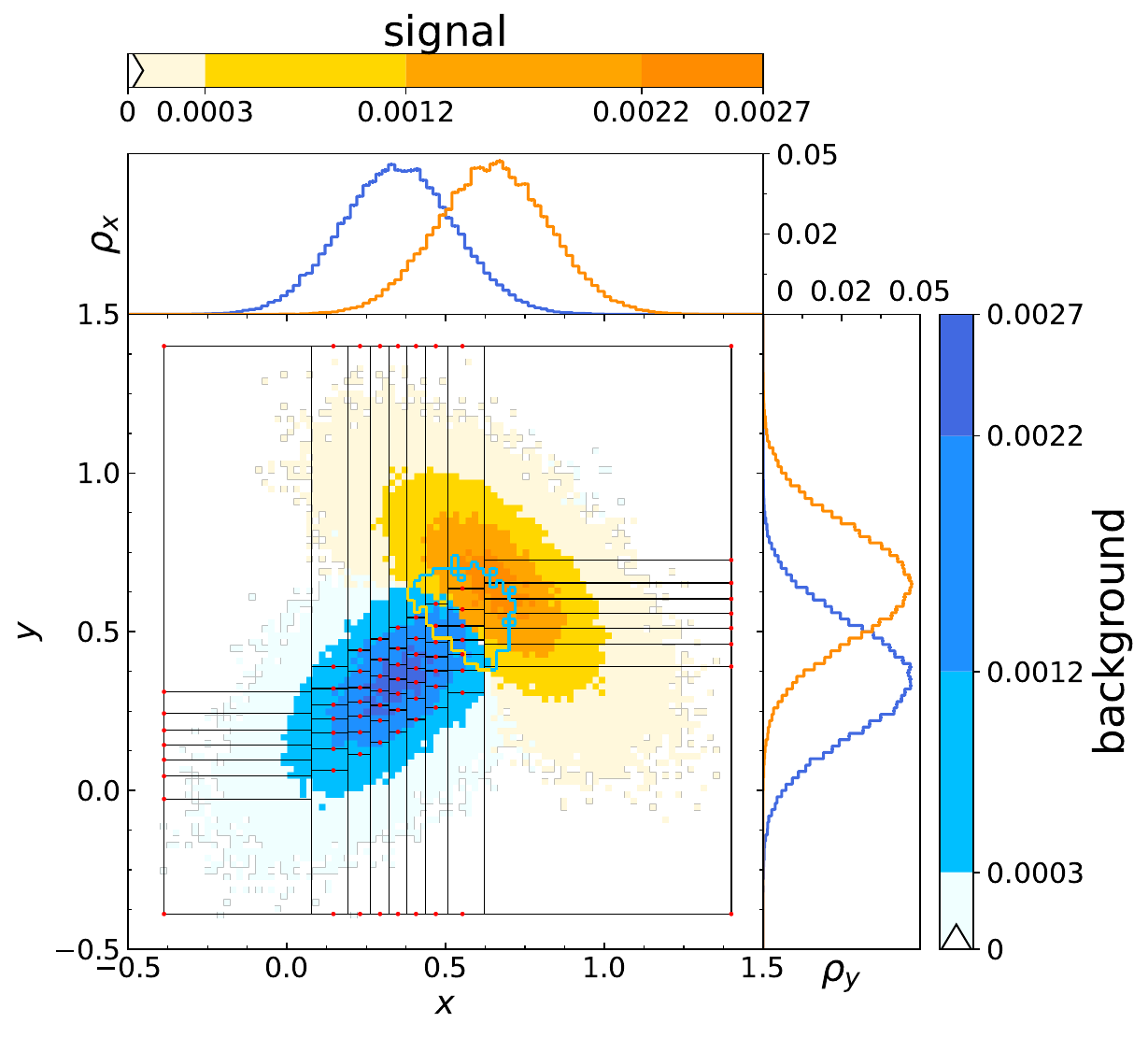}
    \caption{Defining background sample with irregular grid for $d = 3$,}
    \end{subfigure}
\quad
    \begin{subfigure}[t]{0.48\textwidth}
    \centering
    \includegraphics[width=0.99\textwidth]{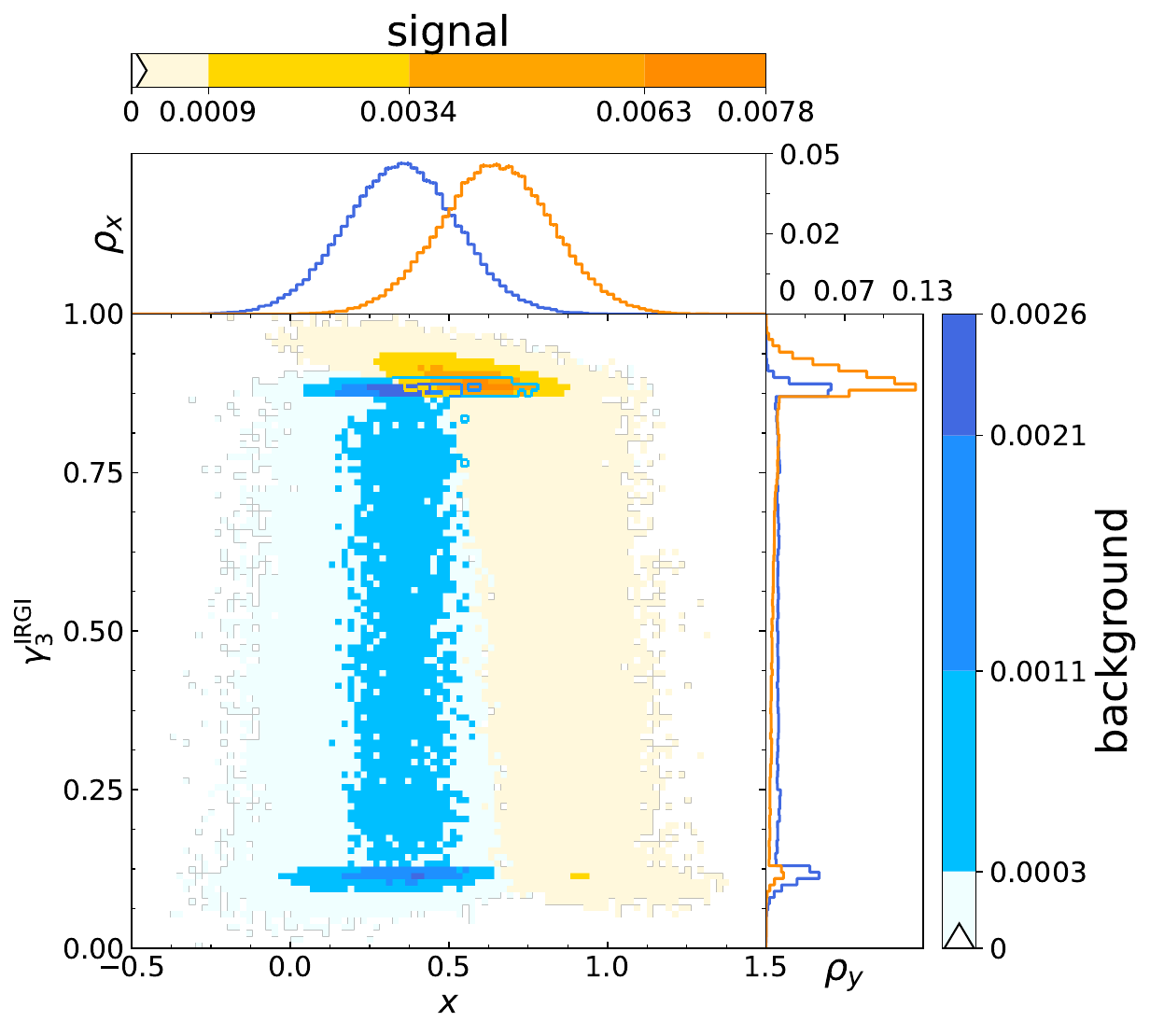}
    \caption{IRGI-modified classifier $\gamma_3$ applied to testing sample.}
    \end{subfigure}
\caption{IRGI technique, $d=3$.}
\label{fig:IRGI3A}
\end{figure}

\begin{figure}
\centering
    \begin{subfigure}[t]{0.48\textwidth}
    \centering
    \includegraphics[width=0.99\textwidth]{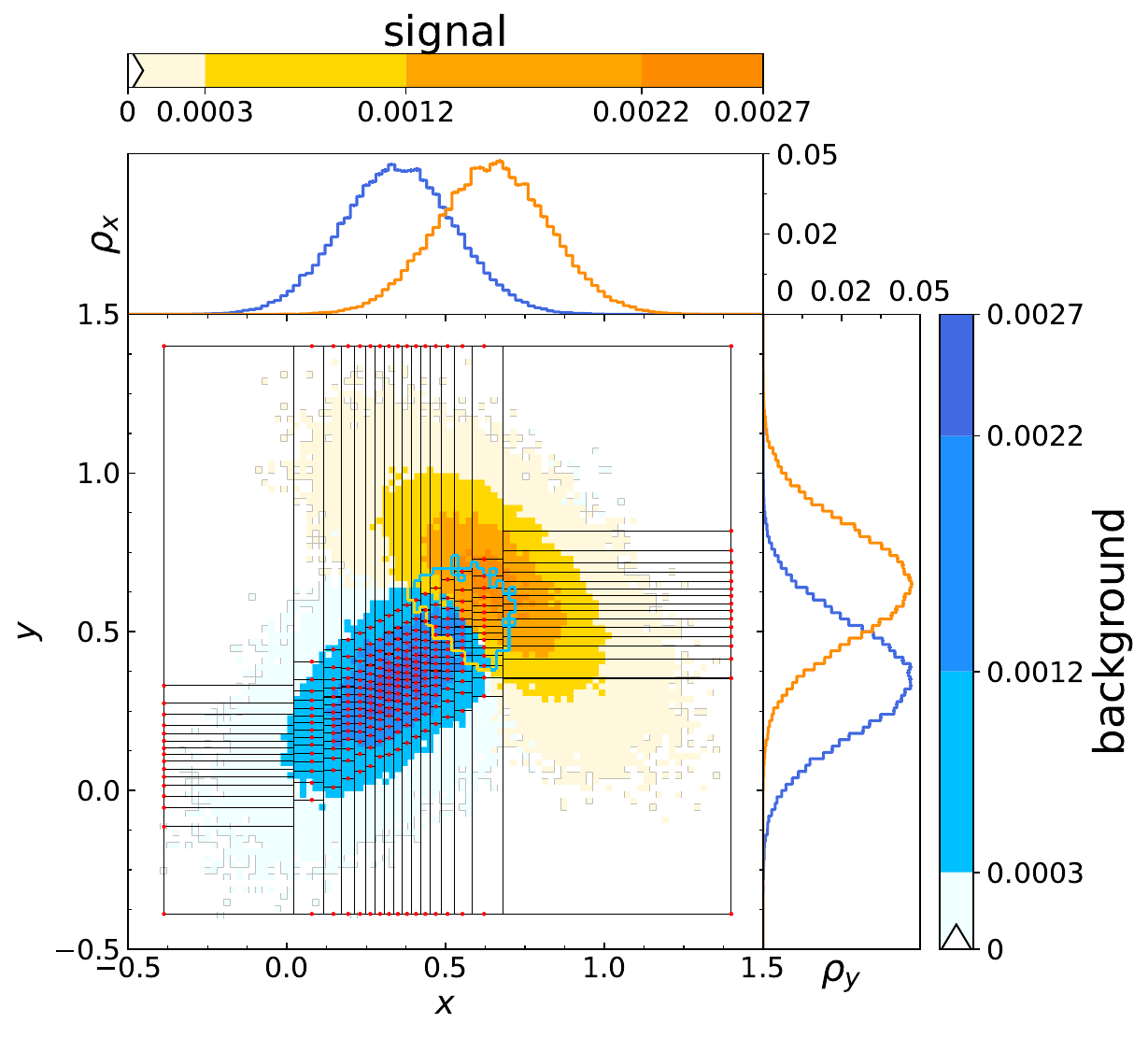}
    \caption{Defining background sample with irregular grid for $d = 4$,}
    \end{subfigure}
\quad
    \begin{subfigure}[t]{0.48\textwidth}
    \centering
    \includegraphics[width=0.99\textwidth]{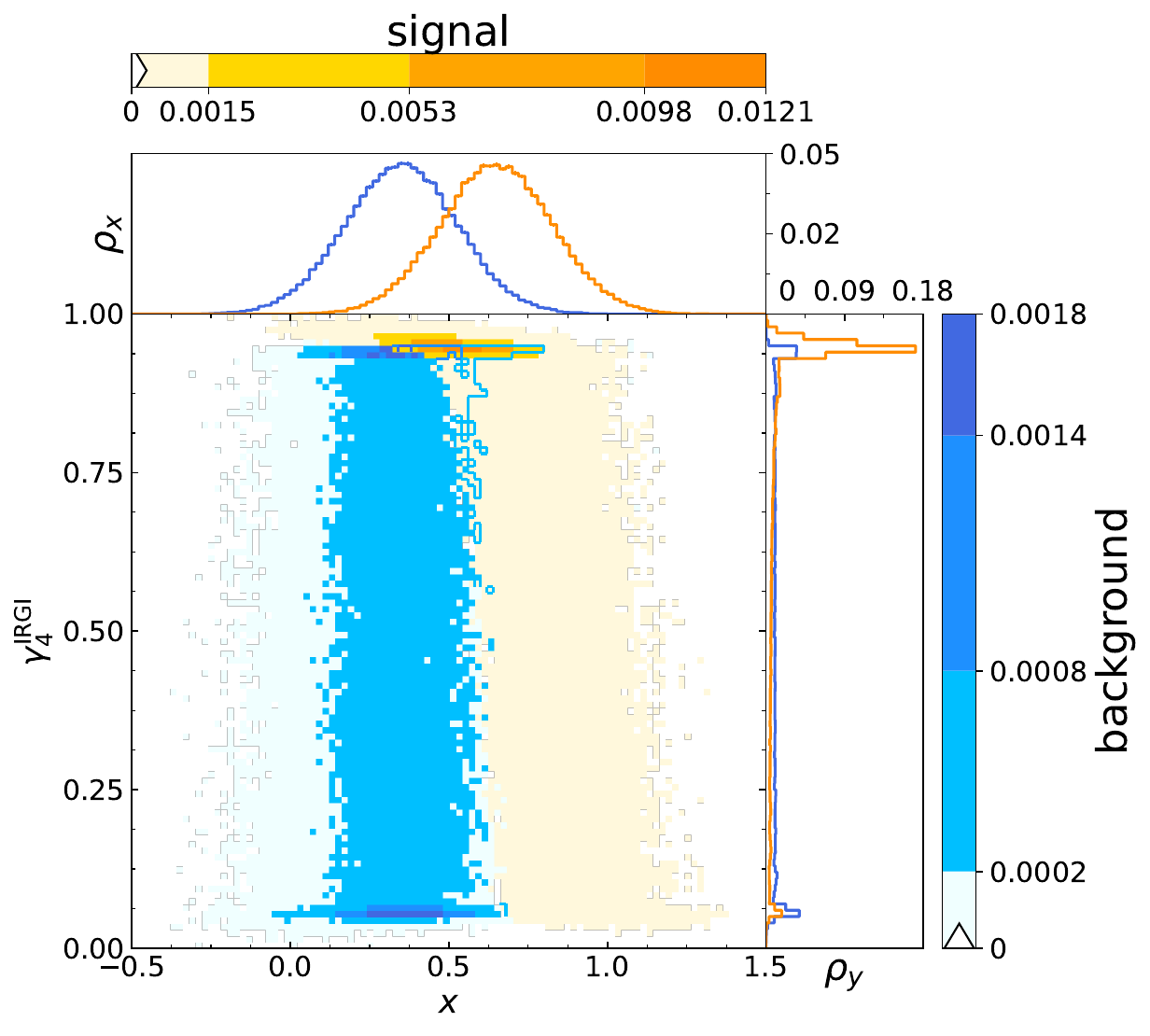}
    \caption{IRGI-modified classifier $\gamma_4$ applied to testing sample.}
    \end{subfigure}
\caption{IRGI technique, $d=4$.}
\label{fig:IRGI4A}
\end{figure}

\begin{figure}
\centering
    \begin{subfigure}[t]{0.48\textwidth}
    \centering
    \includegraphics[width=0.99\textwidth]{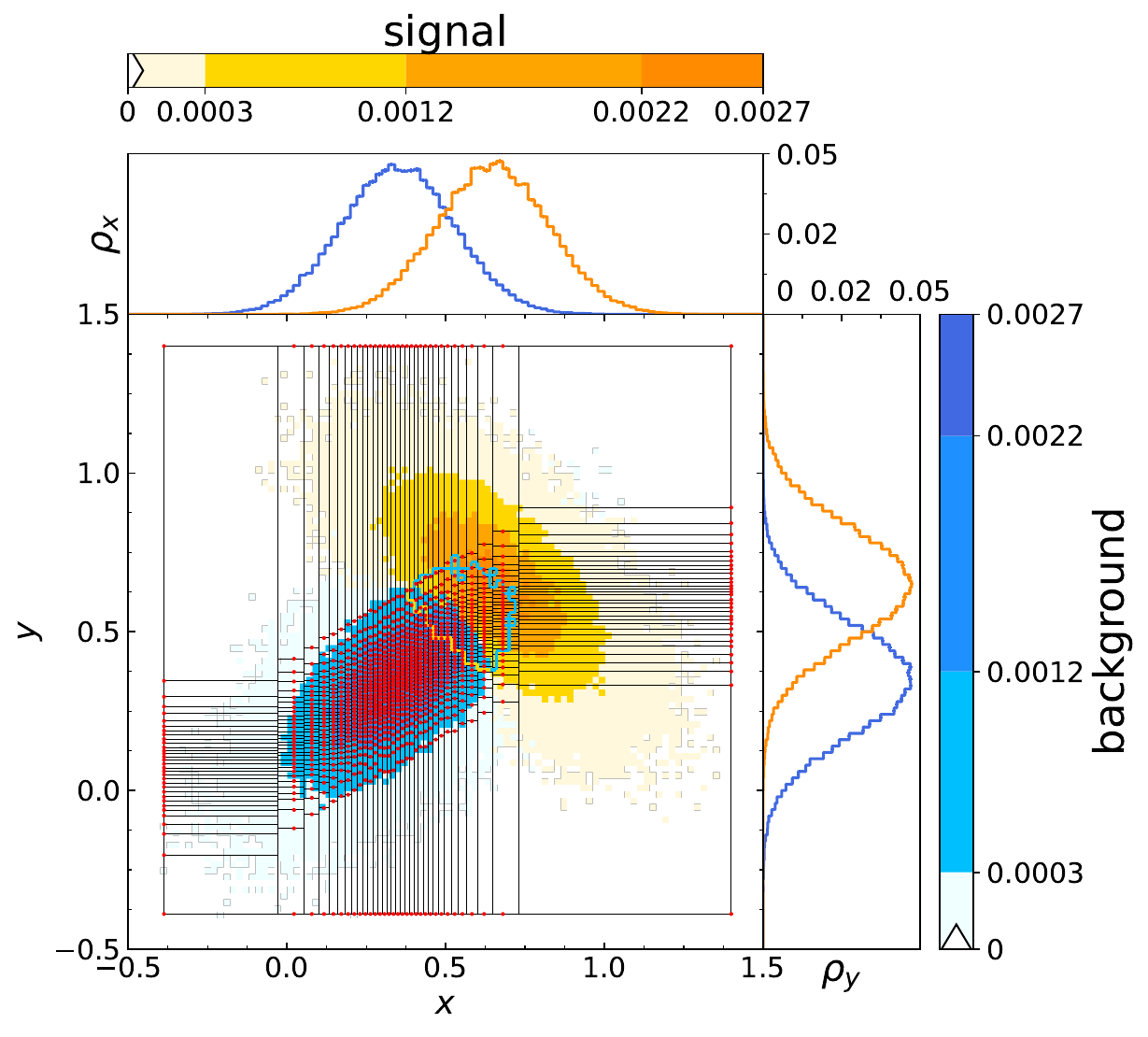}
    \caption{Defining background sample with irregular grid for $d = 5$,}
    \end{subfigure}
\quad
    \begin{subfigure}[t]{0.48\textwidth}
    \centering
    \includegraphics[width=0.99\textwidth]{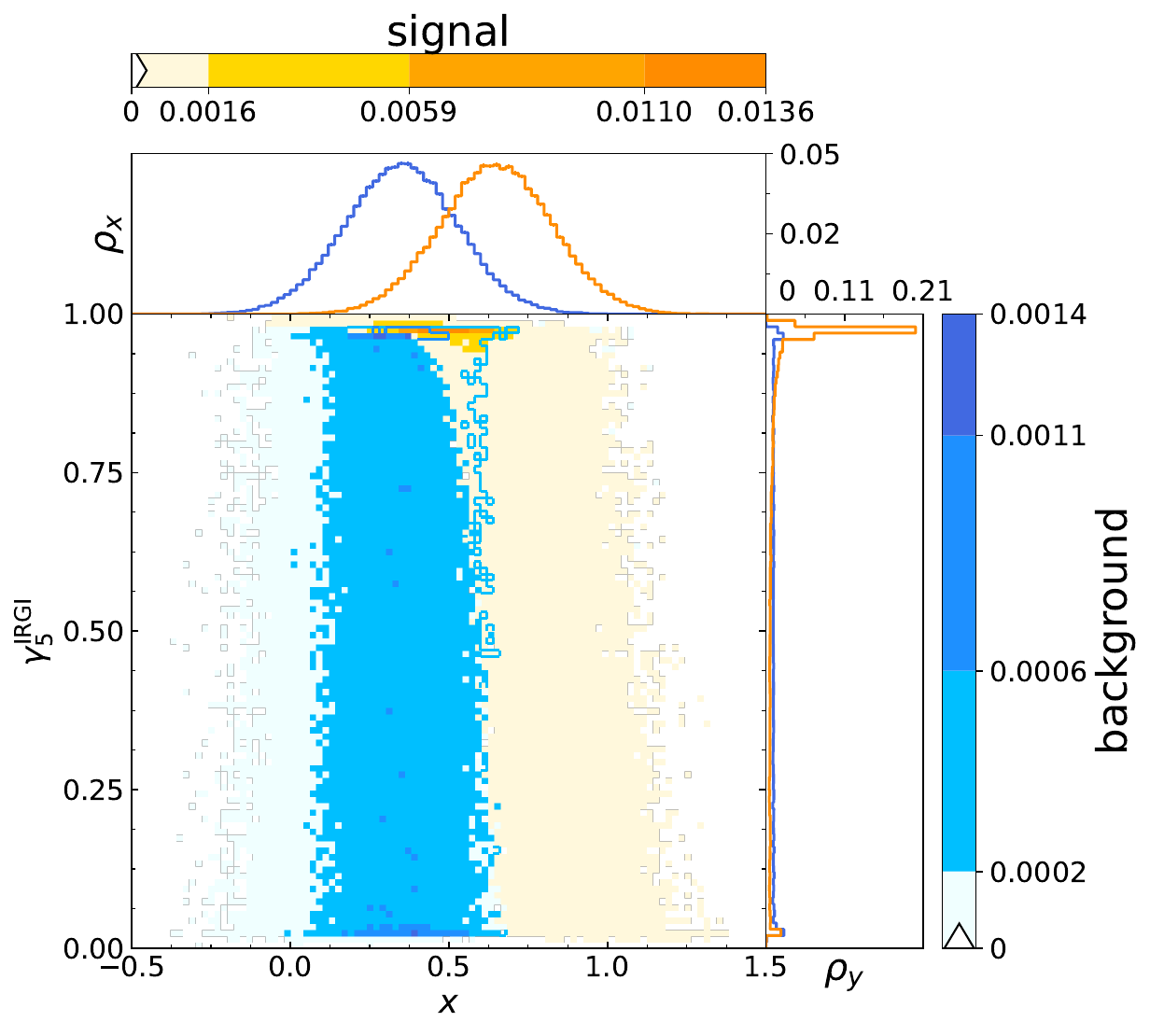}
    \caption{IRGI-modified classifier $\gamma_5$ applied to testing sample.}
    \end{subfigure}
\caption{IRGI technique, $d=5$.}
\label{fig:IRGI5A}
\end{figure}

\begin{figure}
\centering
    \begin{subfigure}[t]{0.48\textwidth}
    \centering
    \includegraphics[width=0.99\textwidth]{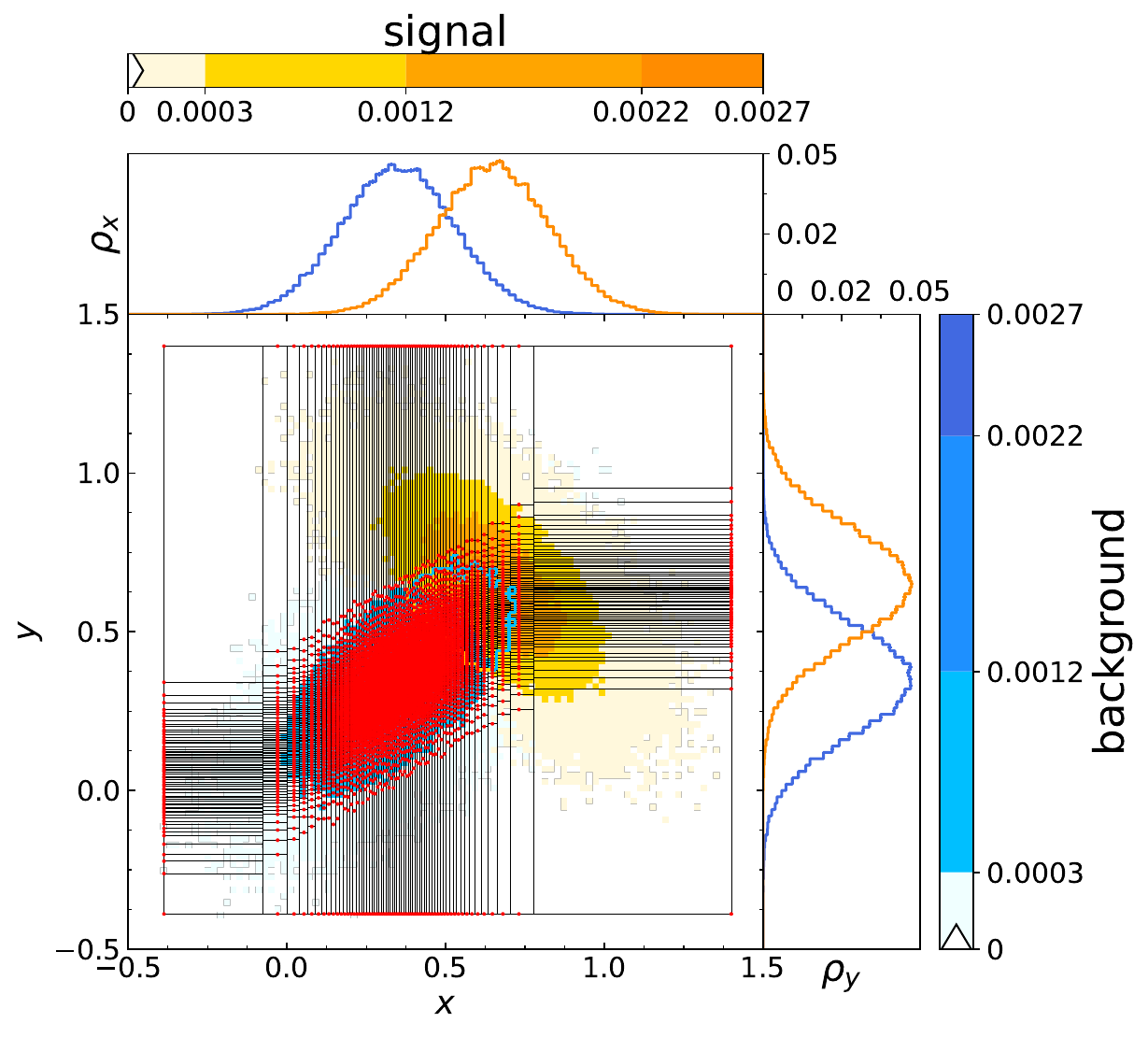}
    \caption{Defining background sample with irregular grid for $d = 6$,}
    \end{subfigure}
\quad
    \begin{subfigure}[t]{0.48\textwidth}
    \centering
    \includegraphics[width=0.99\textwidth]{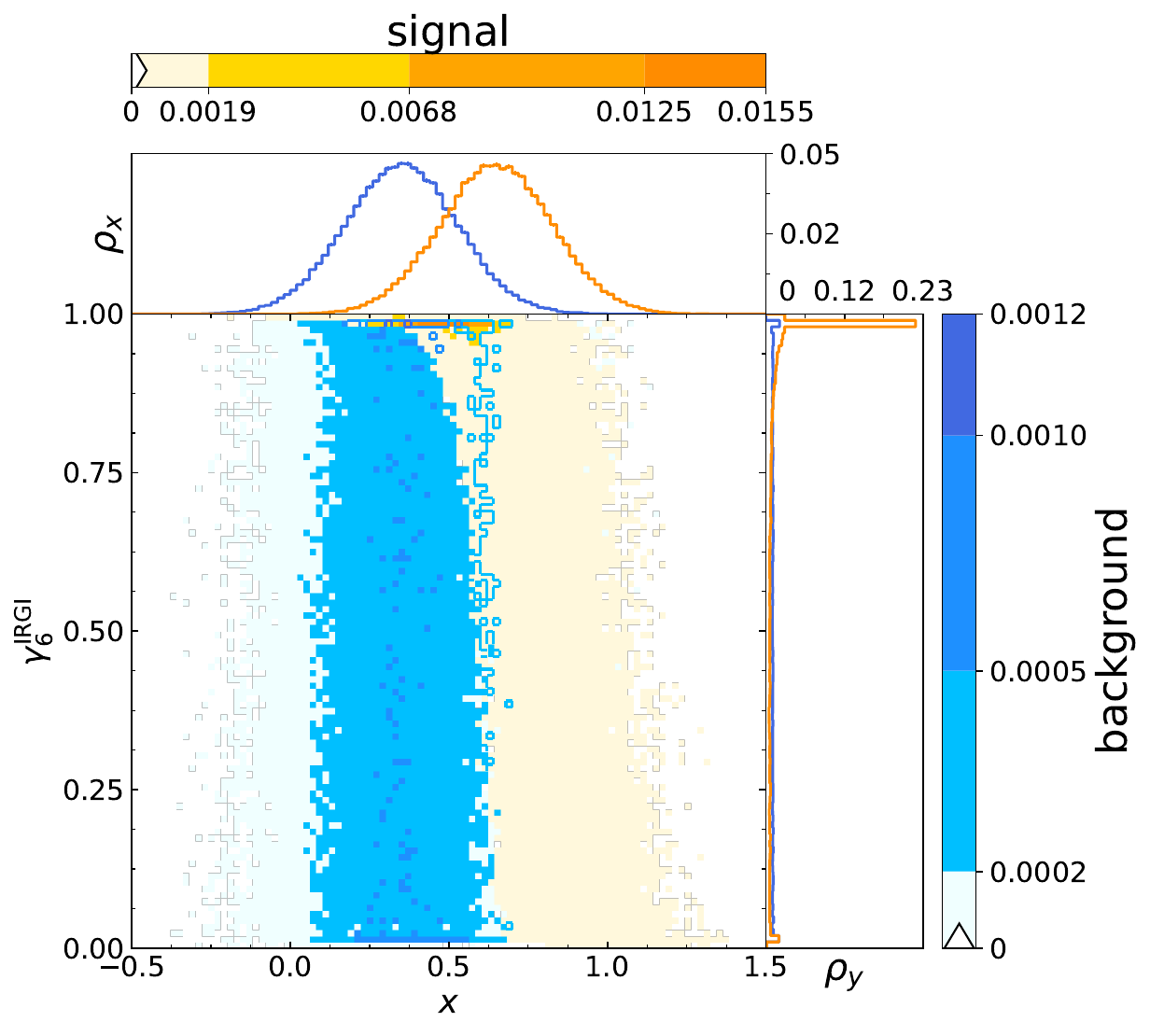}
    \caption{IRGI-modified classifier $\gamma_6$ applied to testing sample.}
    \end{subfigure}
\caption{IRGI technique, $d=6$.}
\label{fig:IRGI6A}
\end{figure}

\begin{figure}
\centering
    \begin{subfigure}[t]{0.48\textwidth}
    \centering
    \includegraphics[width=0.99\textwidth]{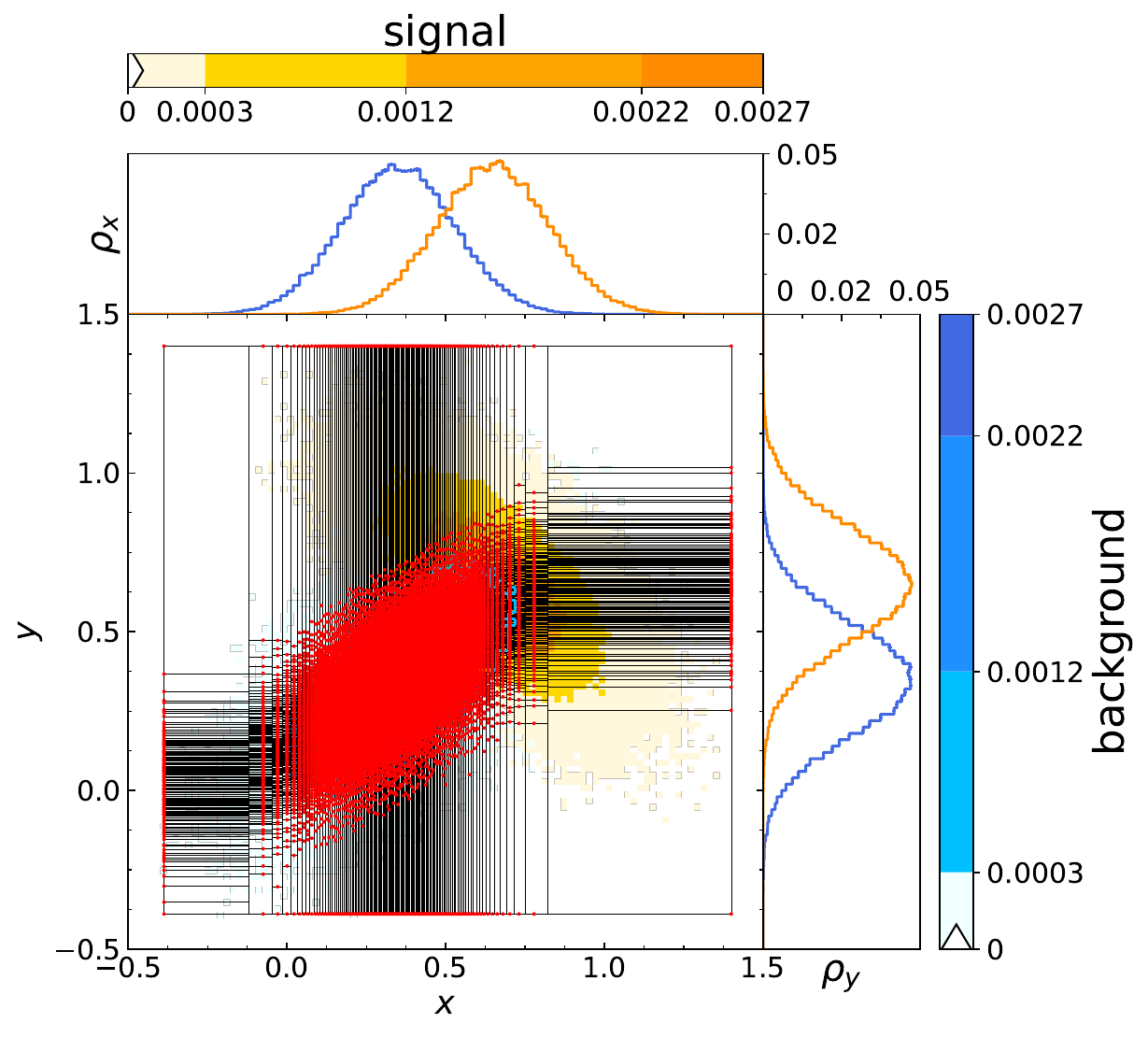}
    \caption{Defining background sample with irregular grid for $d = 7$,}
    \end{subfigure}
\quad
    \begin{subfigure}[t]{0.48\textwidth}
    \centering
    \includegraphics[width=0.99\textwidth]{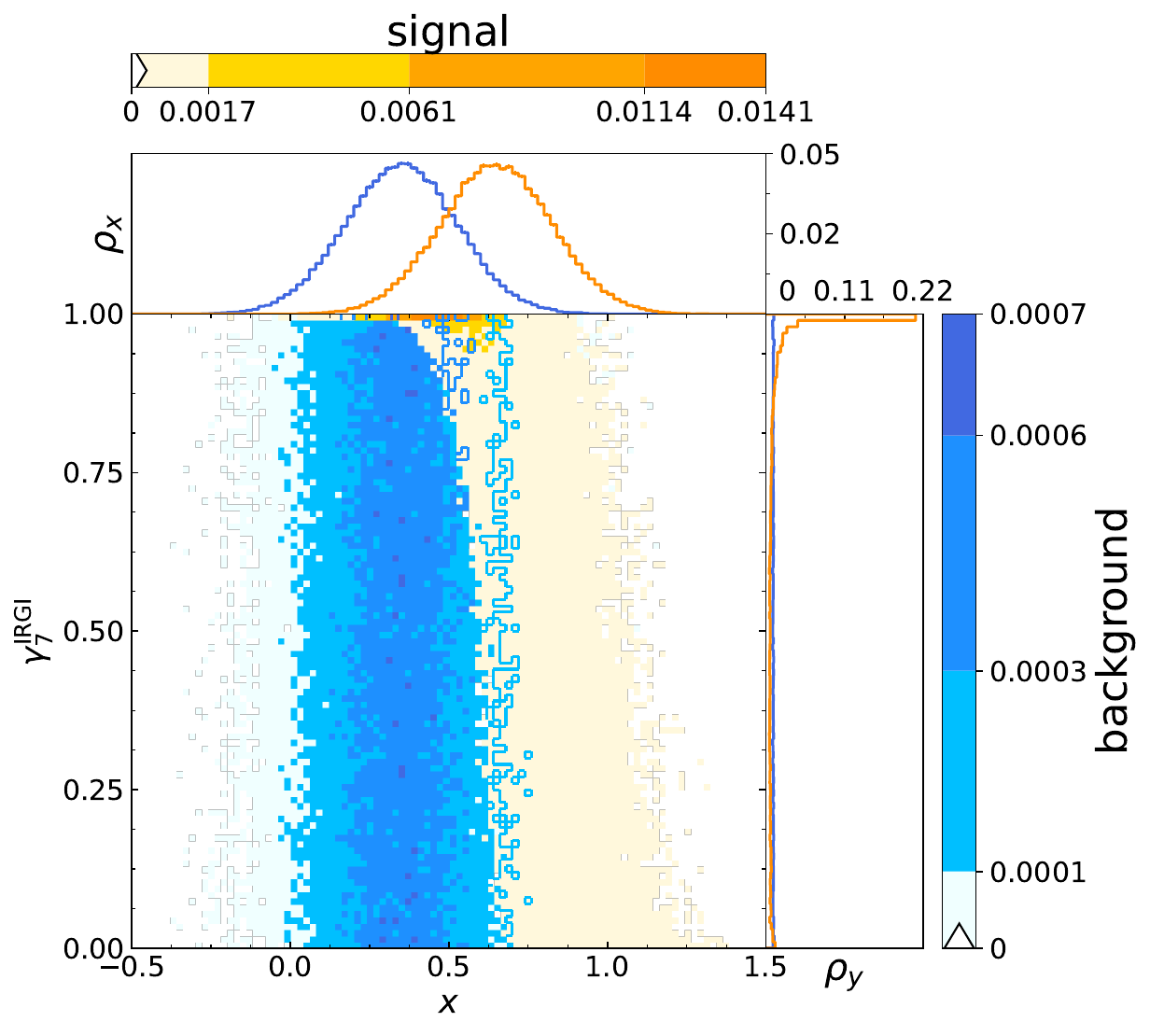}
    \caption{IRGI-modified classifier $\gamma_7$ applied to testing sample.}
    \end{subfigure}
\caption{IRGI technique, $d=7$.}
\label{fig:IRGI7A}
\end{figure}
\end{document}